\documentclass[aps,prd,showpacs,floatfix,twocolumn,nofootinbib,superscriptaddress]{revtex4-1} %
\usepackage{dcolumn}
\usepackage{amsmath,amsfonts,amssymb,mathrsfs,times,bm}
\usepackage{extarrows}
\usepackage{graphicx}
\usepackage{float}
\usepackage{booktabs}
\usepackage{graphicx,epstopdf}
\usepackage{dcolumn}
\usepackage{exscale}
\usepackage{relsize}
\usepackage[colorlinks=true,breaklinks=true,linkcolor=blue,citecolor=blue,urlcolor=blue]{hyperref}
\newcommand{\nn}{\nonumber}

\begin{document}

\title{Kerr-Newman black hole lensing of relativistic massive particles in the weak field limit}

\author{Guansheng He\hspace*{0.6pt}\href{https://orcid.org/0000-0002-6145-0449}{\includegraphics[scale=0.035]{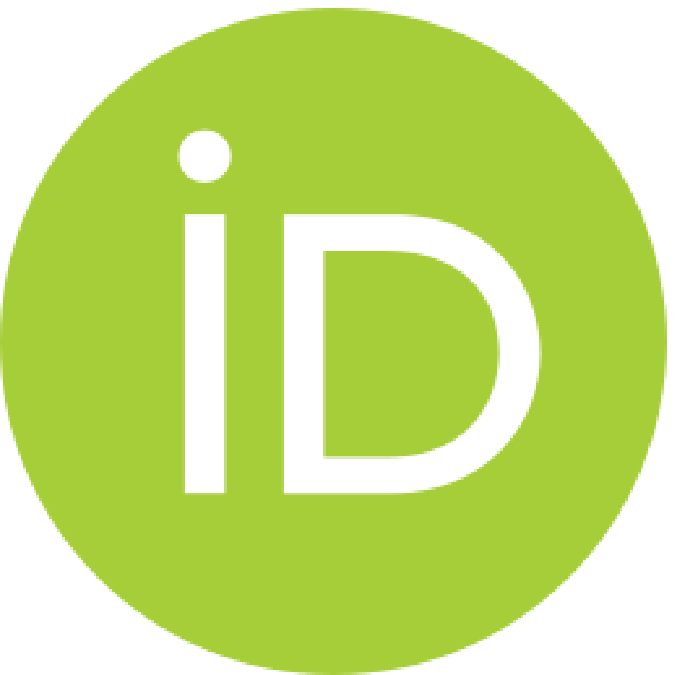}}\hspace*{0.6pt}}
\affiliation{School of Mathematics and Physics, University of South China, Hengyang, 421001, China}
\affiliation{Hunan Key Laboratory of Mathematical Modeling and Scientific Computing, Hengyang, 421001, China}
\author{Wenbin Lin\hspace*{0.6pt}\href{https://orcid.org/0000-0002-4282-066X}{\includegraphics[scale=0.035]{ORCID.eps}}\hspace*{0.6pt}}
\email{lwb@usc.edu.cn}
\affiliation{School of Mathematics and Physics, University of South China, Hengyang, 421001, China}
\affiliation{Hunan Key Laboratory of Mathematical Modeling and Scientific Computing, Hengyang, 421001, China}
\affiliation{School of Physical Science and Technology, Southwest Jiaotong University, Chengdu, 610031, China}

\date{\today}

\begin{abstract}
The gravitational lensing of relativistic neutral massive particles caused by a Kerr-Newman black hole is investigated systematically in the weak-field limit. Based on the Kerr-Newman metric in Boyer-Lindquist coordinates, we first derive the analytical form of the equatorial gravitational deflection angle of a massive particle in the third post-Minkowskian approximation. The resulting bending angle, which is found to be consistent with the result in the previous work, is adopted to solve the popular Virbhadra-Ellis lens equation. The analytical expressions for the main observable properties of the primary and secondary images of the particle source are thus obtained beyond the weak-deflection limit, within the framework of standard perturbation theory. The observables include the positions, magnifications, and gravitational time delays of the individual images, the differential time delay, and the total magnification and centroid position. The explicit forms of the correctional effects induced by the deviation of the initial velocity of the massive particle from the speed of light on the observables of the lensed images are then achieved. Finally, serving as an application of the formalism, the supermassive black hole at the Galactic center, Sagittarius A$^{\ast}$, is modeled to be a Kerr-Newman lens. The magnitudes of the velocity-induced correctional effects on the practical lensing observables as well as the possibilities to detect them in this scenario are also analyzed.

\end{abstract}


\maketitle

\section{Introduction}
Gravitational lensing is one of the most powerful tools in modern astrophysics and cosmology. It provides extensive astronomical applications (e.g., testing gravity theories~\cite{KP2005,KP2006a,KP2006b,Liu2016,MWS2020} and the cosmic censorship conjecture~\cite{WP2007}, determining the Hubble constant~\cite{Refsdal1964}, detecting dark matter~\cite{Wambsganss1998,JS2019} and dark energy~\cite{Hu2002,SKA2010,CCZ2012}, and constraining neutrino mass~\cite{LM2019,MLMHBN2019}), and has attracted much attention since the discovery of the first doubly imaged quasar in 1979~\cite{WCW1979}. Due to the traditional advantages of electromagnetic signals in astronomical observations, the previous works have been devoted mainly to the investigation of gravitational lensing phenomena of light by means of different approaches in the weak-field limit
(see, e.g.,~\cite{Sereno2004,WS2004,TKNA2014,JBO2019,CX2021}, and references therein) or the strong-field limit (see, for instance,~\cite{VE2000,BCIS2001,ERT2002,Bozza2002,VE2002,VK2008,Vi2009,CJ2009,WLFY2012,ZX2016,TG2017,LX2019,ZX2020}).

Actually, with the coming of multi-messenger astronomy, a full theoretical consideration of the gravitational lensing phenomena of a massive particle with a nonzero rest mass also deserves our effort, for which two reasons are responsible. The first one lies in the fact that the lensing effect of a massive particle (e.g., a neutrino or cosmic-ray particle) caused by a gravitational system may be more evident than the lightlike counterpart under the same circumstances. This is because the decrease of the velocity at infinity (the initial velocity) of a test particle leads to the increase of the total deflection angle for a given lens system~\cite{Silverman1980,AR2002}. This property of gravitational lensing of massive particles is of great significance to two aspects, which include increasing efficiently the opportunities to observe gravitational lensing events and making the consideration of the first-order, second-order, and even higher-order contributions to the lensing observables non-trivial. A second reason is that one can expect that the study of gravitational lensing of massive particles may speed up the advancement of joint multi-messenger observations (such as the joint neutrino and electromagnetic detection~\cite{MFHM2019,Keivani2018,IceCube2018}), since all of the messengers emitted by an astrophysical source may experience different gravitational bending processes before approaching their detectors.

To our knowledge, the previous literatures focused on the gravitational lensing of massive particles appear to be relatively rare, and most of them have been dedicated to the study of the weak- or strong-field gravitational deflection angle in various geometries (see, e.g.,~\cite{Silverman1980,AR2002,AR2003,WS2004,AP2004,BSN2007,PNH2014,HL2016,HL2017,CG2018,J2018,CGV2019,CGJ2019,JBGA2019,LZLH2019,LHZ2020,LJ2020a,LO2020}), which serves as one of the main parts of gravitational lensing. For example, Accioly and Ragusa~\cite{AR2002,AR2003} computed the gravitational deflection angle of a relativistic massive particle propagating in the Schwarzschild field, in the third post-Minkowskian (PM) approximation. It was not until recent years that the lens equation of massive particles was solved to obtain the observable properties of the lensed images. In 2016, Liu {\it et al.}~\cite{LYJ2016} based on the exact formula for the Schwarzschild deflection angle of a general massive particle~\cite{Tsupko2014} and solved the small angles lens equation~\cite{SEF1992} in the weak- and strong-field limits, respectively, to obtain the approximate angular positions and signed magnifications of the lensed images. The leading-order correctional effects caused by the deviation of the initial velocity of a massive particle from the speed of light on the deflection angle, angular image positions, and the magnifications for both ultrarelativistic and nonrelativistic particles were also discussed. The procedure of Ref.~\cite{LYJ2016} was later generalized to Reissner-Nordstr\"{o}m spacetime~\cite{PJ2019}. The authors of Ref.~\cite{PJ2019} obtained the timelike deflection angle in terms of an elliptical function, and investigated the first-order velocity-induced correctional effects on the deflection angle as well as the approximate image positions and magnifications for ultrarelativistic and nonrelativistic particles in the weak- and strong-field limits. Recently, the timelike time delay in Schwarzschild geometry was studied in detail in Ref.~\cite{JL2019}, where the differential time delay of the lensed images of a particle source in the 1PM approximation and the first-order velocity effect on it were discussed, on the basis of the exact total coordinate time of a massive particle. More recently, the series expansion form of the total propagating time of a test particle in a stationary axisymmetric spacetime, as well as the leading-order timelike differential time delay of the images and the first-order velocity effect on it, were derived in Ref.~\cite{LJ2020b}. There are also other works devoted to the study of some of the observable properties of the lensed images of a massive-particle source (e.g.,~\cite{BT2017,JH2021,LLJ2021}).

However, it seems fair to mention that further work is necessary with respect to the issue of gravitational lensing of massive particles. A first reason is that there is still a lack of a systematical consideration of the first-order, second-order, and even higher-order contributions to all of the main observables of the images in the lensing scenario of massive particles. It is of interest, since the velocity-induced effect on an image observable in or beyond the weak-deflection limit may be so evident that its magnitude is much larger than that of the corresponding null observable, while the general behaviors of the velocity effects on the lensing observables have not been analyzed up to now. In fact, the consideration of the velocity effects themselves~\cite{WS2004} is also a significant component of gravitational lensing of massive particles. Furthermore, we know that rapid progress in techniques of position, time, angular, and photometric measurements has been made in the past decades. The high-accuracy angular measurement in astronomical projects is nowadays at the level of $1\!\sim\!10\,\mu$arcsec ($\mu$as) or even better~\cite{Perryman2001,Prusti2016,SN2009,Chen2014,Malbet2012,Malbet2014,Trippe2010}. Especially, the planned Nearby Earth Astrometric Telescope (NEAT) mission~\cite{Malbet2012,Malbet2014} aims at an unprecedented space-borne accuracy of $0.05\,\mu$as. Additionally, the recent photometric precision has been at the level of about $10\,\mu$mag or better~\cite{Koch2010,Chapellier2011,BK2018,Kurtz2005}. For instance, the original \emph{Kepler} Mission has an extreme photometric precision of a few $\mu$mag~\cite{Koch2010,BK2018}, although it ended prematurely due to the failure of one of four reaction wheels in 2013. It has been renamed as the \emph{K2} mission with new purposes and a lower photometric precision (within a factor of two of the nominal \emph{Kepler} performance)~\cite{VJ2014,AHILR2015,Huber2016}. Moreover, the present precision of Very Long Baseline Interferometry (VLBI) technique~\cite{Burke1969,Rogers1970,TMS1986,Hirabayashi1998,Ma1998,HKS2000,K2001} in measuring the differential time delay is at the level of $10^{-12}$\,s (ps) at least. The proposed delay precision of the next-generation VLBI system is 4ps~\cite{Petra2009,SB2012,Niell2018}. It can be expected that the first- and second-order contributions (even higher-order contributions) to the observable properties of the lensed images, as well as the velocity effects on them, may be detectable in current (or near future) high-accuracy astronomical measurements.

In present work, we adopt the standard perturbative analysis to investigate in detail the weak-field gravitational lensing of relativistic massive particles induced by a Kerr-Newman (KN) black hole, which acts as a natural extension of the previous works~\cite{KP2005,AKP2011}. First, we calculate analytically the gravitational deflection angle of a massive particle propagating in the equatorial plane of the KN source up to the 3PM order in Boyer-Lindquist coordinates, via an approach which is different from that in Ref.~\cite{HJ2020}. The deflection angle is then utilized to solve the popular Virbhadra-Ellis lens equation~\cite{VE2000}. The explicit forms for the main observable properties of the primary and secondary images, which include the positions, magnifications, and gravitational time delays of the individual images, the sum and difference relations of the image positions or magnifications, the differential time delay, along with the magnification-weighted centroid position, are thus achieved beyond the weak-deflection limit. The analytical expressions of the velocity effects on the zeroth-, first-, or second-order contribution to the image observables are also obtained in the weak-field limit. As an application of the formalism, we model the supermassive black hole at the Galactic center (i.e., Sagittarius A$^{\ast}$)~\cite{EG1997,NMGPG1998,BS1999} as a KN lens, and analyze in detail the magnitudes of the velocity effects on the practical lensing observables and the possibilities of their detection. Our discussions are restricted in the weak-field, small-angle, and thin-lens approximation~\cite{WS2004,Wambsganss1998}.

The organization of this paper is as follows. Section~\ref{sect2} gives the basic notations and assumptions used in this work. In Section~\ref{sect3}, we first review the KN metric in Boyer-Lindquist coordinates, and then derive the gravitational deflection angle of a relativistic massive particle propagating in the equatorial plane of the lens up to the 3PM order. Section~\ref{sect4} is devoted to obtaining the weak-field expressions of the timelike observable properties of the lensed images via solving the Virbhadra-Ellis lens equation, on the basis of the standard perturbation theory analysis. Section~\ref{sect5} presents the analytical forms of the velocity effects induced by the deviation of the initial velocity of the particle from the speed of light on the observables of the lensed images beyond the weak-deflection limit. In Section~\ref{sect6}, the Galactic supermassive black hole is modeled to be a KN lens, and the magnitudes of the velocity effects as well as the possibilities to detect them are analyzed. A summary is given in Section~\ref{sect7}. Conventionally, Greek indices run over $0,~1,~2$, and $3$.

\section{Notations and assumptions} \label{sect2}
In this paper, geometrized units where $G\!=\!c\!=\!1$ and the metric signature $(+,~-,~-,~-)$ are used. $\{\bm{e}_1,~\bm{e}_2,~\bm{e}_3\}$ denotes the orthonormal basis of a three-dimensional Cartesian coordinate system $(x,~y,~z)$, whose origin is located at the barycenter of the central body. For the sake of simplicity, the massive particle is assumed to be neutral in this work.

\begin{figure}[t]
\centering
\includegraphics[width=\linewidth]{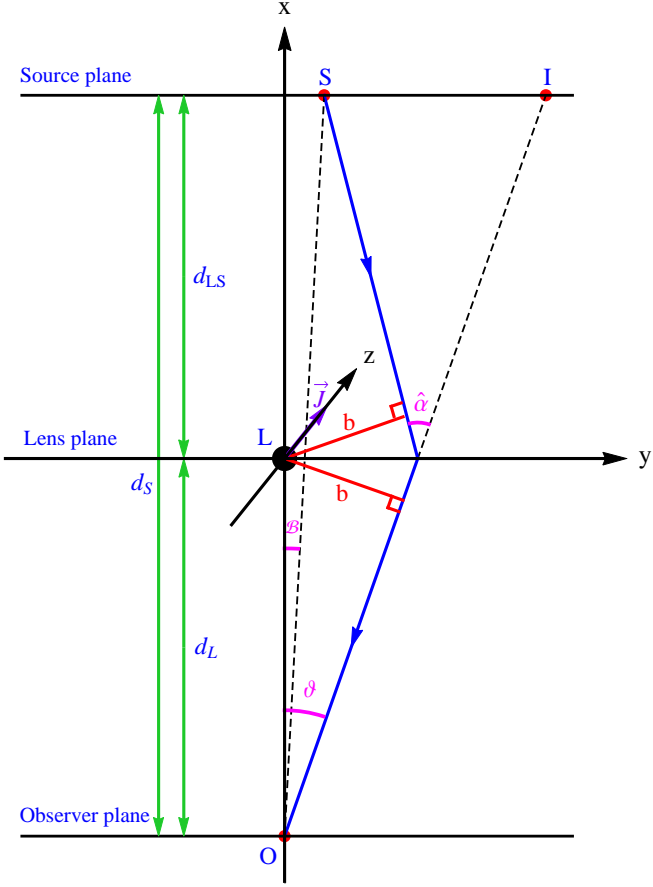}
\caption{The lens diagram of a Kerr-Newman black hole. The positions of the source, lens, observer, and image are given by $S,~L,~O$, and $I$, respectively. All of them are situated in the equatorial plane ($x\!-\!y$ plane) of the lens. The $x$ axis is assumed to be the optic axis $OL$ which joins the lens and observer. $d_{S}$ and $d_{L}$ are the angular diameter distances of the source and lens from the observer, respectively, and $d_{LS}$ is the angular diameter distance of the source from the lens. $\mathcal{B}$ and $\vartheta$ denote respectively the angular source and image positions. $\hat{\alpha}$ is the gravitational deflection angle of the massive particle. $b~(=d_L\sin\vartheta)$ denotes the impact parameter. Without loss of generality, the intrinsic angular momentum vector $\bm{J}=J\,\bm{e}_3$ of the gravitational lens is assumed to be along the positive $z$ axis ($J>0$).  }   \label{Figure1}
\end{figure}

We focus on the scenario where a relativistic massive particle with an initial velocity $w~(>0)$, emitted by the source, is deflected by the lens and propagates to the observer without looping around the lens (i.e., no relativistic images appear). The lens diagram is shown in Fig.~\ref{Figure1}, where the notations for the main lens quantities are given. In the weak-field and thin-lens approximation mentioned above, we can assume the deflection effect takes place in a cosmologically small region around the lens. Thus, the observer and source are regarded to be located in the asymptotically flat region of the KN geometry, and the propagating path of the test particle is approximated by its two asymptotes (the blue lines in Fig.~\ref{Figure1})~\cite{Wambsganss1998}. Furthermore, as done in Ref.~\cite{KP2005}, we adopt the assumption that the angular positions of the lensed images are positive. It implies that the position $\mathcal{B}$ of the source is positive when the image is on the same side of the lens as the source, and negative when the image is on the opposite side.

\section{Weak gravitational deflection of massive particles} \label{sect3}
In this section, we consider the gravitational deflection of a relativistic massive particle propagating in the equatorial plane of a KN black hole, within the 3PM approximation.

\subsection{The Kerr-Newman metric}
The metric of the KN spacetime in Boyer-Lindquist coordinates $(t,~r,~\zeta,~\varphi)$ is given by~\cite{BL1967,AN2014}
\begin{eqnarray}
&&\nn ds^2=\frac{\Delta}{\rho^2}\left(dt-a\sin^2\zeta\,d\varphi\right)^2\!-\!\frac{\sin^2\zeta}{\rho^2}\!\left[(r^2+a^2)d\varphi-a dt\right]^2  \\
&&\hspace*{28pt}-\frac{\rho^2}{\Delta}dr^2-\rho^2d\zeta^2~,  \label{Metric}
\end{eqnarray}
where $\Delta=r^2+a^2-2Mr+Q^2$ and $\rho^2=r^2+a^2\cos^2\zeta$. $M$, $Q$, and $a\equiv J/M~(>0)$ denote the rest mass, electrical charge, and angular momentum per unit mass of the KN black hole, respectively. We use the relation $a^2+Q^2\leq M^2$ to avoid the naked singularity of the black hole.

\subsection{Equations of motion}
The geodesic equation of a test particle in a given spacetime geometry is equivalent to the Euler-Lagrangian equation with the Lagrangian $\mathcal{L}=\frac{1}{2}g_{\mu\nu}\dot{x}^\mu\dot{x}^\nu$~\cite{St1984}, which reads for the equatorial motion ($\zeta=\pi/2$) in KN spacetime:
\begin{eqnarray}
&&\nn 2\mathcal{L}=\left(1-\frac{2Mr-Q^2}{r^2}\right)\dot{t}^2-\frac{r^2}{\Delta}\dot{r}^2   \\
&&\hspace*{28pt} -\frac{\left(r^2+a^2\right)^2-a^2\Delta}{r^2}\dot{\varphi}^2+\frac{2\,a\left(2Mr-Q^2\right)}{r^2}\dot{t}\,\dot{\varphi}~,~~~~\label{L1}
\end{eqnarray}
where a dot denotes the derivative with respect to the affine parameter $\hat{\xi}$ which describes the trajectory~\cite{WS2004,We1972}. Along the particle's orbit, we have $2\mathcal{L}=1$. Two constants of motion can be then obtained from Eq.~\eqref{L1} as follows~\cite{AR2002}:
\begin{eqnarray}
&&\nn E\equiv\frac{\partial\mathcal{L}}{\partial\dot{t}}=\left(1-\frac{2Mr-Q^2}{r^2}\right)\dot{t}+\frac{a\left(2Mr-Q^2\right)}{r^2}\,\dot{\varphi}  \\
&&\hspace*{39pt}=\frac{1}{\sqrt{1-w^2}}~, \label{C1}  \\
&&\nn \hat{L}\equiv-\frac{\partial\mathcal{L}}{\partial\dot{\varphi}}=\frac{\left(r^2\!+\!a^2\right)^2\!-\!a^2\Delta}{r^2}\,\dot{\varphi}-\frac{a\left(2Mr-Q^2\right)}{r^2}\,\dot{t}  \\
&&\hspace*{46pt}=\frac{s\,w\,b}{\sqrt{1-w^2}}~.~~~~\label{C2}
\end{eqnarray}
Here, $E$ and $\hat{L}$ represent the conserved orbital energy and angular momentum per unit mass, respectively. The impact parameter $b$ is defined by $|\hat{L}|/E\equiv w\,b$~\cite{LYJ2016,CG2018,JBGA2019},
which is in accord with its definition $b\equiv|\hat{L}|/E$ for null geodesics $(w=1)$. Moreover, for a given intrinsic angular momentum $J~(>0)$ of the KN source, $\hat{L}$ is positive when the massive particle takes prograde motion relative to the rotation of the lens, while it is negative for retrograde motion of the particle. We thus follow the idea of Ref.~\cite{AKP2011} to define the sign of $\hat{L}$ by the sign parameter $s$ as follows:
\begin{eqnarray}
\hspace*{10pt} s\equiv \hbox{sign}(\hat{L})=\left\{
\begin{array}{ll}
+1, & \hbox{for prograde motion}      \\
-1, & \hbox{for retrograde motion}
\end{array}
\right. .~~~
\end{eqnarray}

According to Eqs.~\eqref{C1} - \eqref{C2}, we have
\begin{eqnarray}
&&\dot{t}=\frac{\left[(r^2+a^2)^2-a^2\Delta\right]E+a\hat{L}\left(Q^2-2Mr\right)}{r^2\Delta}~, \label{dot-t}  \\
&&\dot{\varphi}=\frac{\left(\Delta-a^2\right)(\hat{L}-aE)+aEr^2}{r^2\Delta}~.~~~~~~\label{dot-varphi}
\end{eqnarray}
The substitution of Eqs.~\eqref{dot-t} - \eqref{dot-varphi} into Eq.~\eqref{L1} yields
\begin{equation}
r^4\dot{r}^2=\left[(r^2+a^2)E-a\hat{L}\right]^2-\left[(\hat{L}-aE)^2+r^2\right]\Delta~.  \label{dot-r}
\end{equation}
Eqs.~\eqref{dot-t} - \eqref{dot-r} are consistent with the results in Refs.~\cite{Sultana2013,Arakida2021} for the case of no electrical charge and $w=1$.

\subsection{Equatorial gravitational deflection angle of a massive particle up to the 3PM order}
We utilize the approach given in Ref.~\cite{KP2005} to perform our calculation of the 3PM equatorial deflection angle of a massive particle. It should be pointed out that the weak-field and small-angle approximation enables us to make the PM series expansion for the deflection angle:
\begin{equation}
\hat{\alpha}=\sum\limits_{i=1}^3 N_i\left(\frac{M}{b}\right)^i+\mathcal{O}(M^4)~, \label{alpha-general}
\end{equation}
where the coefficients $N_i$ are the unknown functions of $w$, $a$, and $Q$, and $M/b\ll1$.

The first thing to obtain the explicit form of $\hat{\alpha}$ is the determination of the 3PM relation between $b$ and $r_0$, with $r_0$ being the distance of closest approach to the lens for the particle. We know $\dot{r}$ in Eq.~\eqref{dot-r} at the distance $r=r_0$ should vanish, and it implies
{\small
\begin{eqnarray}
&&\nn b=\frac{r_0}{w\left(1-\frac{2M}{r_0}+\frac{Q^2}{r_0^2}\right)}\!\times\!\Bigg{\{}\!\!\left(1-\frac{2M}{r_0}+\frac{a^2+Q^2}{r_0^2}\right)^{1/2}  \\
&&\hspace*{16.5pt}\times\!\left[w^2\!+\!(1\!-\!w^2)\!\left(\frac{2M}{r_0}\!-\!\frac{Q^2}{r_0^2}\right)\right]^{1/2}
\!\!-\!\frac{sa}{r_0}\!\left(\frac{2M}{r_0}\!-\!\frac{Q^2}{r_0^2}\right)\!\!\Bigg{\}}~,~~~~~~ \label{b-r0}
\end{eqnarray}}
where we have omitted the other solution which is nonphysical, and $M/r_0$ is much smaller than $1$ to guarantee a weak field.
By defining
\begin{equation}
h\equiv M/r_0~,~~~~\hat{a}\equiv a/M~,~~~~\hat{Q}\equiv Q/M~,~~ \label{haQ}
\end{equation}
and using the series expansion of Eq.~\eqref{b-r0} in $h$, we find up to 3PM order
\begin{equation}
b=r_0\left[1+A_1h+A_2h^2+A_3h^3+\mathcal{O}(M^4)\right]~, \label{b-r0-2}
\end{equation}
where
\begin{eqnarray}
&&A_1=\frac{1}{w^2}~,                                                             \label{A1}  \\
&&A_2=\frac{4w^2-1-4w^3s\,\hat{a}+w^4\hat{a}^2-w^2\hat{Q}^2}{2w^4}~,              \label{A2}  \\
\nn &&A_3=\frac{1}{2w^6}\Big{[}1-4w^2+8w^4-8w^5s\,\hat{a}+(1+2w^2)w^4\hat{a}^2                \\
&&\hspace*{26pt}+(1-4w^2)w^2\hat{Q}^2+2w^5s\,\hat{a}\,\hat{Q}^2\Big{]}~.          \label{A3}
\end{eqnarray}
However, we want to express the deflection angle in terms of the invariant impact parameter. In order to express $r_0$ in terms of $b$, we guess reasonably that the series expansion of $r_0$ in $M/b$
takes the following form:
\begin{eqnarray}
r_0=b\!\left[1+a_1\frac{M}{b}\!+a_2\!\left(\frac{M}{b}\right)^2\!\!+a_3\!\left(\frac{M}{b}\right)^3\!\!+\!\mathcal{O}(M^4)\right] ,~~~~~~ \label{b-r0-3}
\end{eqnarray}
with $a_i~(i=1,~2,~3)$ being undetermined coefficients. By substituting Eq.~\eqref{b-r0-3} into Eq.~\eqref{b-r0-2} conversely and requiring the first- and higher-order terms on the right-hand side of Eq.~\eqref{b-r0-2} to vanish, we find
\begin{eqnarray}
&&a_1=-\frac{1}{w^2}~,                                                             \label{a1}  \\
&&a_2=-\frac{4w^2-1-4w^3s\,\hat{a}+w^4\hat{a}^2-w^2\hat{Q}^2}{2w^4}~,              \label{a2}  \\
\nn&&a_3=-\frac{2w(2-\hat{Q}^2)\!-\!s\,\hat{a}\!\left[2+w^2(4-\hat{Q}^2)\right]\!+\!(w+w^3)\hat{a}^2}{w^3}~. \\  \label{a3}
\end{eqnarray}

We now turn to the exact expression of the bending angle, which can be written via Eqs.~\eqref{dot-varphi} - \eqref{dot-r} as follows~\cite{We1972,KP2005}:
\begin{widetext}
\begin{eqnarray}
\nn&&\hat{\alpha}=2\int_{r_0}^{+\infty}\left|\frac{d\varphi}{dr}\right|dr-\pi       \\
&&\hspace*{9pt} =2\int_{r_0}^{+\infty}\!\!\frac{\left(1-\frac{2M}{r}+\frac{Q^2}{r^2}\right)\left(w-\frac{sa}{b}\right)+\frac{sa}{b}}
{r^2\left(1-\frac{2M}{r}+\frac{a^2+Q^2}{r^2}\right)\sqrt{\frac{1}{b^2}\left(1+\frac{a^2}{r^2}-\frac{sawb}{r^2}\right)^2
-\left(1-\frac{2M}{r}+\frac{a^2+Q^2}{r^2}\right)\!\left[\frac{1-w^2}{b^2}+\frac{1}{r^2}\left(w-\frac{sa}{b}\right)^2\right]}}dr-\pi~.~~~~~   \label{alpha-exact-1}
\end{eqnarray}
Eq.~\eqref{alpha-exact-1} can be rewritten by defining a new variable $x\equiv r_0/r~(0\leq x\leq1)$ in the form

\begin{eqnarray}
\hat{\alpha}=2\int_{0}^{1}\frac{\left(1-2hx+\hat{Q}^2h^2x^2\right)\left(\frac{wb}{r_0}-s\hat{a}h\right)+s\hat{a}h}{\left[1-2hx+\left(\hat{a}^2+\hat{Q}^2\right)h^2x^2\right]\sqrt{H}}dx-\pi~, \label{alpha-exact-2}
\end{eqnarray}
where $h,~\hat{a}$, and $\hat{Q}$ have been defined above, and $H$ and $b/r_0$ are given as follows:
\begin{eqnarray}
&&H=\left(1+\hat{a}^2h^2x^2-\frac{wb}{r_0}s\hat{a}hx^2\right)^2-\left[1-2hx+\left(\hat{a}^2+\hat{Q}^2\right)h^2x^2\right]\left[1-w^2+\left(\frac{wb}{r_0}-s\hat{a}h\right)^2x^2\right] ~,~~~~ \label{H}   \\
&&\frac{b}{r_0}=\frac{\sqrt{\left[1-2h+\left(\hat{a}^2+\hat{Q}^2\right)h^2\right]\left[w^2+\left(1-w^2\right)\left(2h-\hat{Q}^2h^2\right)\right]}
-s\hat{a}h^2\left(2-\hat{Q}^2h\right)}{w\left(1-2h+\hat{Q}^2h^2\right)}~.  \label{b-r0-4}
\end{eqnarray}
By performing the series expansion of the factor $1/\sqrt{H}$ on the right-hand side of Eq.~\eqref{alpha-exact-2} in $h$, we have
\begin{eqnarray}
\nn&&\frac{1}{\sqrt{H}}=\frac{1}{w\sqrt{1-x^2}}\Bigg{\{}1+2\left[\frac{1}{w^2(1+x)}-1\right]xh-\frac{4-4ws\hat{a}-w^2(1+x)\hat{Q}^2}{w^2(1+x)}x^2h^2    \\
\nn&&\hspace*{1.22cm}-\frac{2\left\{w\!\left[4-(2+x)\hat{Q}^2+(1+w^2)\hat{a}^2\right]-s\hat{a}\left[2+w^2\left(4-(1+x)\hat{Q}^2\right)\right]\right\}}{w^3(1+x)}x^2h^3+\mathcal{O}(M^4)\Bigg{\}}^{-\frac{1}{2}}~~ \\
\nn&&\hspace*{22pt}=\frac{1}{w\sqrt{1\!-\!x^2}}\Bigg{\{}\!1\!+\!\left[1-\frac{1}{w^2(1+x)}\right]\!xh+\frac{3-w^2(1+x)\!\left[2+4ws\hat{a}-w^2(1+x)\left(3-\hat{Q}^2\right)\right]}{2w^4(1+x)^2}x^2h^2 \\
\nn&&\hspace*{33pt}+\,\frac{1}{2w^4(1+x)^2}\Big{\{}3x-\frac{5x}{w^2(1+x)}-2ws\hat{a}\!\left[2(1-2x)+w^2(4+10x+6x^2)-w^2(1+x)^2\hat{Q}^2\right]    \\
   &&\hspace*{33pt}+\,w^2(1+x)\!\left[8-3x+2(1+w^2)\hat{a}^2-(4-x)\hat{Q}^2+w^2x(1+x)\left(5-3\hat{Q}^2\right)\right]\!\Big{\}}x^2h^3\!+\!\mathcal{O}(M^4)\!\Bigg{\}}~.~~~~~~ \label{H2}
\end{eqnarray}
After substituting Eq.~\eqref{H2} into the integrand of Eq.~\eqref{alpha-exact-2}, we then use the power series expansion of the integrand in $h$, integrate it over $x$, and find
\begin{eqnarray}
&&\nn\hat{\alpha}=2\left(1+\frac{1}{w^2}\right)h+\left[\frac{3\pi}{4}\left(1+\frac{4}{w^2}\right)-\frac{2}{w^2}\left(1+\frac{1}{w^2}\right)-\frac{4s\hat{a}}{w}-\frac{\pi\hat{Q}^2}{4}\left(1+\frac{2}{w^2}\right)\right]h^2  \\
&&\nn\hspace*{18pt}+\,\Bigg{\{}\frac{10}{3}+\frac{26}{w^2}+\frac{9}{w^4}+\frac{7}{3w^6}-\frac{3\pi}{2w^2}\left(1+\frac{4}{w^2}\right)-\frac{2s\hat{a}}{w}\left[3\pi-2+\frac{2(\pi-3)}{w^2}\right]  \\
&&\hspace*{18pt}+\left(1+\frac{1}{w^2}\right)\hat{a}^2-\left(2+\frac{22-\pi}{2w^2}-\frac{\pi-1}{w^4}\right)\hat{Q}^2+\frac{\pi s\hat{a}\hat{Q}^2}{w}\Bigg{\}}h^3+\mathcal{O}(M^4)~.  \label{alpha-3PM}
\end{eqnarray}

Finally, the explicit form of the equatorial gravitational deflection angle of a relativistic massive particle up to the 3PM order can be obtained by plugging Eqs.~\eqref{haQ}, and \eqref{b-r0-3} - \eqref{a3} into Eq.~\eqref{alpha-3PM} as
\begin{eqnarray}
&&\nn\hat{\alpha}_{\text{KN}}=2\left(1+\frac{1}{w^2}\right)\frac{M}{b}+\frac{3\pi}{4}\left(1+\frac{4}{w^2}\right)\frac{M^2}{b^2}-\frac{4saM}{wb^2}
-\frac{\pi}{4}\left(1+\frac{2}{w^2}\right)\frac{Q^2}{b^2}+\frac{2}{3}\left(5+\frac{45}{w^2}+\frac{15}{w^4}-\frac{1}{w^6}\right)\frac{M^3}{b^3}    \\
&&\hspace*{29pt}-2\pi\left(\frac{3}{w}+\frac{2}{w^3}\right)\frac{saM^2}{b^3}+2\left(1+\frac{1}{w^2}\right)\frac{a^2M}{b^3}
-2\left(1+\frac{6}{w^2}+\frac{1}{w^4}\right)\frac{Q^2M}{b^3}+\frac{\pi saQ^2}{wb^3}+\mathcal{O}(M^4)~,~~~~~~  \label{alpha-3PM-Final}
\end{eqnarray}
or equivalently,
\begin{equation}
\hat{\alpha}_{\text{KN}}=N_1(w)\frac{M}{b}+N_2(w,~\hat{a},~\hat{Q})\frac{M^2}{b^2}+N_3(w,~\hat{a},~\hat{Q})\frac{M^3}{b^3}+\mathcal{O}(M^4)~, \label{alpha-3PM-Final-2}
\end{equation}
with
\end{widetext}
\begin{eqnarray}
&&N_1(w)=2\left(\!1+\frac{1}{w^2}\!\right)~, \label{N1}  \\
&&N_2(w,~\hat{a},~\hat{Q})=\frac{3\pi}{4}\!\left(\!1\!+\!\frac{4}{w^2}\!\right)\!-\!\frac{4s\hat{a}}{w}\!-\!\frac{\pi}{4}\left(\!1\!+\!\frac{2}{w^2}\!\right)\!\hat{Q}^2~,~~~~~~ \label{N2}
\end{eqnarray}
\begin{eqnarray}
\nn&&N_3(w,~\hat{a},~\hat{Q})=\frac{2}{3}\left(\!5\!+\!\frac{45}{w^2}\!+\!\frac{15}{w^4}\!-\!\frac{1}{w^6}\!\right)\!-\!2\pi\left(\!\frac{3}{w}\!+\!\frac{2}{w^3}\!\right)s\hat{a} ~~ \\
&&\hspace*{22pt} +\,2\left(\!1\!+\!\frac{1}{w^2}\!\right)\!\hat{a}^2\!-\!2\left(\!1+\frac{6}{w^2}\!+\!\frac{1}{w^4}\!\right)\!\hat{Q}^2\!+\!\frac{\pi s\hat{a}\hat{Q}^2}{w}~.   \label{N3}
\end{eqnarray}

The comparison of Eq.~\eqref{alpha-3PM-Final} with the results presented in the previous works is made as follows. It is interesting to find that Eq.~\eqref{alpha-3PM-Final} is in agreement with the result derived by means of a different method in Ref.~\cite{HJ2020}, after replacing our sign parameter $s$ by $-s$ (in their notation). When the black hole's spin vanishes and the initial velocity of the particle reaches the speed of light (i.e., $a=0,~w=1$), Eq.~\eqref{alpha-3PM-Final} is reduced to the third-order Reissner-Nordstr\"{o}m deflection angle of light
\begin{equation}
\hat{\alpha}_{\text{RN}}=\frac{4M}{b}+\frac{15\pi M^2}{4b^2}-\frac{3\pi Q^2}{4b^2}+\frac{128M^3}{3b^3}-\frac{16MQ^2}{b^3}~,   \label{alpha-RN}
\end{equation}
which is in accordance with Eq.\,(53) of Ref.~\cite{KP2005} and Eq.\,(8.22) of Ref.~\cite{CCDR2015}. For the case of no electrical charge of the lens and $w=1$, up to the 3PM order,
Eq.~\eqref{alpha-3PM-Final} becomes~\cite{AKP2011}
\begin{eqnarray}
\nn&&\hat{\alpha}_{\text{Kerr}}=\frac{4M}{b}+\frac{15\pi M^2}{4b^2}-\frac{4saM}{b^2}+\frac{128M^3}{3b^3}~~~ \\
&&\hspace*{33.5pt}-\,\frac{10\pi saM^2}{b^3}+\frac{4a^2M}{b^3}~.   \label{alpha-Kerr}
\end{eqnarray}
If both the spin and electrical charge of the black hole disappear simultaneously, Eq.\,\eqref{alpha-3PM-Final} can be simplified to the third-order Schwarzschild deflection angle of massive particles~\cite{AR2002,AR2003,LZLH2019}
\begin{eqnarray}
\nn&&\hat{\alpha}_{\text{S}}=2\left(1+\frac{1}{w^2}\right)\frac{M}{b}+\frac{3\pi}{4}\left(1+\frac{4}{w^2}\right)\frac{M^2}{b^2} ~~\\
&&\hspace*{23pt}+\,\frac{2}{3}\left(5+\frac{45}{w^2}+\frac{15}{w^4}-\frac{1}{w^6}\right)\frac{M^3}{b^3}~.  \label{alpha-Schwar}
\end{eqnarray}
Moreover, Eq.\,\eqref{alpha-3PM-Final} is also consistent with the result for the second-order KN deflection of massive particles derived via different approaches~\cite{HL2016,HL2017} when the third-order contributions on the right-hand side of Eq.\,\eqref{alpha-3PM-Final} are dropped.

With respect to the spin-induced terms on the right-hand side of Eq.~\eqref{alpha-3PM-Final}, it should be pointed out that the second-order spin-induced contribution is negative and positive for the particle's prograde $(s=+1)$ and retrograde $(s=-1)$ motions relative to the rotation of the lens, respectively. This conclusion also holds qualitatively for the total of the third-order spin-induced contributions, although a special spin-dependent term whose contribution is always positive is present on the right-hand side of Eq.~\eqref{alpha-3PM-Final}.

\section{Lensing observables} \label{sect4}
In this section, we solve the Virbhadra-Ellis lens equation~\cite{VE2000} and discuss the timelike observable properties of the lensed images (i.e., the primary and secondary images) beyond the weak-deflection limit, in the framework of the weak-field, small-angle, and thin-lens approximation.

\subsection{Lens equation}
According to the lens diagram in Fig.~\ref{Figure1}, we can obtain the Virbhadra-Ellis lens equation, which reads~\cite{VE2000}:
\begin{equation}
\tan\mathcal{B}=\tan\vartheta-D\left[\tan\vartheta+\tan(\hat{\alpha}-\vartheta)\right]~,  \label{LE}  \vspace*{7pt}
\end{equation}
with $D=d_{LS}/d_{S}$.

We apply the analysis of the standard perturbation theory to solving Eq.~\eqref{LE}. For the sake of a convenient discussion, we use the scaled variables via the following definitions~\cite{KP2005,KP2006a,KP2006b}:
\begin{equation}
\beta\equiv\frac{\mathcal{B}}{\vartheta_E}~,~~~~~~\theta\equiv\frac{\vartheta}{\vartheta_E}~,~~~~~~\varepsilon\equiv\frac{\vartheta_{\bullet}}{\vartheta_E}=\frac{\vartheta_E}{4D}~.   \label{variables}
\end{equation}
Here, $\vartheta_E\equiv\sqrt{4DM/d_L}$ is the angular Einstein ring radius of light in the weak-deflection limit. $\vartheta_{\bullet}\equiv\arctan\left(M_\bullet/d_L\right)$ denotes the angle subtended by the special gravitational radius which is defined as $M_\bullet\equiv GM/c^2$ (equal to the lens' mass $M$ in geometrized units) and different from the conventional one~\cite{O1939,AT1968}. $\varepsilon$ serves as the new expansion parameter for analyzing the observable characteristics of the lensed images. It is worth mentioning that we don't adopt the angular Einstein ring radius of massive particles but $\vartheta_E$ as the natural scale in Eq.~\eqref{variables}, since the scale factor should be constant for a given lensing scenario and independent on the initial velocity of the massive particle. This treatment guarantees that all of the possible velocity effects on the angular image position are absorbed by the scaled variable $\theta$. Moreover, since $\vartheta_E$ is of the same order of magnitude as $D\varepsilon$, Eq.~\eqref{LE} can be reduced to the small angles lens equation $\vartheta=\mathcal{B}+\alpha$~\cite{SEF1992,Wambsganss1998} by defining a reduced deflection angle $\alpha\equiv D\hat{\alpha}$, when the third- and higher-order contributions in $\varepsilon$ to $\mathcal{B}$, $\vartheta$, and $\hat{\alpha}$ are omitted.

The perturbation analysis enables us to assume the series expansion of the scaled angular position of the image in $\varepsilon$
\begin{equation}
\theta=\theta_0+\theta_1\varepsilon+\theta_2\varepsilon^2+\mathcal{O}(\varepsilon^3)~,   \label{theta}
\end{equation}
where $\theta_0~(>0)$ denotes its zeroth-order value in the weak-deflection limit, while $\theta_1$ and $\theta_2$ are the unknown coefficients of the first- and second-order contributions to the angular image position, respectively.

Now we turn our attention to the solution of the lens equation. Substituting Eqs.~\eqref{alpha-3PM-Final-2}, \eqref{variables}, \eqref{theta} and the relation $b=d_L\sin\vartheta$ into Eq.~\eqref{LE}, up to the third order of $\varepsilon$, we have
\begin{eqnarray}
\nn&&0=D\left(4\beta-4\theta_0+\frac{N_1}{\theta_0}\right)\varepsilon+\frac{D\left[N_2-\left(N_1+4\theta_0^2\right)\theta_1\right]}{\theta_0^2}\varepsilon^2      \\
\nn&&+\,\frac{D}{3\theta_0^3}\!\left[N_1^3\!+\!3N_3\!-\!12DN_1^2\theta_0^2\!+\!N_1\left(56D^2\theta_0^4\!+\!3\theta_1^2\!-\!3\theta_0\theta_2\right)  \right.   \\
&&\left. +\,64D^2\theta_0^3\left(\beta^3-\theta_0^3\right)-6N_2\theta_1-12\theta_0^3\theta_2\right]\!\varepsilon^3+\mathcal{O}(\varepsilon^4)~,   \label{LE-2}
\end{eqnarray}
which is the same as Eq.\,(65) of Ref.~\cite{KP2005} for the case of $w=1$ and $a=Q=0$.

\begin{widetext}
\subsection{Image positions}  \label{IP}
The requirement for the disappearance of the first- and higher-order corrections on the right-hand side of Eq.~\eqref{LE-2} leads to
\begin{eqnarray}
&&\theta_0=\frac{1}{2}\left[\sqrt{\beta^2+2\left(1+\frac{1}{w^2}\right)}+\beta\right]~,    \label{theta0} \\
&&\theta_1=\frac{N_2}{N_1+4\theta_0^2}
=\frac{3\pi\left(4+w^2\right)-16ws\hat{a}-\pi\left(2+w^2\right)\hat{Q}^2}{8\left(1+w^2+2w^2\theta_0^2\right)}~,   \label{theta1} \\
&&\theta_2=\frac{1}{3\theta_0\left(N_1\!+\!4\theta_0^2\right)}\!\left(N_1^3+3N_3-12DN_1^2\theta_0^2+64D^2\beta^3\theta_0^3
+56D^2N_1\theta_0^4-64D^2\theta_0^6-6N_2\theta_1+3N_1\theta_1^2\right)~.   \label{theta2}
\end{eqnarray}
By means of Eq.~\eqref{theta0} which indicates $\beta=\theta_0-\frac{N_1}{4\theta_0}$, it can be seen that Eqs.~\eqref{theta1} - \eqref{theta2} are consistent with Eqs.\,(32) - (33) of Ref.~\cite{AKP2011}, respectively, when the lens' electrical charge vanishes $(Q=0)$ and $w=1$ is assumed.

With the consideration of the last assumption made in Sect.~\ref{sect2} and the general form of the scaled image position given in Eqs.~\eqref{theta0} - \eqref{theta2}, the angular positions (denoted by $\theta^+$ and $\theta^-$, respectively) of the positive- and negative-parity images can be expressed explicitly in terms of the angular source position $\beta$ as
\begin{equation}
\theta^{\pm}=\theta_0^{\pm}+\theta_1^{\pm}\varepsilon+\theta_2^{\pm}\varepsilon^2+\mathcal{O}(\varepsilon^3)~,   \label{theta-PN}
\end{equation}
where
\begin{eqnarray}
&&\theta_0^{\pm}=\frac{1}{2}\left[\sqrt{\beta^2+2\left(1+\frac{1}{w^2}\right)}\pm|\beta|\right]~,    \label{theta0-PN}  \\
&&\theta_1^{\pm}=\frac{3\pi\left(4+w^2\right)-16ws^{\pm}\hat{a}-\pi\left(2+w^2\right)\hat{Q}^2}{16\left(1+w^2\right)}\!\left[1\mp\frac{|\beta|}{\sqrt{\beta^2+2\left(1+\frac{1}{w^2}\right)}}\right]~, \label{theta1-PN} \\
\nn&&\theta_2^{\pm}=\frac{\left[2(1+6w^2+w^4)-\pi w^3s^{\pm}\hat{a}\right]\left(1+3w^2-w^2\hat{Q}^2\right)+w^3(1+w^2)\left(2w\hat{a}^2-3\pi s^{\pm}\hat{a}\right)-\frac{8D^2}{3}(1+w^2)^3}
{w^6\sqrt{\beta^2+2\left(1+\frac{1}{w^2}\right)}\left[\sqrt{\beta^2+2\left(1+\frac{1}{w^2}\right)}\pm|\beta|\right]^2}  \\
\nn&&\hspace*{26pt}-\frac{\left[3\pi(4+w^2)-16ws^{\pm}\hat{a}-\pi(2+w^2)\hat{Q}^2\right]^2}
{128w^2(1\!+\!w^2)\sqrt{\beta^2\!+\!2\left(1\!+\!\frac{1}{w^2}\right)}\left[\sqrt{\beta^2\!+\!2\left(1\!+\!\frac{1}{w^2}\right)}\pm|\beta|\right]^2}
\!\left[1\mp\frac{|\beta|}{\sqrt{\beta^2+2\left(1+\frac{1}{w^2}\right)}}\right]\!\!\left[3\pm\frac{|\beta|}{\sqrt{\beta^2+2\left(1+\frac{1}{w^2}\right)}}\right]   \\
&&\hspace*{26pt}-\frac{4D}{w^4\sqrt{\beta^2+2\left(1+\frac{1}{w^2}\right)}}\left\{(1+w^2)^2(1-D)-\frac{(1+w^2)w^2D}{12}\left[\sqrt{\beta^2+2\left(1+\frac{1}{w^2}\right)}\pm|\beta|\right]^2\right\}~.  \label{theta2-PN}
\end{eqnarray}
\end{widetext}
Here, we have used $s^{+}$ and $s^{-}$ to denote the sign parameter of the positive- and negative-parity images~\cite{AKP2011}, respectively. Notice that the value of $s^{+}$ is $+1$ for prograde motion of the particle, and $-1$ for its retrograde motion. Simultaneously, the relation $s^{+}=-s^{-}$ always holds in our scenario. Note also that $\beta$ is positive and negative ($|\beta|=-\beta$) when the image and source are on the same and opposite sides of the optic axis, respectively. Additionally, Eqs.~\eqref{theta0-PN} - \eqref{theta1-PN} are consistent with Eq.\,(72) in Ref.~\cite{PJ2019} when $w=1$ and $\hat{a}=0$ are assumed.

\begin{figure*}
\centering \hspace*{7pt}
\begin{minipage}[b]{7.7cm}
\includegraphics[width=7.7cm]{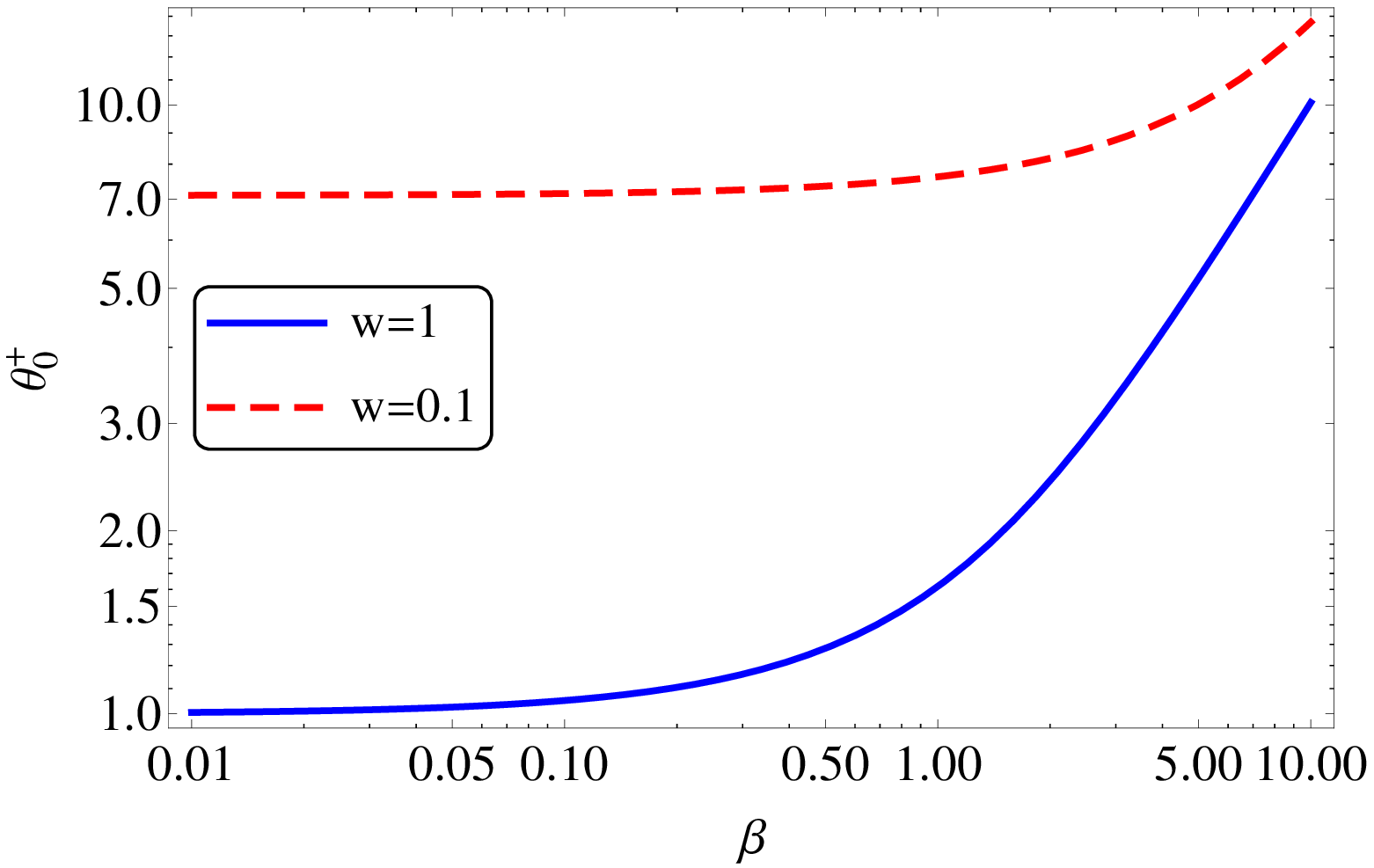} \vspace{-5pt}
  \centerline{(a) $\theta_0^{+}(\beta)$ }
\end{minipage} \hspace*{26pt}
\begin{minipage}[b]{7.7cm}
\includegraphics[width=7.7cm]{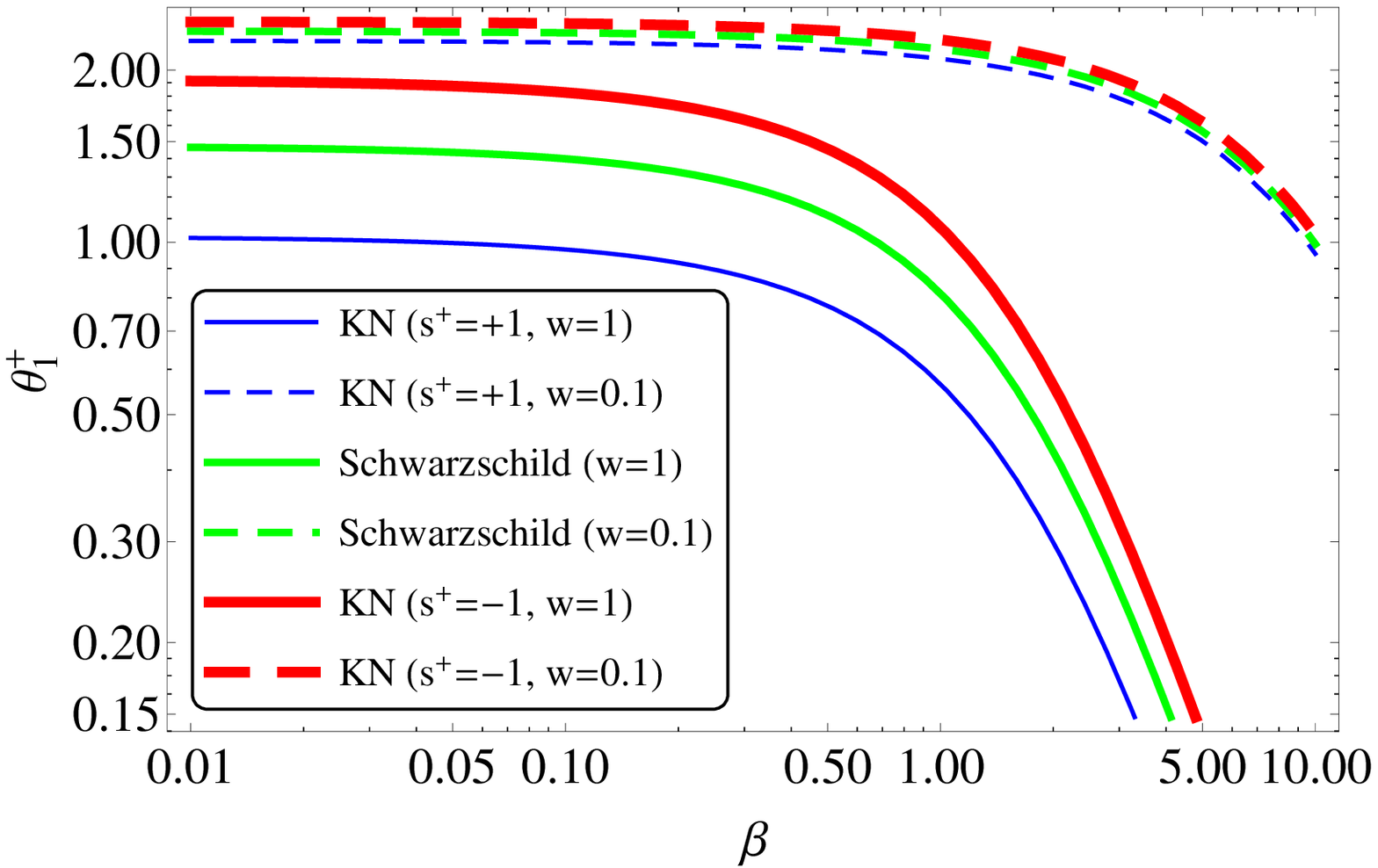} \vspace{-4pt}
  \centerline{(b) $\theta_1^{+}(\beta)$ }
\end{minipage} \\ \hspace*{3pt}
\begin{minipage}[b]{7.8cm}
\includegraphics[width=7.8cm]{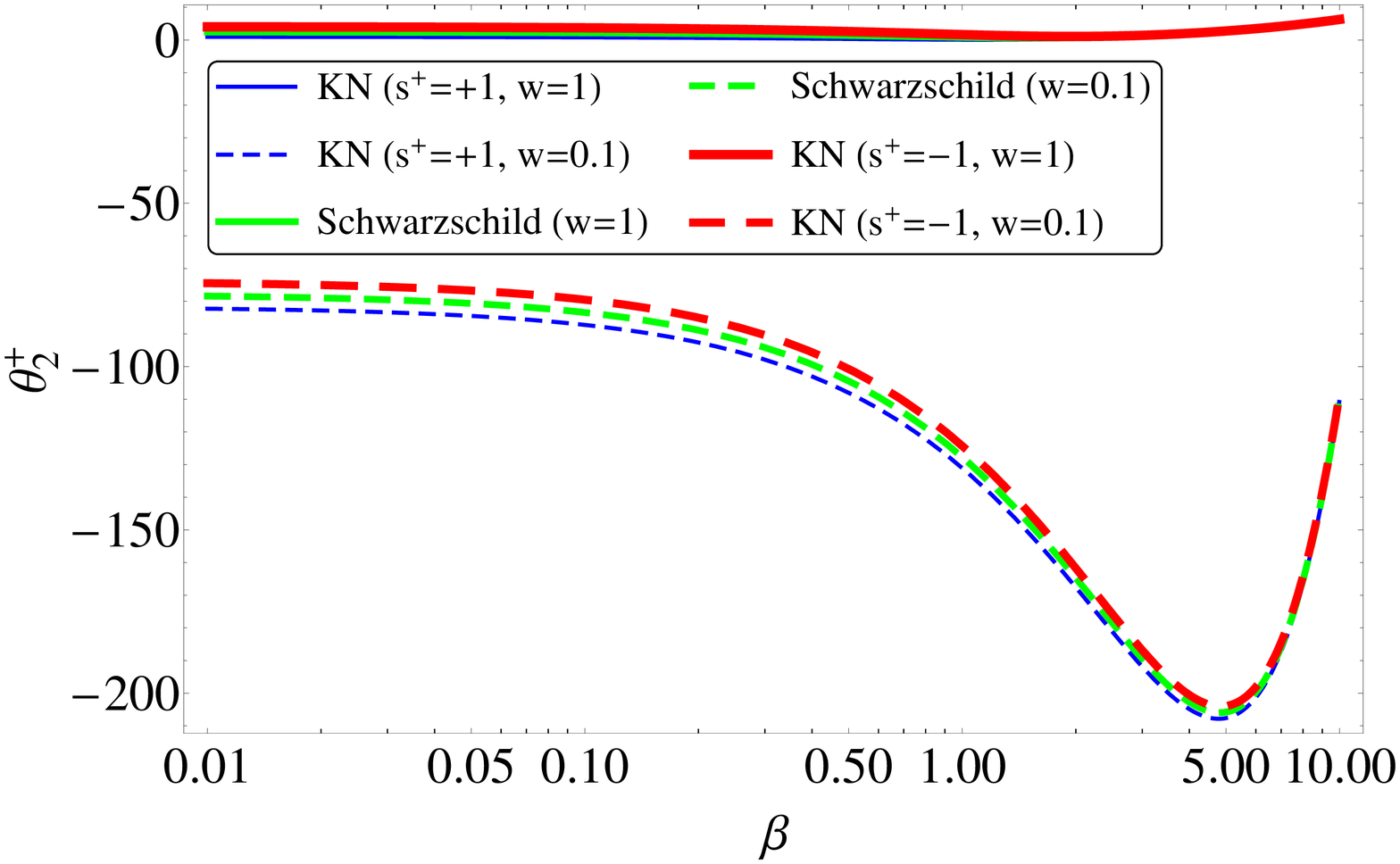} \vspace{-1pt}
  \centerline{(c) $\theta_2^{+}(\beta)$ }
\end{minipage}   \hspace*{27pt}
\begin{minipage}[b]{7.7cm}
\includegraphics[width=7.7cm]{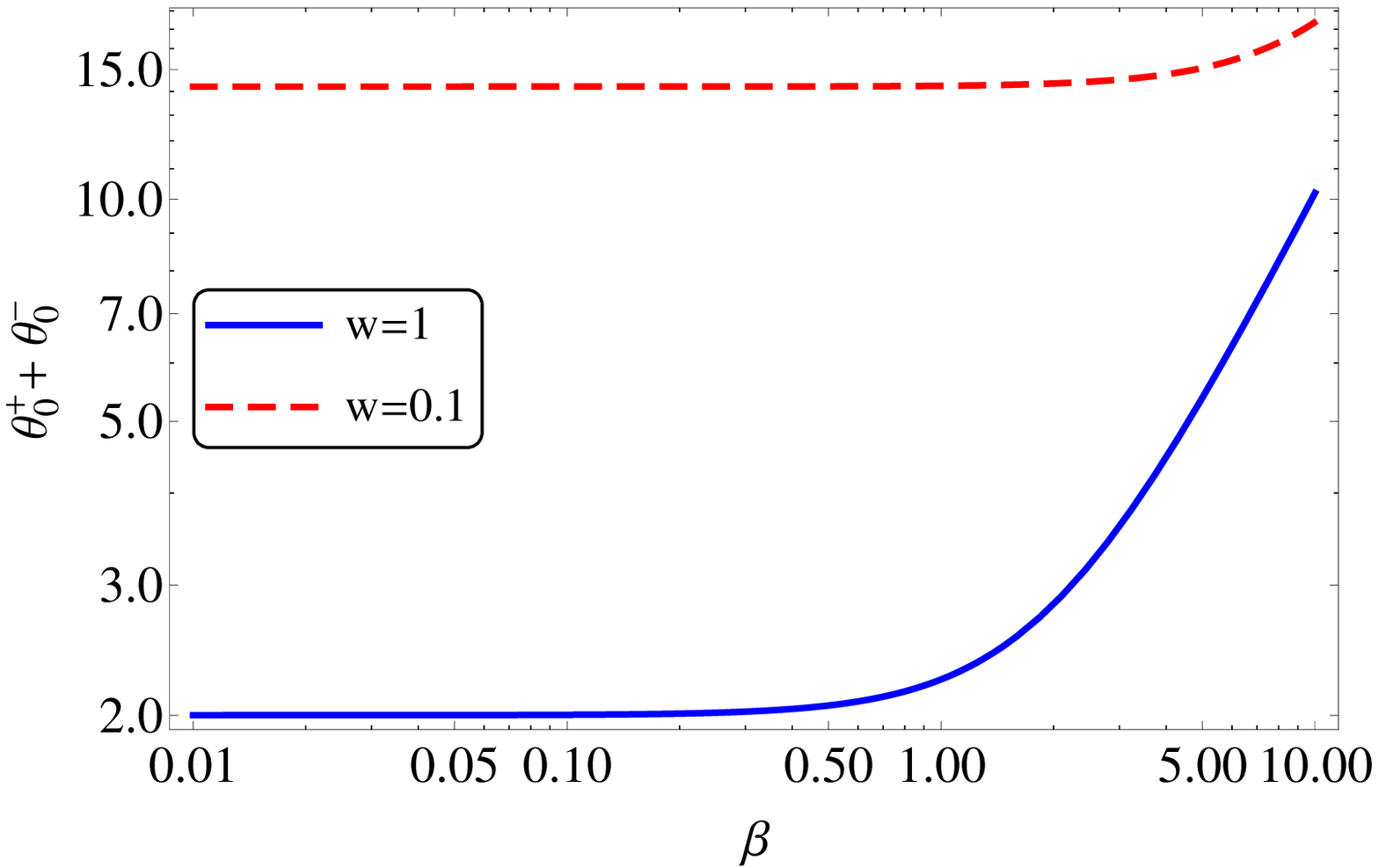} \vspace{-1.5pt}
  \centerline{(d) $\theta_0^{+}(\beta)+\theta_0^{-}(\beta)$ }
\end{minipage} \\ \hspace*{6pt}
\begin{minipage}[b]{7.58cm} \hspace*{-5pt}
\includegraphics[width=7.58cm]{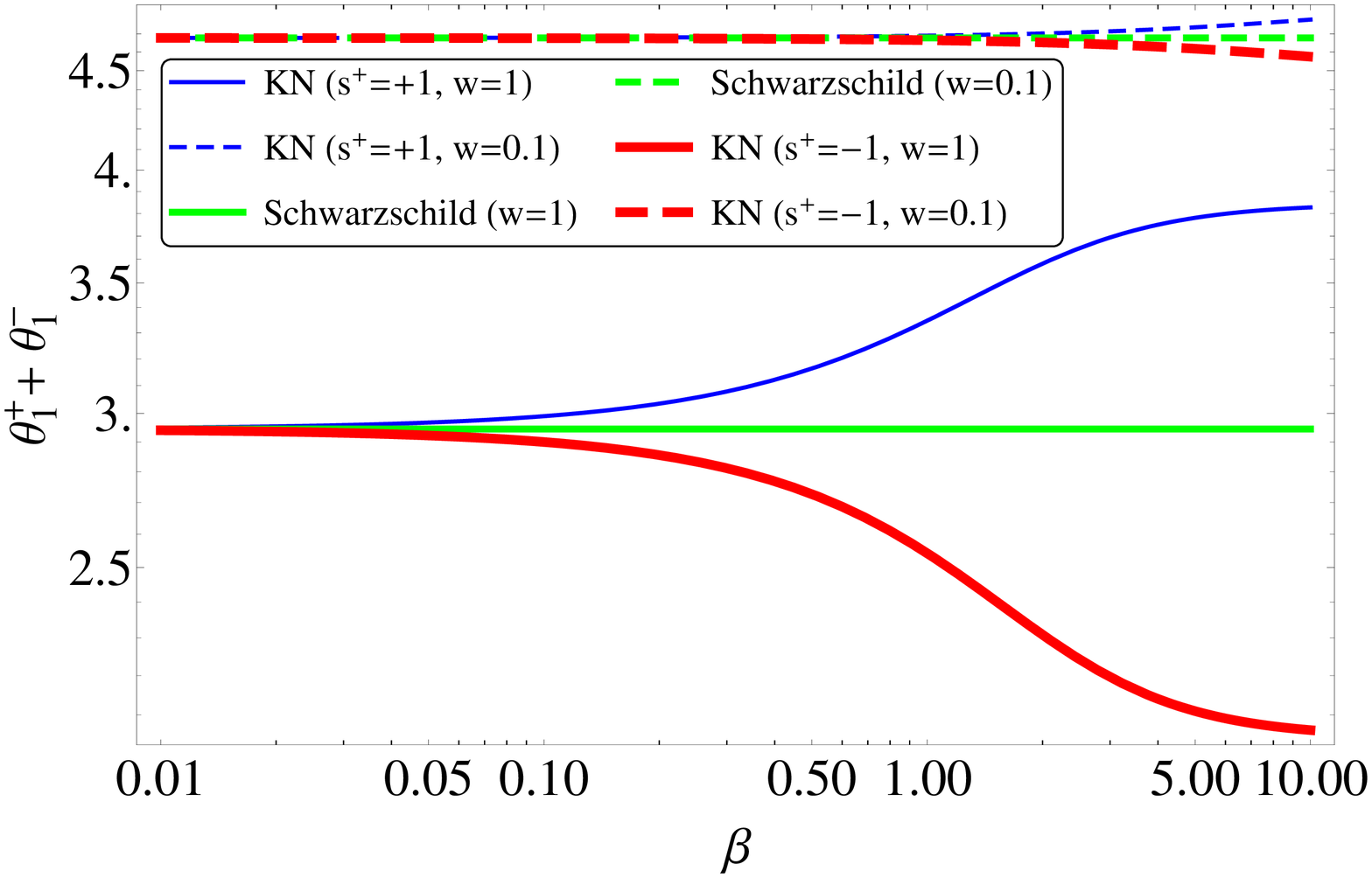} \vspace{-3pt}
  \centerline{(e) $\theta_1^{+}(\beta)+\theta_1^{-}(\beta)$ }
\end{minipage} \hspace*{30pt}
\begin{minipage}[b]{7.4cm}
\includegraphics[width=7.4cm]{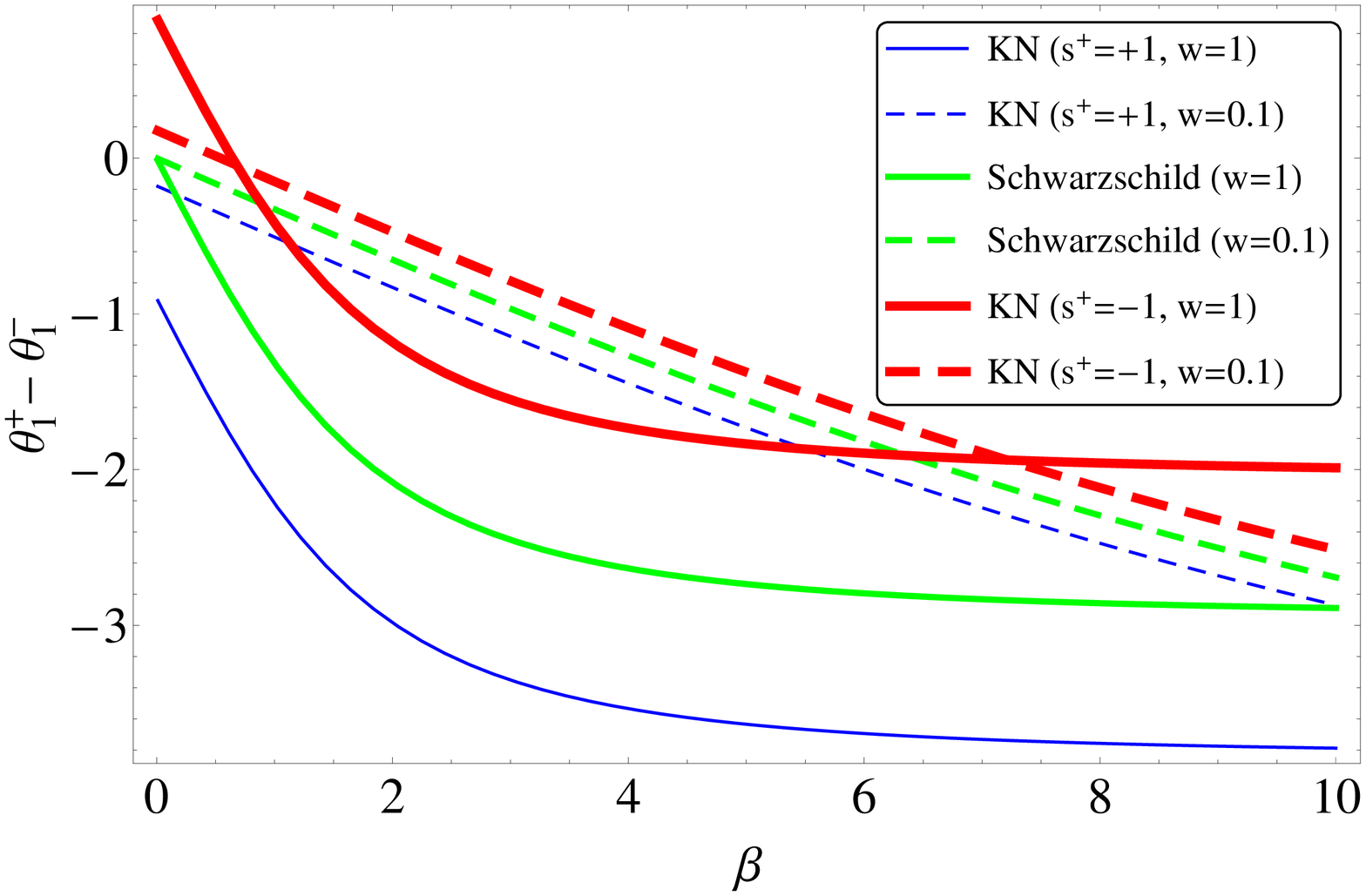}
  \centerline{(f) $\theta_1^{+}(\beta)-\theta_1^{-}(\beta)$ }
\end{minipage} \\ \hspace*{3pt}
\begin{minipage}[b]{7.9cm}
\includegraphics[width=7.9cm]{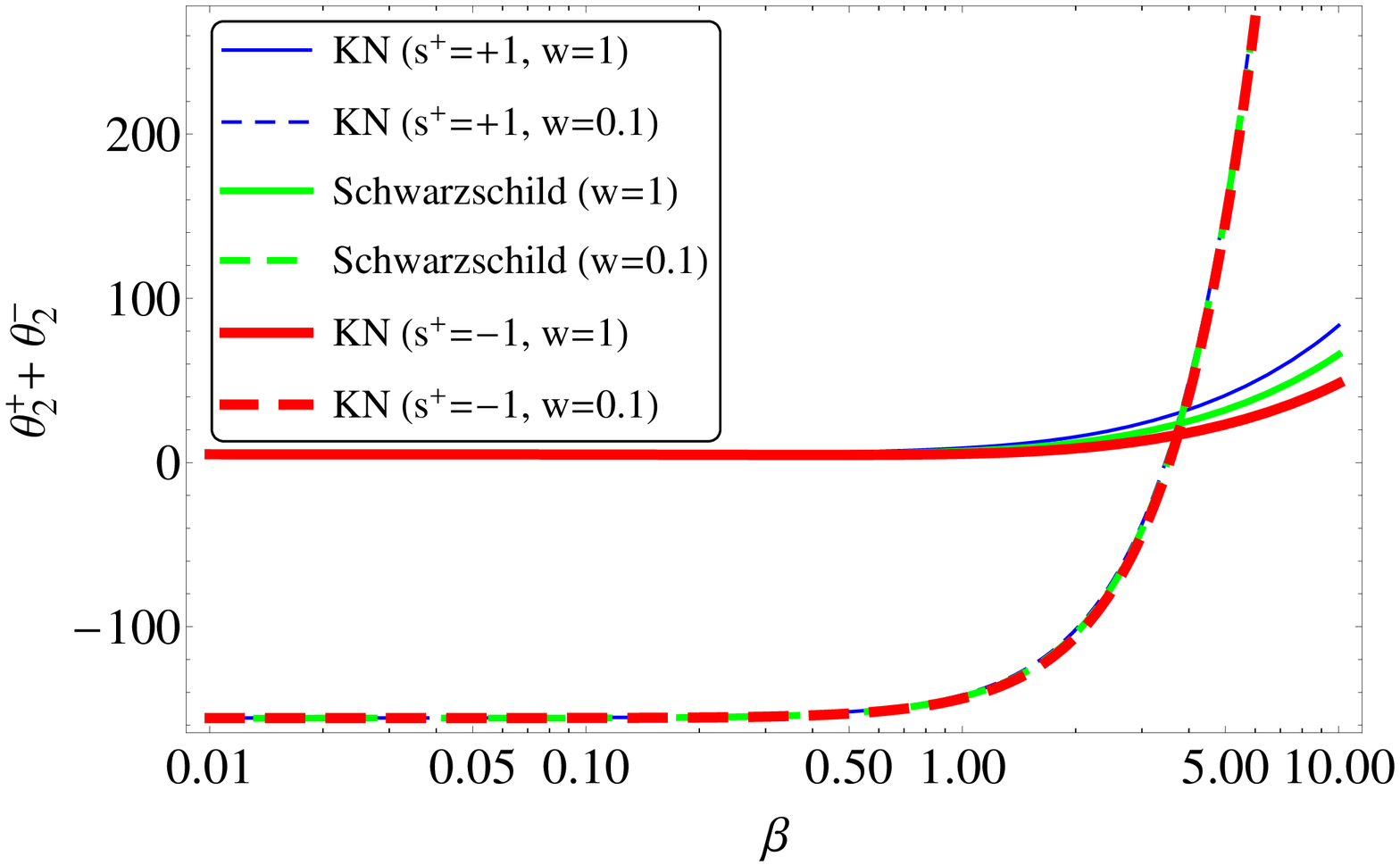} \vspace{-1pt}
  \centerline{(g) $\theta_2^{+}(\beta)+\theta_2^{-}(\beta)$ }
\end{minipage} \hspace*{19pt}
\begin{minipage}[b]{8.cm}
\includegraphics[width=8.cm]{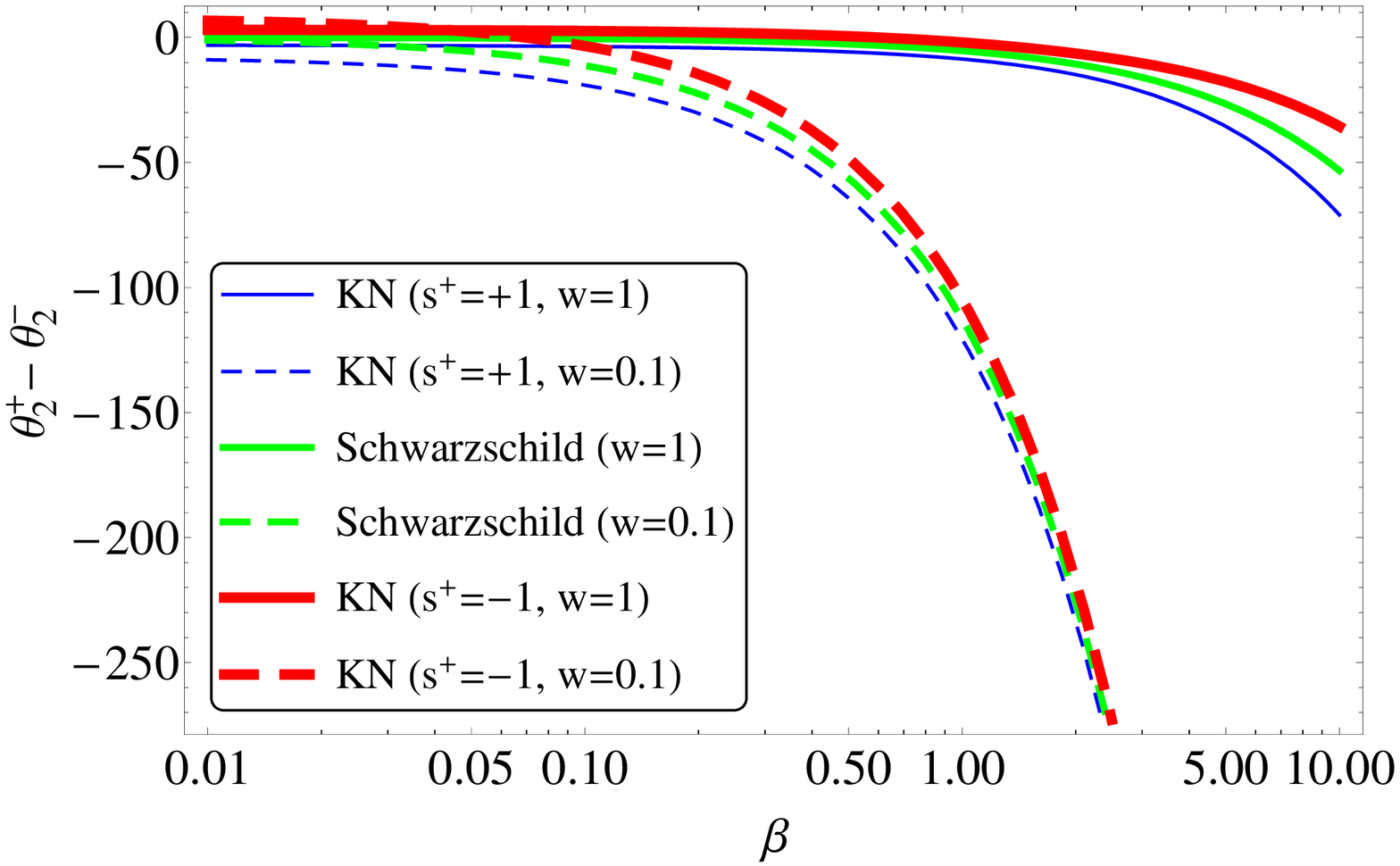}
  \centerline{(h) $\theta_2^{+}(\beta)-\theta_2^{-}(\beta)$ }
\end{minipage}

\caption{$\theta_0^{+}$, $\theta_1^{+}$, $\theta_2^{+}$, $\theta_0^{+}+\theta_0^{-}$, $\theta_1^{+}\pm\theta_1^{-}$, and $\theta_2^{+}\pm\theta_2^{-}$
plotted as the functions of $\beta~(\in[\,0.01,~10\,])$ for prograde ($s^{+}=+1$) or retrograde ($s^{+}=-1$) motion of the particle. Here and in the following figures of this section, we assume $w=0.1$, $\hat{a}=0.9,~\hat{Q}=0.01$, and $D=0.5$, as an example of our scenario of the KN lensing of massive particles. Additionally, the cases of the KN lensing of light ($w=1$) as well as the Schwarzschild lensing $(a=Q=0)$ of light and massive particles are also presented for comparison. } \label{Figure2}
\end{figure*}
From Eqs.~\eqref{theta0-PN} - \eqref{theta2-PN}, we can obtain the observable product, sum, and difference relations for the coefficients of the scaled angular image positions:
\begin{eqnarray}
&&\theta_0^{+}\theta_0^{-}=\frac{1}{2}\left(1+\frac{1}{w^2}\right)~, \label{NewR-1}  \\
&&\theta_0^{+}+\theta_0^{-}=\sqrt{\beta^2+2\left(1+\frac{1}{w^2}\right)}~,~~~~ \label{NewR-2}  \\
&&\theta_0^{+}-\theta_0^{-}=|\beta|~, \label{NewR-3}
\end{eqnarray}
\begin{widetext}
\begin{eqnarray}
&&\theta_1^{+}+\theta_1^{-}=\frac{3\pi(4+w^2)-\pi(2+w^2)\hat{Q}^2}{8(1+w^2)}+\frac{2w}{1+w^2}\frac{s^{+}\hat{a}|\beta|}{\sqrt{\beta^2+2\left(1+\frac{1}{w^2}\right)}}~, \label{NewR-4}  \\
&&\theta_1^{+}-\theta_1^{-}=-\frac{3\pi(4+w^2)-\pi(2+w^2)\hat{Q}^2}{8(1+w^2)}\frac{|\beta|}{\sqrt{\beta^2+2\left(1+\frac{1}{w^2}\right)}}-\frac{2w s^{+}\hat{a}}{1+w^2}  ~, \label{NewR-5}  \\
&&\nn\theta_2^{+}+\theta_2^{-}=\frac{1}{192\,w^2\,(1\!+\!w^2)^3}\Bigg{\{}\frac{384w^3\!\left\{2\!+\!2w^2(1\!+\!w^4\!+\!w^6)
\!+\!w^2(1\!-\!w^2)^2\!\left[3\!+\!w^2(3\!+\!\beta^2)\right]\beta^2\right\}\hat{a}^2}{(2+2w^2+w^2\beta^2)^{3/2}}+\frac{1}{w\,(2\!+\!2w^2\!+\!w^2\beta^2)^{3/2}}     \\
&&\nn\hspace*{1.75cm}\times\Big{\{}\!(1\!+\!w^2)^2\Big{[}768(1\!+\!10w^2)\!+\!48(448\!-\!27\pi^2)w^4\!+\!24(704\!-\!27\pi^2)w^6\!+\!9(256\!-\!9\pi^2)w^8\!-\!512(6\!-\!5D)D(1\!+\!w^2)^4  \\
&&\nn\hspace*{1.75cm}-\,6w^2\!\left(128+8(112-9\pi^2)w^2+(896-54\pi^2)w^4+(128-9\pi^2)w^6\right)\hat{Q}^2\!-\!9\pi^2w^4(2+w^2)^2\hat{Q}^4\Big{]}\!-6w^2(1+w^2)     \\
&&\nn\hspace*{1.75cm}\times\Big{[}128(2-D)D(1+w^2)^4-3\left(64(1+10w^2)+16(112-9\pi^2)w^4+8(176-9\pi^2)w^6+3(64-3\pi^2)w^8\right)     \\
&&\nn\hspace*{1.75cm}+\,6w^2\left(32+8(28-3\pi^2)w^2+2(112-9\pi^2)w^4+(32-3\pi^2)w^6\right)\hat{Q}^2+3\pi^2w^4(2+w^2)^2\hat{Q}^4\Big{]}\beta^2     \\
&&\nn\hspace*{1.75cm}-2w^4\Big{[}27\pi^2w^4(4+w^2)^2-192(1+10w^2+28w^4+22w^6+3w^8)+128D^2(1+w^2)^4  \\
&&\nn\hspace*{1.75cm}+w^2\left(192(1+7w^2+7w^4+w^6)-18\pi^2w^2(8+6w^2+w^4)\right)\hat{Q}^2+3\pi^2w^4(2+w^2)^2\hat{Q}^4\Big{]}\beta^4\Big{\}}     \\
&&   \hspace*{1.75cm}+192\pi w^3\,|\beta|\,s^{+}\hat{a}\left(4-2w^2+3w^4+w^2\hat{Q}^2\right)\Bigg{\}}~,   \label{NewR-6}   \\
&&\nn\theta_2^{+}-\theta_2^{-}=-\frac{1}{32w^2(1\!+\!w^2)^3}\Bigg{\{}\!\Big{\{}64(1\!+\!10w^2)+16(112-9\pi^2)w^4+8(176-9\pi^2)w^6+3(64-3\pi^2)w^8\!-\!128D^2(1+w^2)^4        \\
&&\nn\hspace*{1.75cm}+\,64w^4(1-w^2)^2\hat{a}^2-2w^2\!\left[32-w^2\!\left(3\pi^2(8+6w^2+w^4)-32(7+7w^2+w^4)\right)\right]\!\hat{Q}^2-\pi^2w^4(2+w^2)^2\hat{Q}^4\Big{\}}|\beta|        \\
&&\nn\hspace*{1.75cm}+\frac{16\pi w^2s^{+}\hat{a}}{(2+2w^2+w^2\beta^2)^{3/2}}\Big{\{}(16\!+\!4w^2\!+\!15w^4)(1+w^2)^2\!+\!6w^2(4\!+\!2w^2\!+\!w^4\!+\!3w^6)\beta^2\!+\!2w^4(4-2w^2+3w^4)\beta^4        \\
&&   \hspace*{1.75cm}+\,w^2\left[2+w^2\left(3-w^4+6(1+w^2)\beta^2+2w^2\beta^4\right) \right]\hat{Q}^2\Big{\}}\Bigg{\}}~.    \label{NewR-7}
\end{eqnarray}
\end{widetext}
There are three points which should be emphasized. First, it is interesting to find that the product of the zeroth-order positions of the positive- and negative-parity images depends on the initial velocity of the massive particle in the weak-deflection limit, which is obviously different from the null case where the value of $\theta_0^{+}\theta_0^{-}$ is always equal to $1$. Since the lens quantities given in Eqs.~\eqref{NewR-2} and \eqref{NewR-4} - \eqref{NewR-7} also depend on $w$, it is possible to study conversely the properties of the particle's source by means of the detection of these observables. Second, due to the presence of the spin-induced contributions, each of the first- and second-order sum and difference relations for the coefficients of the image positions ($\theta_1^{+}\pm\theta_1^{-}$ and $\theta_2^{+}\pm\theta_2^{-}$) appears differently for prograde and retrograde motions of the massive particle. However, this is not the case for the zeroth-order relations $\theta_0^{+}\pm\theta_0^{-}$. Thirdly, it shows that the first- and second-order sum and difference relations for the positional coefficients depend not only on $a$ but also on the electrical charge $Q$ of the black hole. Thus, for a given timelike lens diagram of a Kerr, Reissner-Nordstr\"{o}m, or KN black hole, we may also constrain the intrinsic spin or electrical charge of the lens in turn by detecting the first-order sum and difference relations ($\theta_1^{+}\pm\theta_1^{-}$).

Finally, the coefficients of the zeroth-, first-, and second-order contributions to the position of a positive-parity image, as well as the sum and difference relations given in Eqs.~\eqref{NewR-2} and \eqref{NewR-4} - \eqref{NewR-7}, are plotted as the functions of the angular source position in Fig.~\ref{Figure2}. The KN lensing scenarios for prograde ($s^{+}=+1$) and retrograde ($s^{+}=-1$) motions of the particle with an initial velocity $w=0.1$ are considered respectively in Fig.~\ref{Figure2}.

\subsection{Magnification relations}
We then discuss the magnification relations of the lensed images, including the signed magnifications, total magnification, and the centroid up to the second order in $\varepsilon$.

\subsubsection{Signed magnifications}
The general form of the magnification $\mu$ of a lensed image for a test particle propagating in the equatorial plane of the central body is given by~\cite{VE2000,ERT2002}
\begin{equation}
\mu(\vartheta)=\left[\frac{\sin\mathcal{B}(\vartheta)}{\sin\vartheta}\frac{d\mathcal{B}(\vartheta)}{d\vartheta}\right]^{-1}~.  \label{mu-Define}
\end{equation}
Note that the sign of the magnification of a lensed image gives the image parity. It implies that the magnification $\mu^+$ of the positive-parity primary image $\theta^+$ is positive, while $\mu^-$ of the negative-parity secondary image $\theta^-$ is negative.

Based on Eqs.~\eqref{alpha-3PM-Final-2} and \eqref{LE}, the magnification $\mu$ can be written in the following form by using the series expansion in the small parameter $\varepsilon$
\begin{equation}
\mu=\mu_0+\mu_1\varepsilon+\mu_2\varepsilon^2+\mathcal{O}(\varepsilon^3)~,   \label{mu-SeriesE}
\end{equation}
where the coefficients of the zeroth-, first-, and second-order contributions to the magnification are given by:
\begin{eqnarray}
&&\mu_0=\frac{4\theta_0^4}{4\theta_0^4-\left(1+\frac{1}{w^2}\right)^2}~,  \label{mu0}  \\
&&\mu_1=-\frac{w^4\!\left[3\pi(4+w^2)-16ws\hat{a}-\pi(2+w^2)\hat{Q}^2\right]\!\theta_0^3}{2(1+w^2+2w^2\theta_0^2)^3}~,~~~~~~~  \label{mu1}
\end{eqnarray}
\begin{widetext}
\begin{eqnarray}
&&\nn\mu_2=-\frac{8\theta_0^2}{3(N_1-4\theta_0^2)(N_1+4\theta_0^2)^5}\Big{\{}D^2N_1^6-\!\left[8(2+6D-9D^2)N_1^5+48N_1^2N_3\right]\!\theta_0^2   \\
&&\hspace*{24pt}-\!\left[32(4+12D-17D^2)N_1^4-576N_2^2+384N_1N_3\right]\!\theta_0^4-\!\left[128(2+6D-9D^2)N_1^3+768N_3\right]\!\theta_0^6+256D^2N_1^2\theta_0^8\Big{\}}~.~~~~~~~  \label{mu2}
\end{eqnarray}
\end{widetext}
Here, the first equality in Eq.~\eqref{theta1} and the relation $\beta=\theta_0-\frac{N_1}{4\theta_0}$ have been used.
\begin{figure*}
\centering  \hspace*{0.5pt}
\begin{minipage}[b]{7.7cm}
\includegraphics[width=7.7cm]{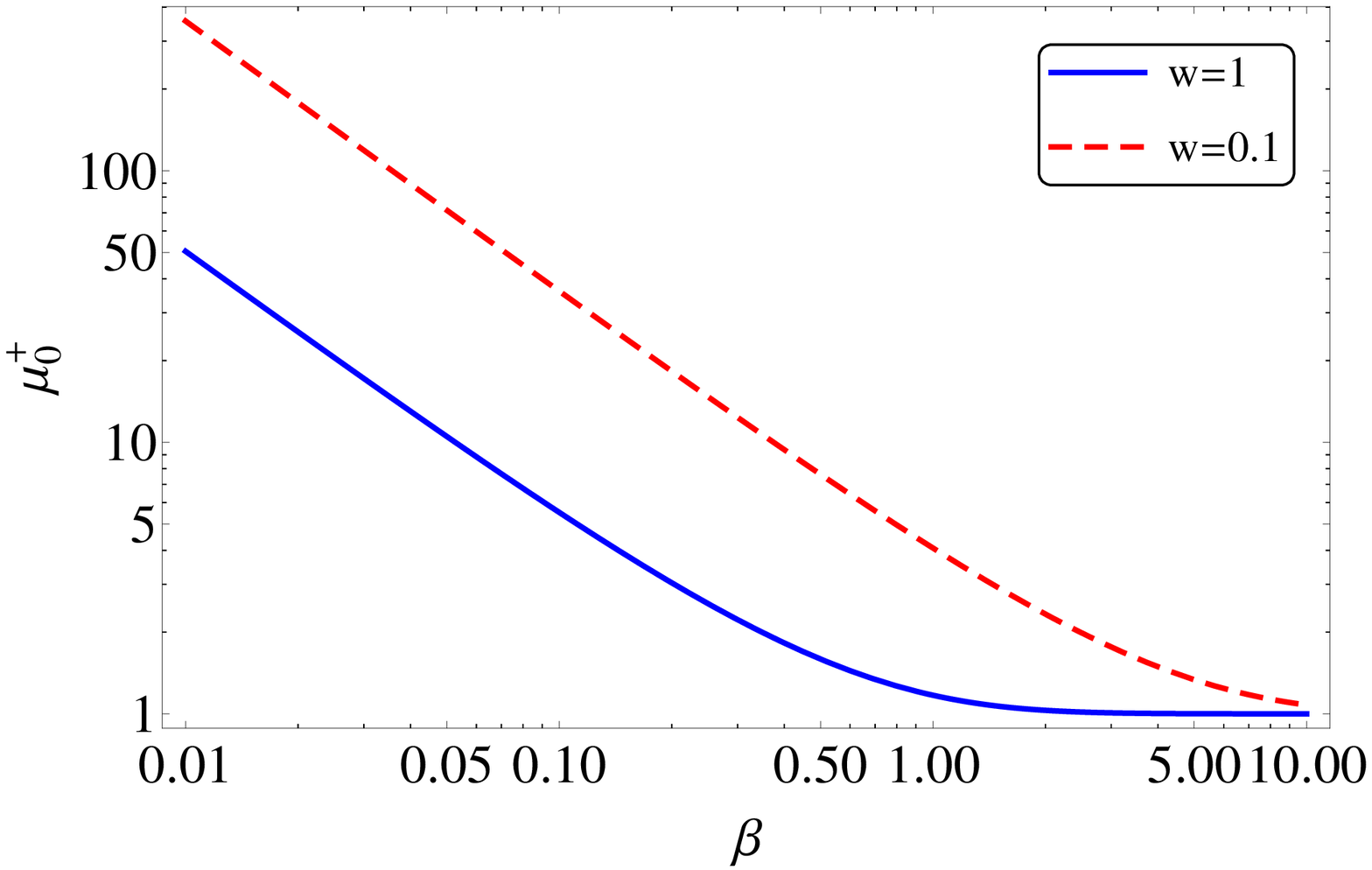} \vspace{-3pt}
  \centerline{(a) $\mu_0^{+}(\beta)$}
\end{minipage} \hspace*{21pt}
\begin{minipage}[b]{7.82cm}
\includegraphics[width=7.82cm]{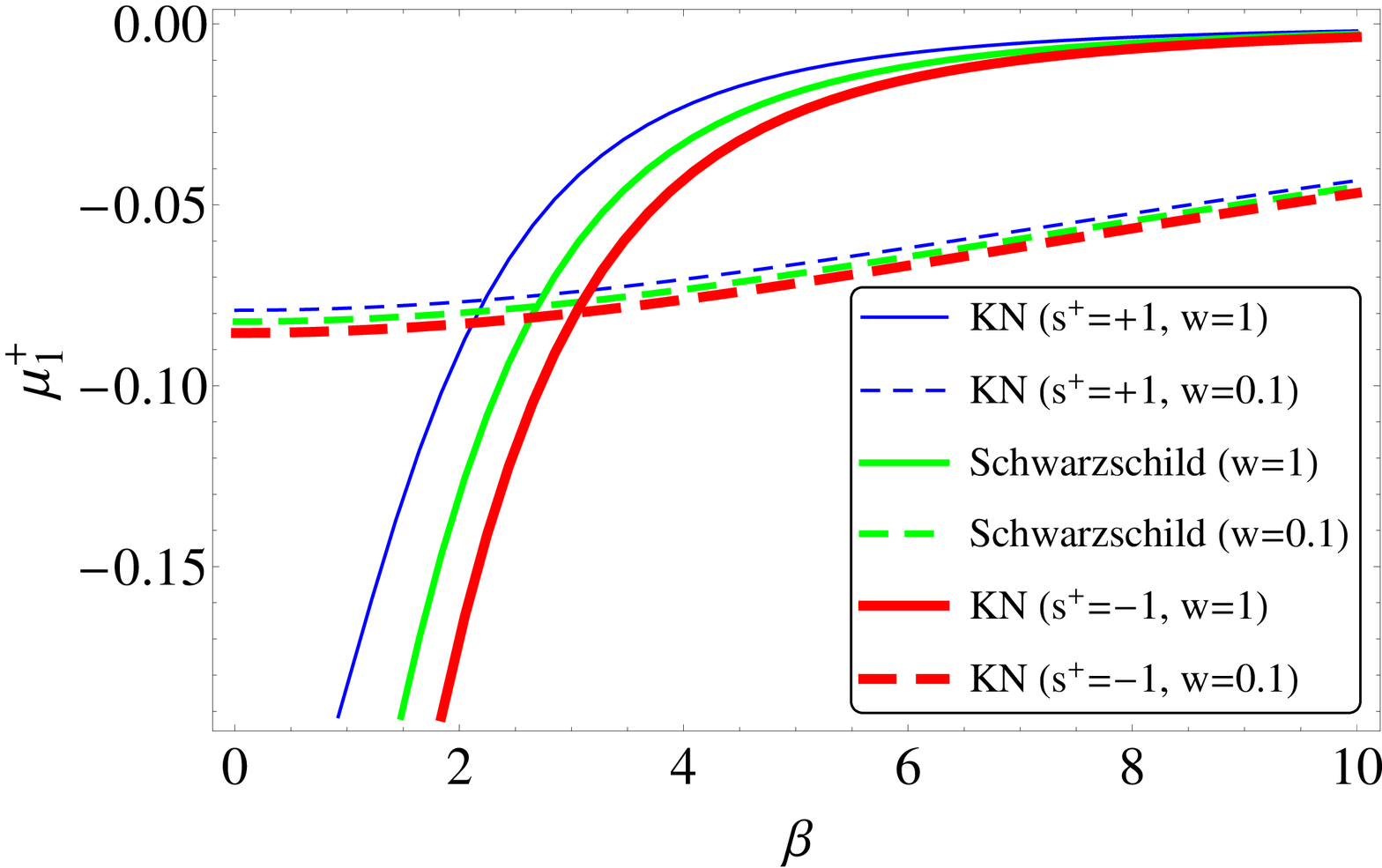}
  \centerline{(b) $\mu_1^{+}(\beta)$}
\end{minipage}   \\
\begin{minipage}[b]{8.1cm}
\includegraphics[width=8.1cm]{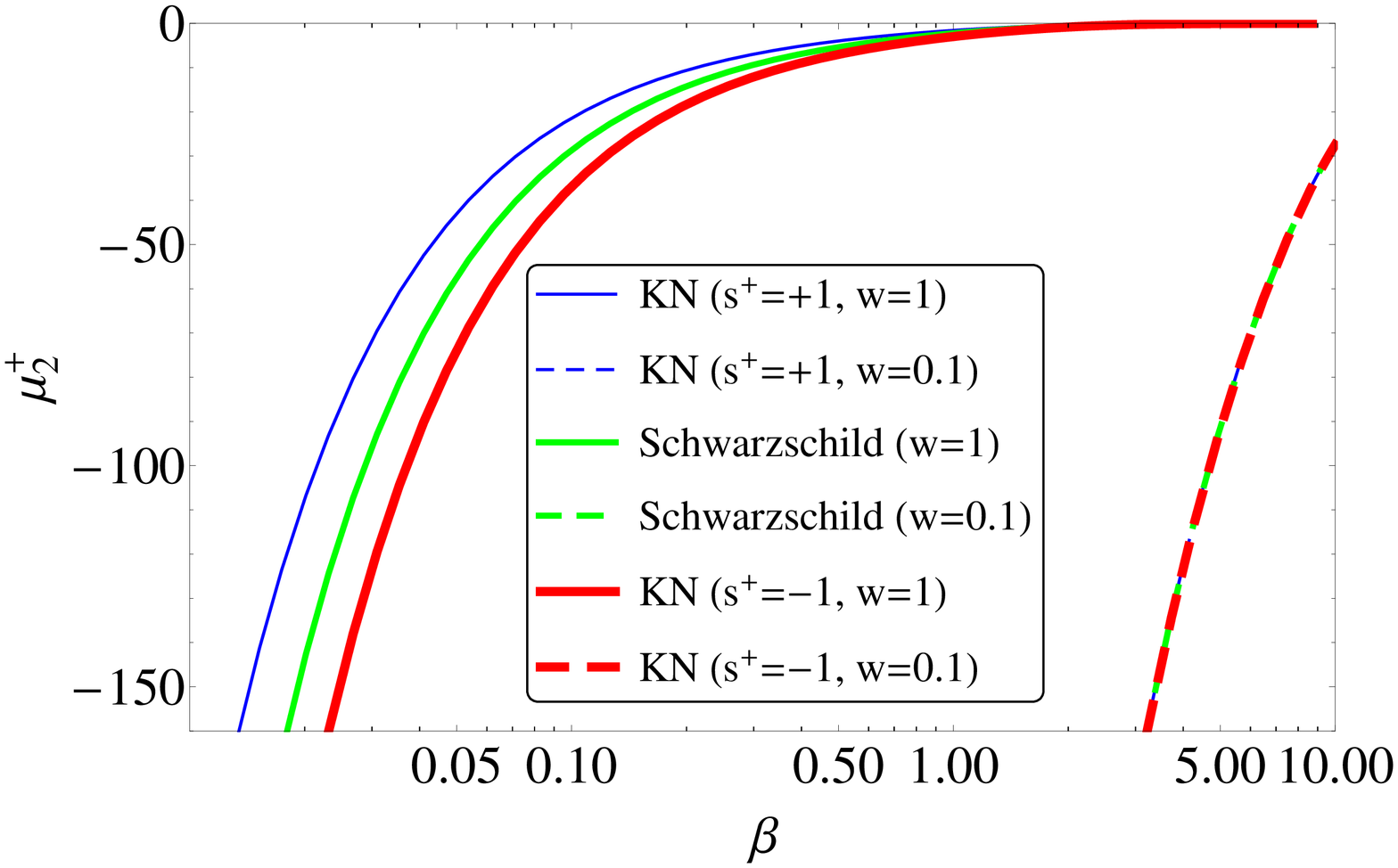} \vspace{-4pt}
  \centerline{(c) $\mu_2^{+}(\beta)$}
\end{minipage} \hspace*{26pt}
\begin{minipage}[b]{7.7cm}
\includegraphics[width=7.7cm]{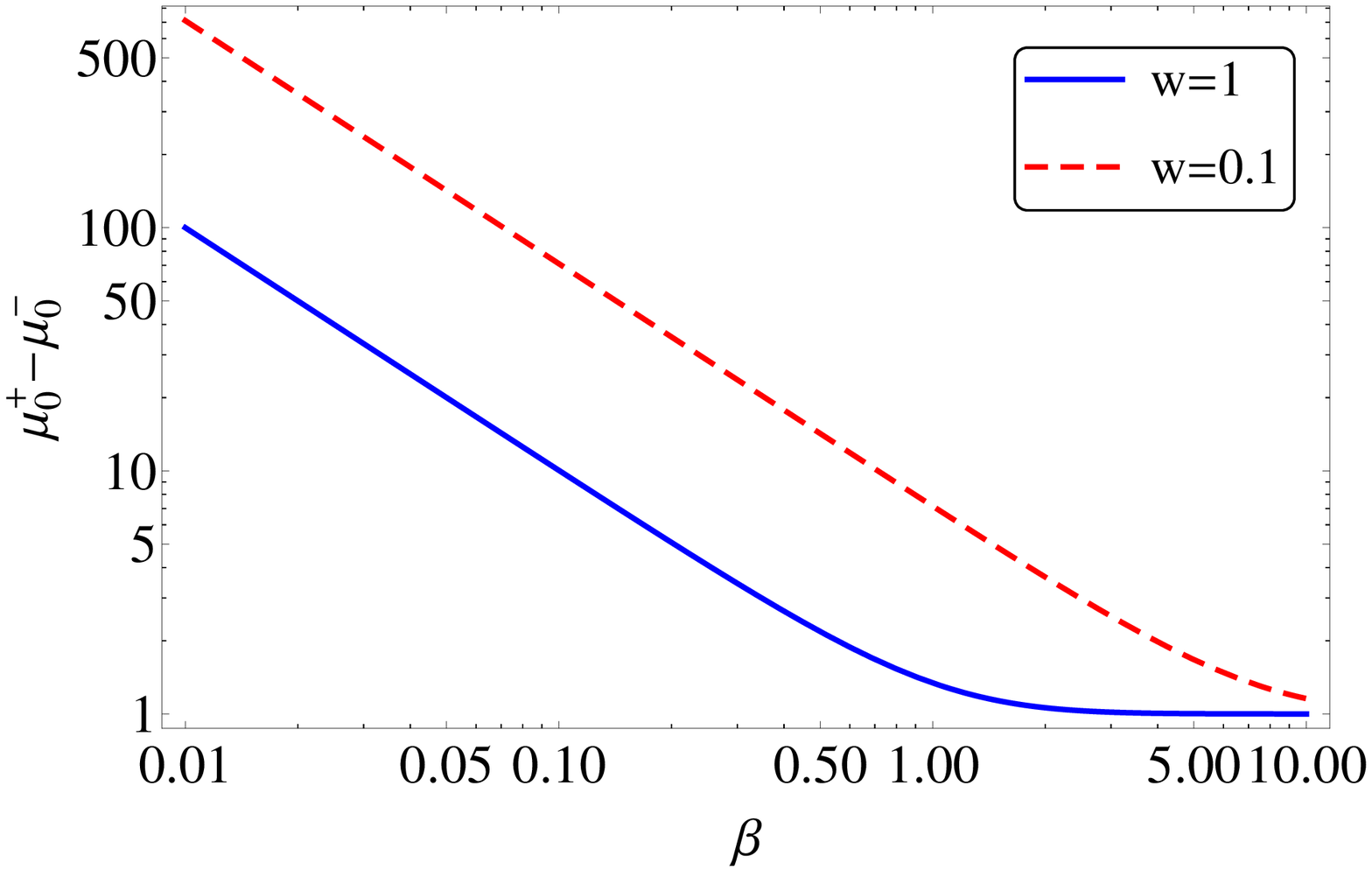}
  \centerline{(d) $\mu_0^{+}(\beta)-\mu_0^{-}(\beta)$}
\end{minipage} \\ \hspace*{-5pt}
\begin{minipage}[b]{8.0cm}
\includegraphics[width=8.0cm]{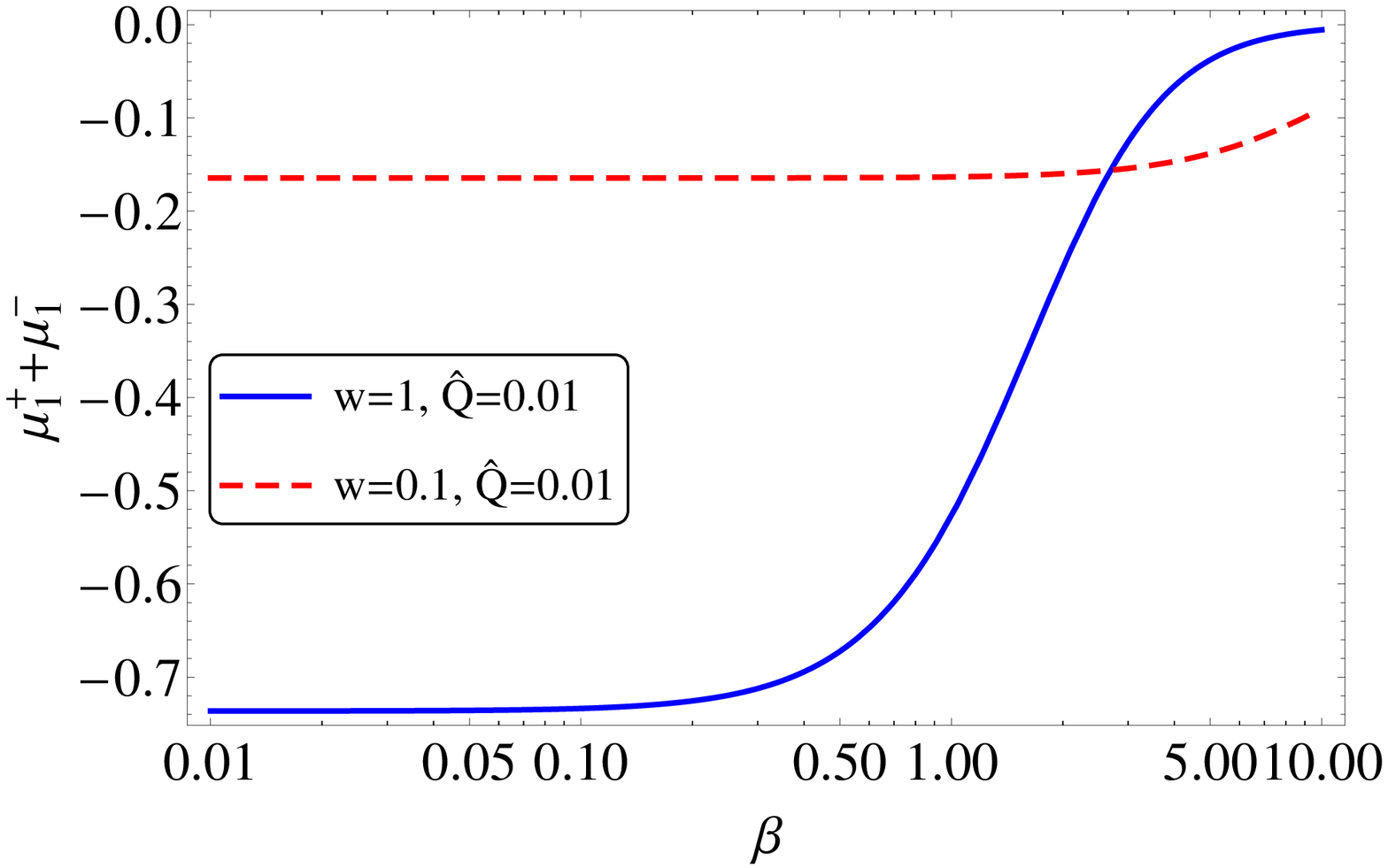} \vspace{-6pt}
  \centerline{(e) $\mu_1^{+}(\beta)+\mu_1^{-}(\beta)$}
\end{minipage} \hspace*{19pt}
\begin{minipage}[b]{7.9cm}
\includegraphics[width=7.9cm]{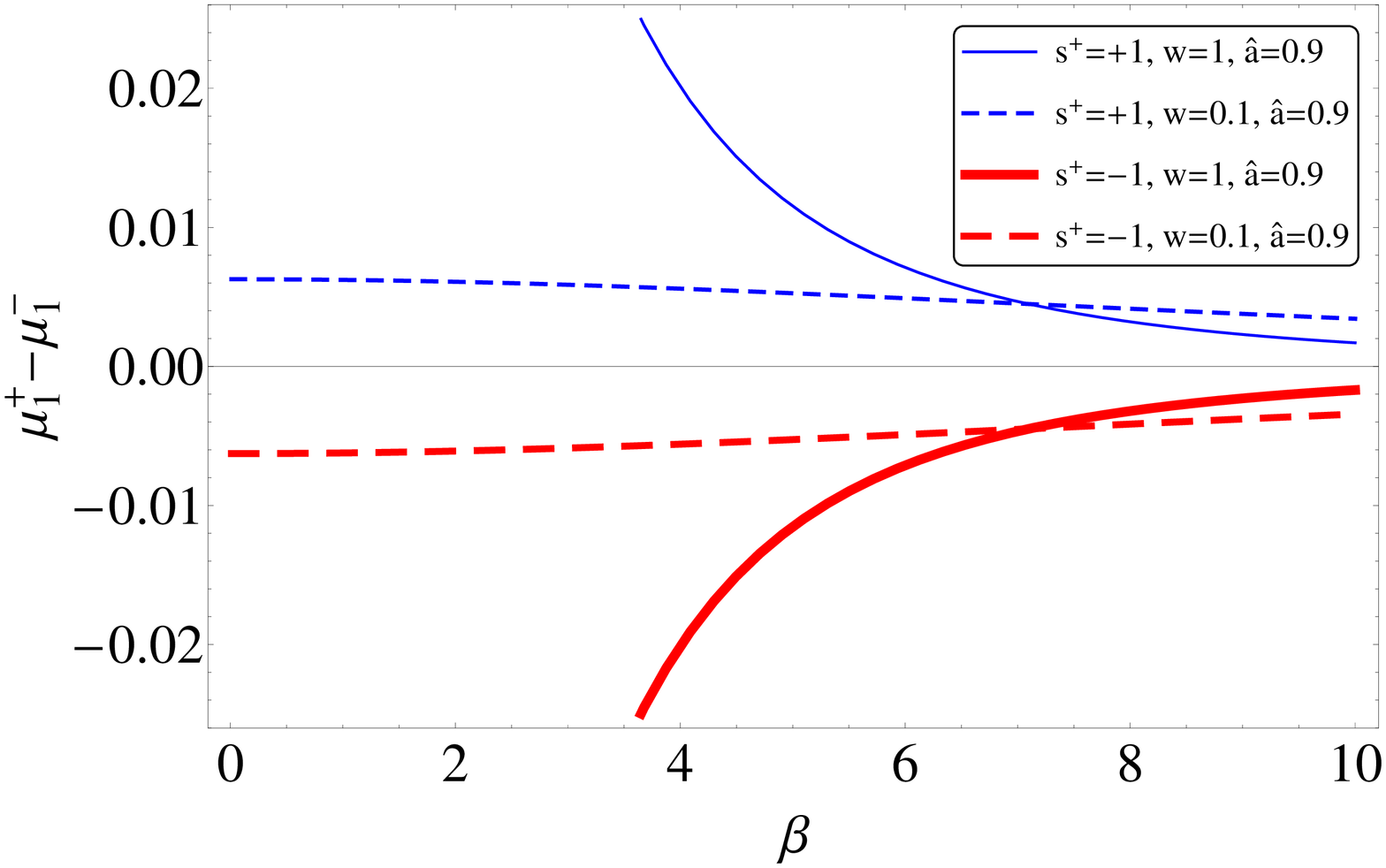}
  \centerline{(f) $\mu_1^{+}(\beta)-\mu_1^{-}(\beta)$}
\end{minipage} \\  \hspace*{1pt}
\begin{minipage}[b]{7.9cm}
\includegraphics[width=7.9cm]{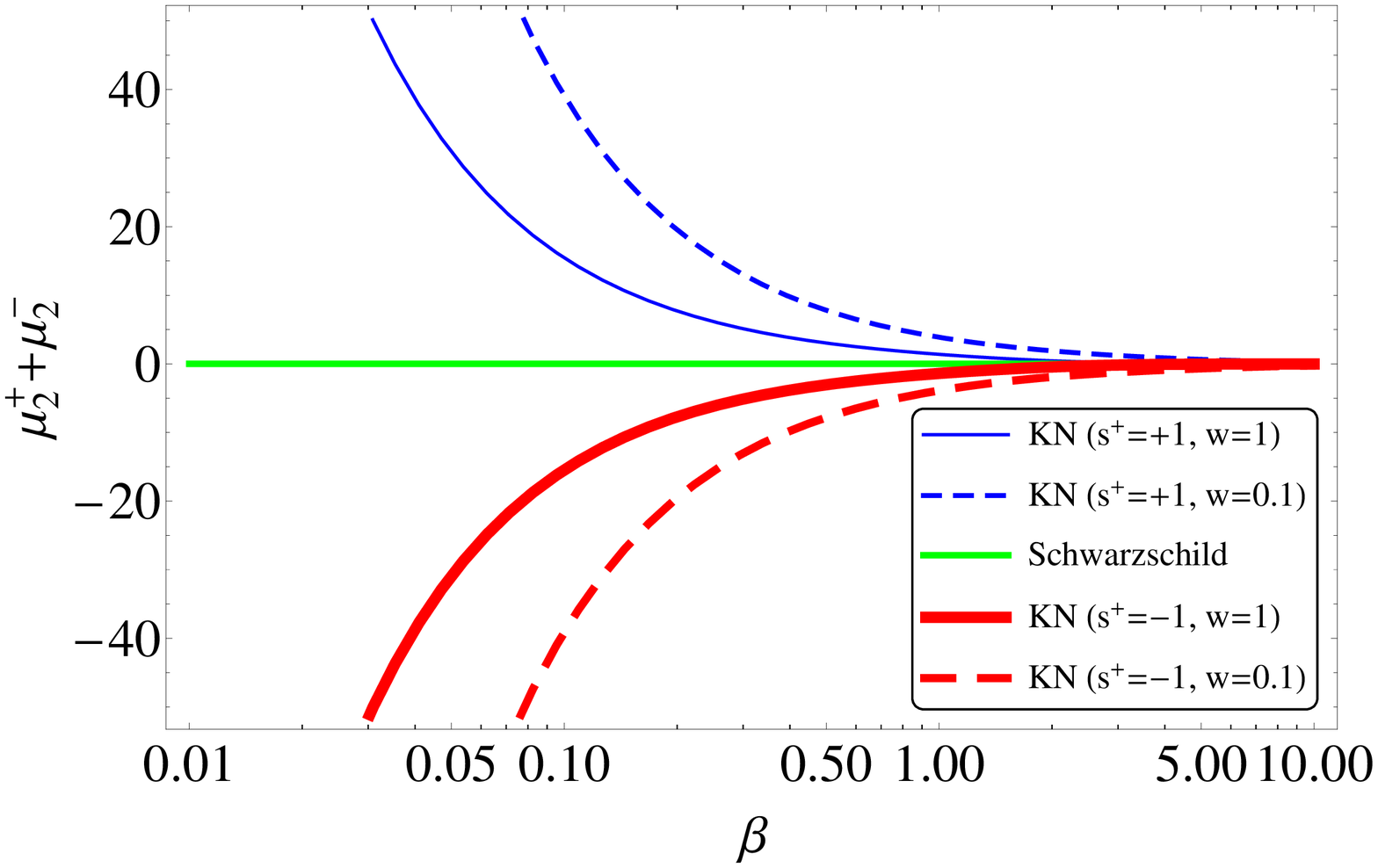} \vspace{-3.5pt}
  \centerline{(g) $\mu_2^{+}(\beta)+\mu_2^{-}(\beta)$}
\end{minipage} \hspace*{12pt}
\begin{minipage}[b]{8.35cm}
\includegraphics[width=8.35cm]{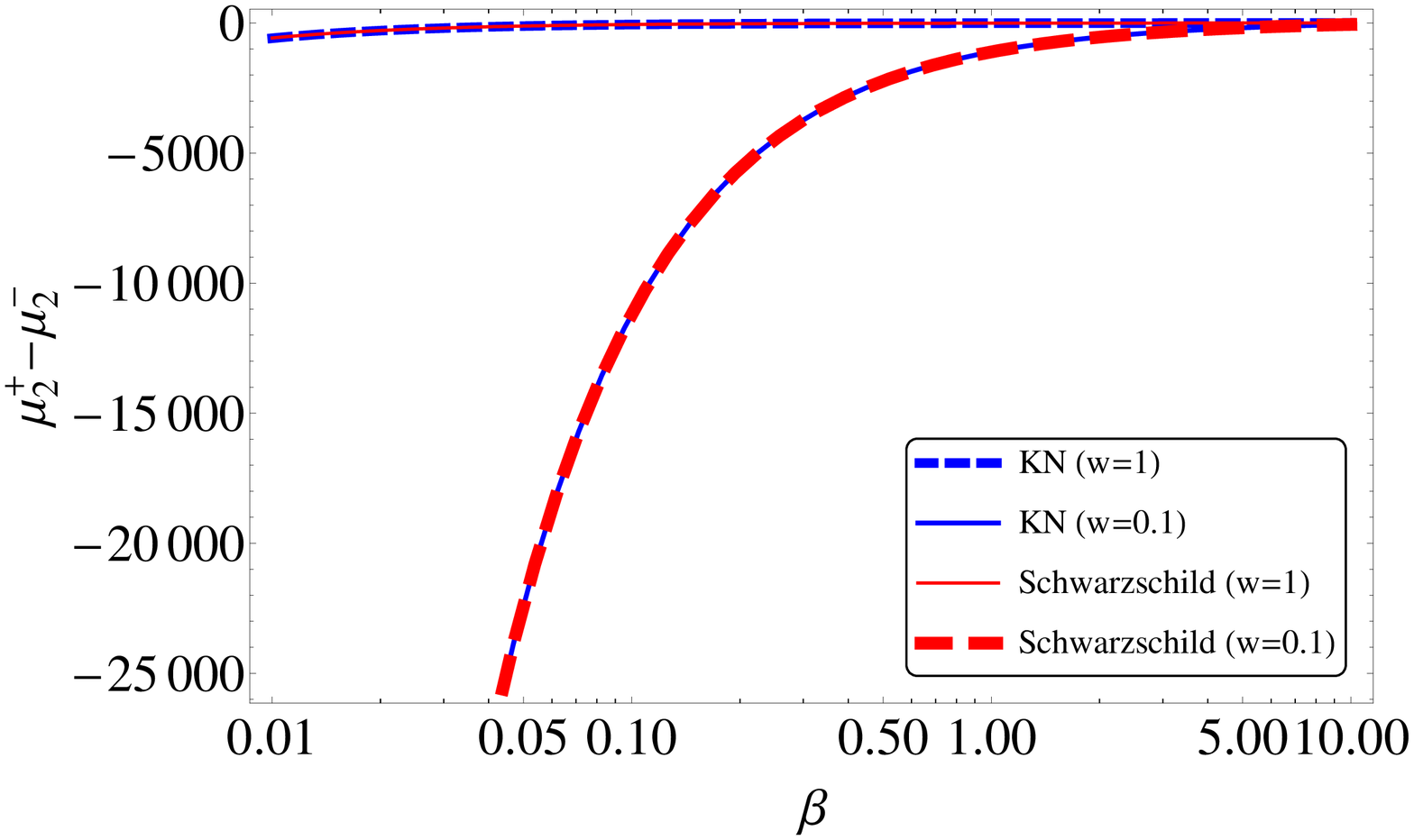}
  \centerline{(h) $\mu_2^{+}(\beta)-\mu_2^{-}(\beta)$}
\end{minipage} \vspace*{-3pt}
\caption{$\mu_0^{+}$, $\mu_1^{+}$, $\mu_2^{+}$, $\mu_0^{+}-\mu_0^{-}$, $\mu_1^{+}\pm\mu_1^{-}$, and $\mu_2^{+}\pm\mu_2^{-}$ plotted as the functions of $\beta$ for prograde ($s^{+}=+1$) or retrograde ($s^{+}=-1$) motion of the massive particle.    } \label{Figure3}
\end{figure*}
With respect to Eqs.~\eqref{mu-SeriesE} - \eqref{mu2}, there are two aspects which are worth pointing out. First, for the case of $w=1$ and no intrinsic angular momentum and electrical charge of the black hole ($a=Q=0$), Eqs.~\eqref{mu0} - \eqref{mu2} are in accord with the null result in Schwarzschild geometry~\cite{KP2005}. Second, as done in Sect.~\ref{IP}, the coefficients of the magnifications ($\mu^+$ and $\mu^-$) of the positive- and negative-parity images can be expressed in terms of the source position as follows:
\begin{eqnarray}
&&\mu_0^{\pm}=\frac{1}{2}\pm\frac{1+\frac{1}{w^2}+\beta^2}{2|\beta|\sqrt{\beta^2+2\left(1+\frac{1}{w^2}\right)}} ~,  \label{mu0-beta}  \\
&&\mu_1^{\pm}=-\frac{3\pi\left(1+\frac{4}{w^2}\right)-\frac{16s^{\pm}\hat{a}}{w}-\pi\left(1+\frac{2}{w^2}\right)\hat{Q}^2}{16\left[\beta^2+2\left(1+\frac{1}{w^2}\right)\right]^{3/2}}~,~~~~  \label{mu1-beta}
\end{eqnarray}
\begin{widetext}
\begin{eqnarray}
\nn&&\mu_2^{\pm}=\frac{\left(1+\frac{1}{w^2}\right)^3+\beta^2\left[2\beta^2+3\left(1+\frac{1}{w^2}\right)\right]^2
\pm\left[3\left(1+\frac{1}{w^2}\right)^2+8\left(1+\frac{1}{w^2}\right)\beta^2+4\beta^4\right]|\beta|\sqrt{\beta^2+2\left(1+\frac{1}{w^2}\right)}}
{3\left[\beta^2\pm|\beta|\sqrt{\beta^2+2\left(1+\frac{1}{w^2}\right)}\,\right]\left[\beta^2+2\left(1+\frac{1}{w^2}\right)\pm|\beta|\sqrt{\beta^2+2\left(1+\frac{1}{w^2}\right)}\,\right]^5}   \\
\nn&&\hspace*{26pt}\times\Bigg{\{}9\!\left[\frac{3\pi}{4}\!\left(\!1\!+\!\frac{4}{w^2}\!\right)\!-\!\frac{4s^{\pm}\hat{a}}{w}\!-\!\frac{\pi}{4}\!\left(\!1\!+\!\frac{2}{w^2}\!\right)\!\hat{Q}^2\right]^2
\!+\!4\!\left[\beta^2\!+\!2\left(1\!+\!\frac{1}{w^2}\right)\!\right]\!\Bigg{\{}\!4D^2\beta^2\!\left(\!1\!+\!\frac{1}{w^2}\!\right)^2\!-\!4(2\!+\!6D\!-\!9D^2)\!\left(\!1\!+\!\frac{1}{w^2}\!\right)^3  \\
&&\hspace*{26pt}-\,3\!\left[\frac{2}{3}\!\left(\!5+\frac{45}{w^2}+\frac{15}{w^4}-\frac{1}{w^6}\!\right)-2\pi\left(\!\frac{3}{w}+\frac{2}{w^3}\!\right)s^{\pm}\hat{a}+2\left(\!1+\frac{1}{w^2}\!\right)\hat{a}^2
-2\left(\!1+\frac{6}{w^2}+\frac{1}{w^4}\!\right)\hat{Q}^2+\frac{\pi s^{\pm}\hat{a}\hat{Q}^2}{w}\right]\!\!\Bigg{\}}\!\Bigg{\}}~.~~~~ \label{mu2-beta}
\end{eqnarray}
\end{widetext}
Notice that the first-order relation $\mu_1^+=\mu_1^-$, which holds well in the static and spherically symmetric spacetime, breaks down in our stationary axisymmetric geometry, because of the presence of the spin-induced contributions to the deflection angle $\hat{\alpha}$. Since both $\mu_1^+$ and $\mu_1^-$ are always negative due to $N_2^{\pm}=N_2(s\!\rightarrow\!s^{\pm})>0$, the magnitudes of the positive- and negative-parity images are corrected by a different amount in the same direction.

Similarly, we are able to obtain the measurable sum and difference relations for the coefficients of the signed magnifications on the basis of Eqs.~\eqref{mu0-beta} - \eqref{mu2-beta} directly:
\begin{eqnarray}
&&\mu_0^{+}+\mu_0^{-}=1~, \label{mu0-PPN}  \\
&&\mu_0^{+}-\mu_0^{-}=\frac{1+\frac{1}{w^2}+\beta^2}{|\beta|\sqrt{\beta^2+2\left(1+\frac{1}{w^2}\right)}}~,  \label{mu0-PMN}  \\
&&\mu_1^{+}+\mu_1^{-}=-\frac{\pi\!\left[3\left(1+\frac{4}{w^2}\right)-\left(1+\frac{2}{w^2}\right)\hat{Q}^2\right]}{8\left[\beta^2+2\left(1+\frac{1}{w^2}\right)\right]^{3/2}}~,~~~~   \label{mu1-PPN} \\
&&\mu_1^{+}-\mu_1^{-}=\frac{2s^{+}\hat{a}}{w\left[\beta^2+2\left(1+\frac{1}{w^2}\right)\right]^{3/2}}~,~~~~~~  \label{mu1-PMN}
\end{eqnarray}
\begin{widetext}
\begin{eqnarray}
&&\mu_2^{+}+\mu_2^{-}=\frac{\pi s^{+}\hat{a}\!\left[16+4w^2+15w^4+4w^2(2+3w^2)\beta^2+w^2(2-w^2-2w^2\beta^2)\hat{Q}^2\right]}{2|\beta|\left[2+w^2(2+\beta^2)\right]^{5/2}}  ~,   \label{mu2-PPN}  \\
\nn&&\mu_2^{+}-\mu_2^{-}=\frac{1}{192w^3|\beta|\left(2\!+\!2w^2\!+\!w^2\beta^2\right)^{5/2}}\Big{\{}\!\!-\!3\Big{[}256(1\!+\!10w^2)\!+\!16(448\!-\!27\pi^2)w^4\!+\!8(704\!-\!27\pi^2)w^6\!+\!3(256\!-\!9\pi^2)w^8 \\
\nn&&\hspace*{1.8cm}+\,512(2-3D)D(1+w^2)^4\Big{]}+128w^2\!\left[2(11D-6)D(1+w^2)^3-3(1+9w^2+19w^4+3w^6)\right]\beta^2   \\
\nn&&\hspace*{1.8cm}+\,256D^2w^4(1\!+\!w^2)^2\beta^4\!-\!384w^4\!\left[2(1\!-\!w^2\!+\!w^4)\!+\!w^2(1\!+\!w^2)\beta^2\right]\hat{a}^2\!+\!6w^2\left[128\!+\!8(112\!-\!9\pi^2)w^2  \right.  \\
&&\hspace*{1.8cm}\left. +\,(896-54\pi^2)w^4+(128-9\pi^2)w^6+64w^2(1+6w^2+w^4)\beta^2\right]\hat{Q}^2+9\pi^2w^4(2+w^2)^2\hat{Q}^4\Big{\}}~.   \label{mu2-PNN}
\end{eqnarray}
It can be seen from Eqs.~\eqref{mu0-PPN} - \eqref{mu2-PNN} that the zeroth-order difference, first- and second-order sum and difference relations for the magnification coefficients are dependent on the initial velocity of the massive particle, in contrast to the zeroth-order sum relation. Moreover, the first-order sum, second-order sum and difference relations for the magnification coefficients depend on the intrinsic electrical charge of the black hole. Different from the case of the second-order difference relation, the terms on the right-hand side of the first-order difference and second-order sum relations will disappear (similar to the null case~\cite{AKP2011}), if the lens' spin is absent.

Fig.\,\ref{Figure3} shows the magnitudes of the coefficients of the zeroth-, first-, and second-order contributions to the magnification of a positive-parity image, as well as those of the sum and difference relations given in Eqs.~\eqref{mu0-PMN} - \eqref{mu2-PNN}, for a given massive particle which takes prograde or retrograde motion.

\subsubsection{Total magnification and centroid}
The total magnification and the magnification-weighted centroid position serve as the important observables, when it is hard to distinguish the angular positions of two images. The total magnification is defined by
\begin{equation}
\mu_{\text{tot}}\equiv|\mu^{+}|+|\mu^{-}|~,  \label{mu-tot}
\end{equation}
which reads up to the second order in $\varepsilon$
\begin{equation}
\mu_{\text{tot}}=\frac{1+\frac{1}{w^2}+\beta^2}{|\beta|\sqrt{\beta^2+2\left(1+\frac{1}{w^2}\right)}}\!+\!\frac{2s^{+}\hat{a}}{w\left[\beta^2+2\left(1+\frac{1}{w^2}\right)\right]^{3/2}}\varepsilon
+\,(\mu_2^{+}-\mu_2^{-})\varepsilon^2+\mathcal{O}(\varepsilon^3)~,   \label{mu-tot-Series}
\end{equation}
with $\mu_2^{+}-\mu_2^{-}$ being given in Eq.~\eqref{mu2-PNN}. In the limit $w\rightarrow1,~a\rightarrow0$, and $Q\rightarrow0$, the total magnification for the case of Schwarzschild lensing of light is recovered~\cite{KP2005}:
\begin{equation}
\mu_{\text{tot-S}}=\frac{2+\beta^2}{|\beta|\sqrt{4+\beta^2}}+\frac{2025\pi^2-1024(4+\beta^2)\left[12(1+D)-D^2(18+\beta^2)\right]}{192|\beta|\left(4+\beta^2\right)^{5/2}}\varepsilon^2+\mathcal{O}(\varepsilon^3)~.  \label{mu-tot-Swar}
\end{equation}

The scaled magnification-weighted centroid position takes the form
\begin{equation}
{\it\Theta}_{\text{cent}}=\frac{\theta^{+}|\mu^{+}|-\theta^{-}|\mu^{-}|}{|\mu^{+}|+|\mu^{-}|}~,  \label{centroid}
\end{equation}
or, in more detail,
\begin{eqnarray}
\nn&&{\it\Theta}_{\text{cent}}=-\frac{N_1^3-64(\theta_0^{+})^6}{4\theta_0^{+}\!\left[N_1^2\!+\!16(\theta_0^{+})^4\right]}
\!+\!\frac{4(N_2^{+}\!-\!N_2^{-})(\theta_0^{+})^2\left[N_1^2\!-\!4N_1(\theta_0^{+})^2\!+\!16(\theta_0^{+})^4\right]}{\left[N_1^2+16(\theta_0^{+})^4\right]^2}\varepsilon
\!-\!\frac{1}{6\theta_0^{+}\!\left[N_1\!+\!4(\theta_0^{+})^2\right]^2\left[N_1^2\!+\!16(\theta_0^{+})^4\right]^3 }    \\
\nn&&\hspace*{20pt}\times\Big{\{}D^2N_1^{10}-24(1-D)DN_1^2\left[N_1^7-16384(\theta_0^{+})^{14}\right](\theta_0^{+})^2+16\left[N_1^3(4+6D-9D^2)+6(N_3^{+}+N_3^{-})\right]  \\
\nn&&\hspace*{20pt}\times\left[N_1^5\!+\!4N_1^4(\theta_0^{+})^2\!-\!256N_1(\theta_0^{+})^8\!-\!1024(\theta_0^{+})^{10}\right](\theta_0^{+})^4\!+\!768\left[(N_2^{-})^2\!+\!N_2^{+}N_2^{-}\!+\!(N_2^{+})^2\right]
\left[64(\theta_0^{+})^6\!-\!N_1^3\right](\theta_0^{+})^6   \\
&&\hspace*{20pt}+\,256N_1\!\left\{6\!\left[(N_2^{-})^2\!+\!4N_2^{+}N_2^{-}\!+\!(N_2^{+})^2\right]\!-\!N_1^4(6\!-\!5D)D\right\}\!\left[N_1\!-\!4(\theta_0^{+})^2\right]\!(\theta_0^{+})^8
\!-\!262144N_1D^2(\theta_0^{+})^{18}\Big{\}}\varepsilon^2\!+\!\mathcal{O}(\varepsilon^3)~,~~~~~~~        \label{centroid-Series}
\end{eqnarray}
where $N_3^{\pm}=N_3(s\rightarrow s^{\pm})$. In terms of the angular source position, Eq.~\eqref{centroid-Series} becomes
\begin{equation}
{\it\Theta}_{\text{cent}}={\it\Theta}_{\text{cent,0}}+{\it\Theta}_{\text{cent,1}}\,\varepsilon+{\it\Theta}_{\text{cent,2}}\,\varepsilon^2+\mathcal{O}(\varepsilon^3)~, \label{centroid-Series-2}
\end{equation}
\begin{figure}
\centering
\begin{minipage}[b]{8.1cm}
\includegraphics[width=8.1cm]{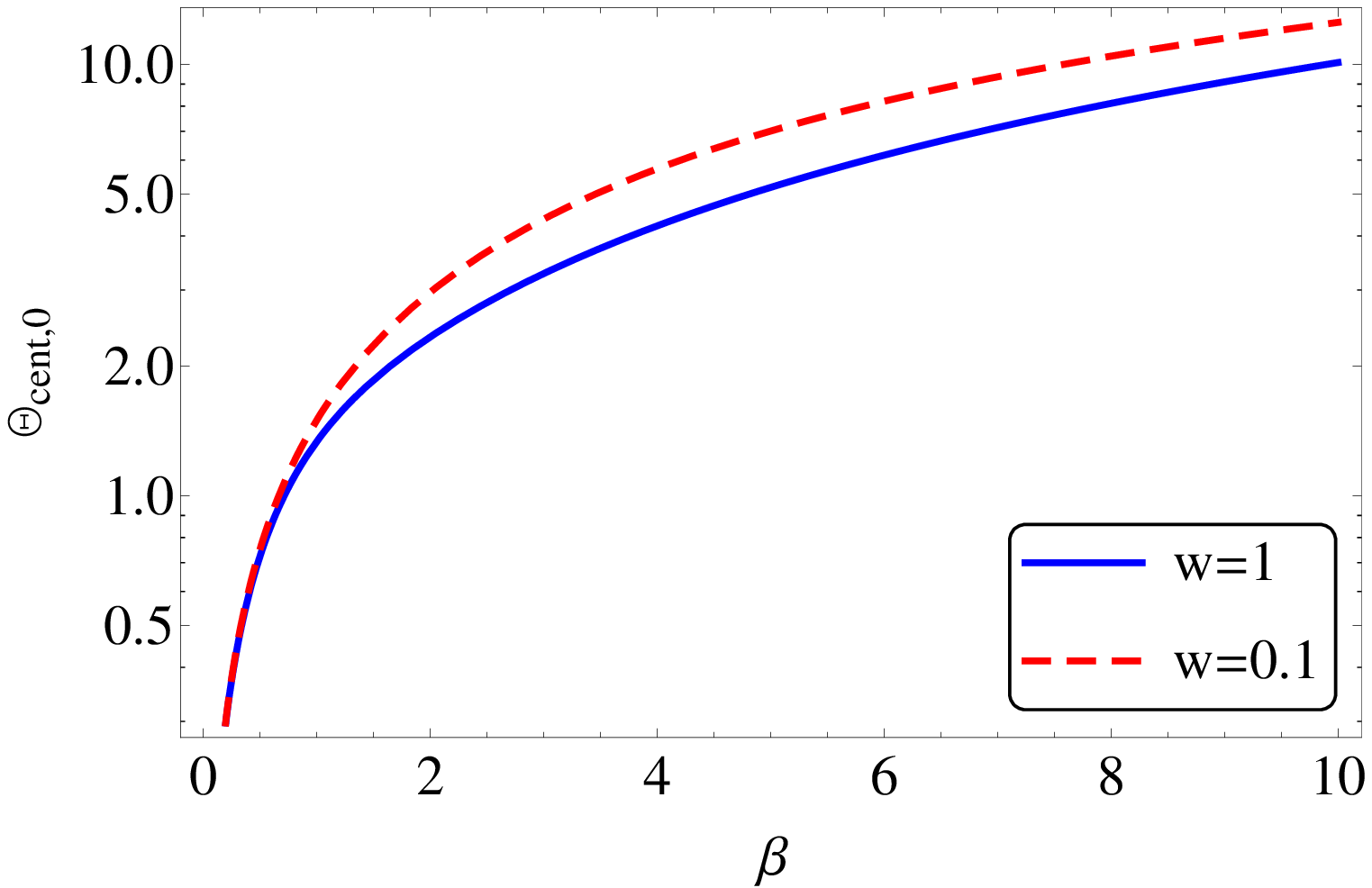} \vspace{-6pt}
  \centerline{(a) ${\it\Theta}_{\text{cent,0}}(\beta)$ }
\end{minipage} \hspace*{22pt}
\begin{minipage}[b]{8.cm}
\includegraphics[width=8.cm]{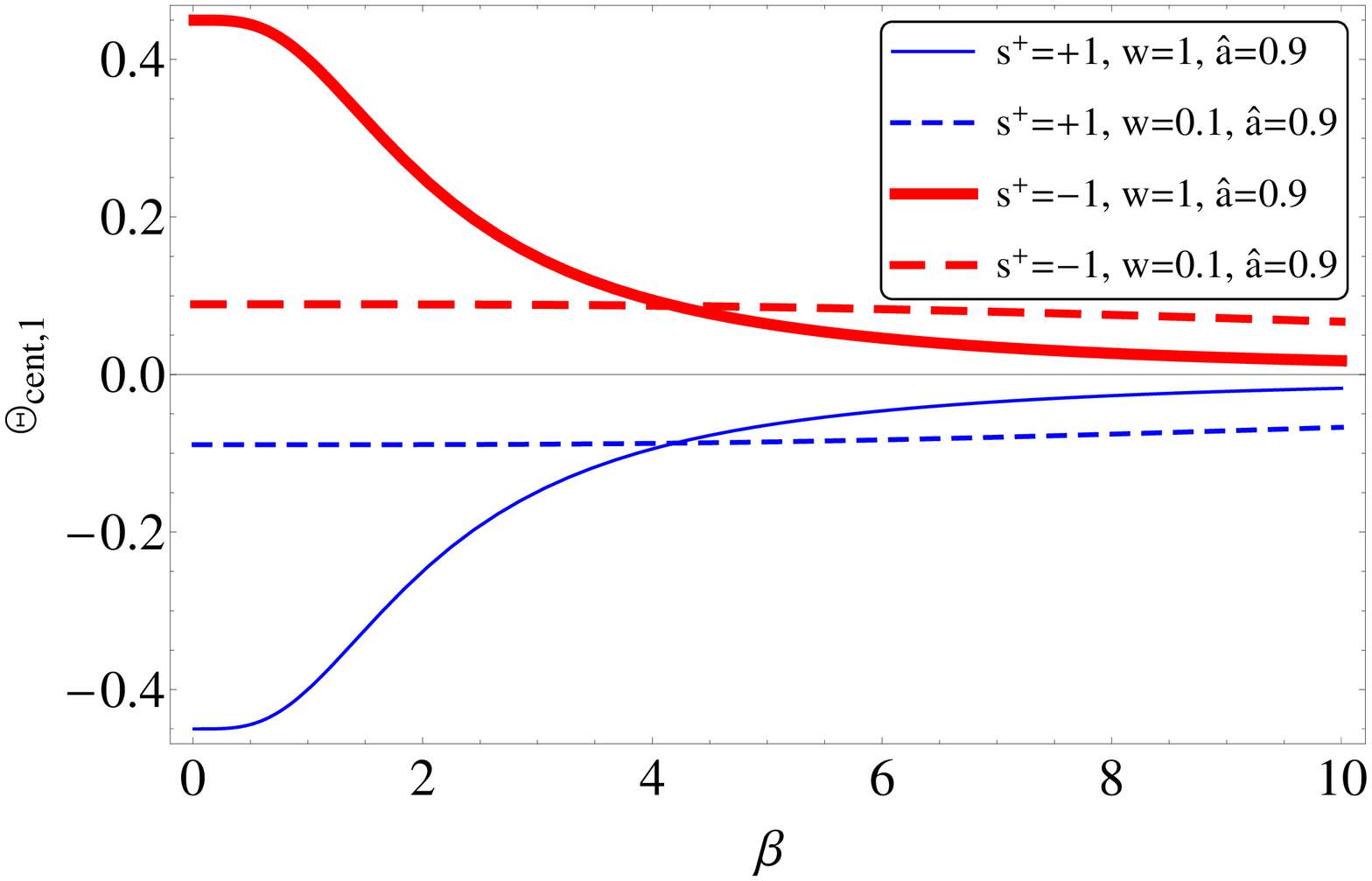}
  \centerline{(b) ${\it\Theta}_{\text{cent,1}}(\beta)$ }
\end{minipage} \vspace*{8pt} \\
\begin{minipage}[b]{8.2cm}
\includegraphics[width=8.2cm]{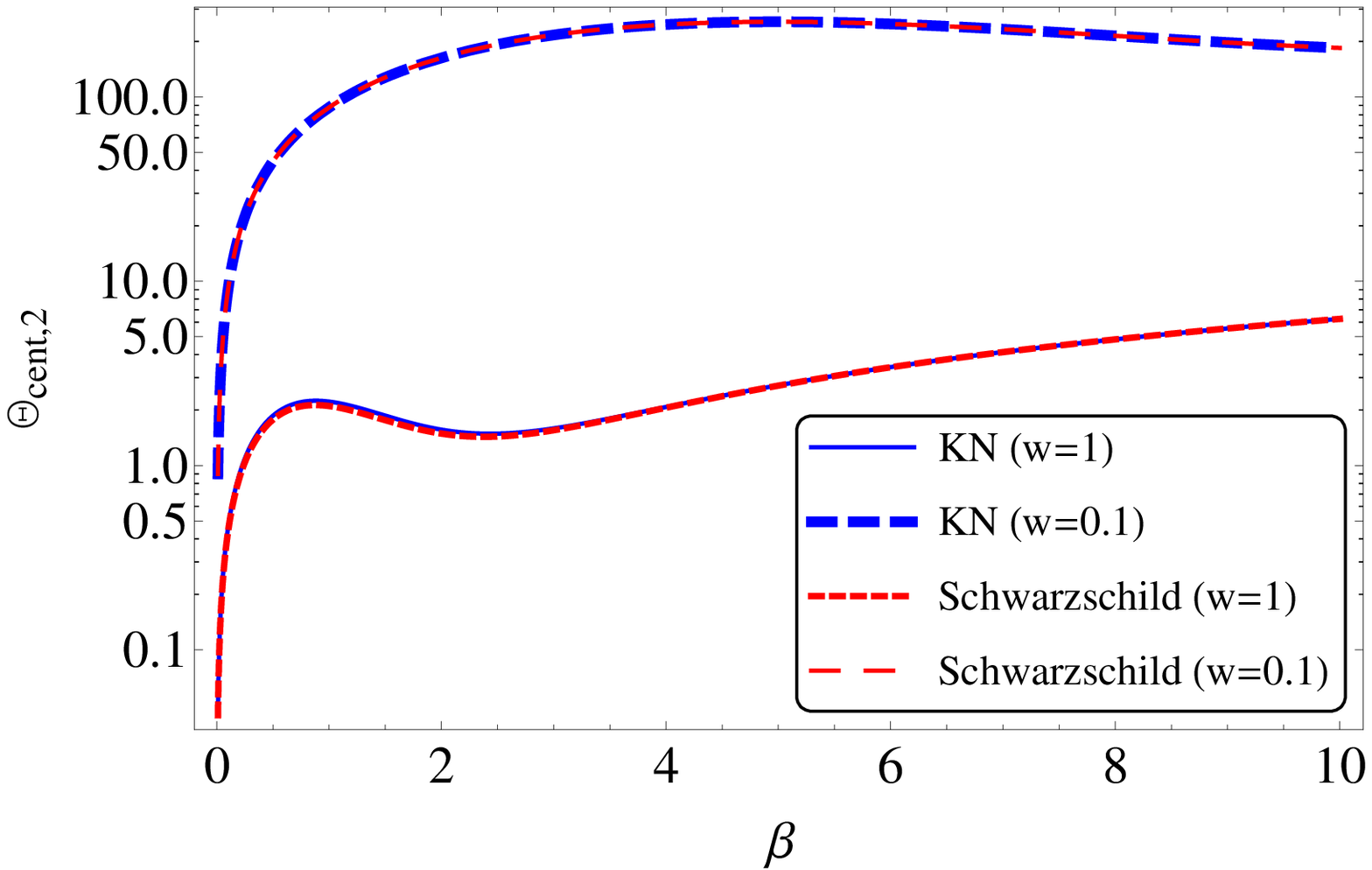}
  \centerline{(c) ${\it\Theta}_{\text{cent,2}}(\beta)$ }
\end{minipage}
\caption{${\it\Theta}_{\text{cent,0}}$, ${\it\Theta}_{\text{cent,1}}$, and ${\it\Theta}_{\text{cent,2}}$ plotted as the functions of $\beta$ for the particle's prograde or retrograde motion. } \label{Figure4}
\end{figure}
with
\begin{eqnarray}
&&{\it\Theta}_{\text{cent,0}}=\frac{|\beta|\left[3\left(1+w^2\right)+2w^2\beta^2\right]}{2\left(1+w^2+w^2\beta^2\right)}~,  \label{centroid-Series-beta-1} \\
&&{\it\Theta}_{\text{cent,1}}= -\frac{ws^{+}\hat{a}\left(1+w^2+2w^2\beta^2\right)}{\left(1+w^2+w^2\beta^2\right)^2}~,   \label{centroid-Series-beta-2} \\
\nn&&{\it\Theta}_{\text{cent,2}}=\frac{32|\beta|}{3w^{18}\!\left[\sqrt{\beta^2\!+\!2\left(1\!+\!\frac{1}{w^2}\right)}\!+\!|\beta|\right]\!\!
\left[2\left(1\!+\!\frac{1}{w^2}\right)\!+\!\left(\!\sqrt{\beta^2\!+\!2\left(1\!+\!\frac{1}{w^2}\right)}\!+\!|\beta|\!\right)^{\!2}\right]^2\!\!
\left[4\left(1\!+\!\frac{1}{w^2}\right)^2\!+\!\left(\!\sqrt{\beta^2\!+\!2\left(1\!+\!\frac{1}{w^2}\right)}\!+\!|\beta|\!\right)^{\!4}\right]^3}  \\
\nn&&\hspace*{30pt}\times\Bigg{\{}|\beta|\left[9(1\!+\!w^2)^4\!+\!60w^2(1\!+\!w^2)^3\beta^2\!+\!108w^4(1\!+\!w^2)^2\beta^4
\!+\!72w^6(1\!+\!w^2)\beta^6\!+\!16w^8\beta^8\right]\!+\!\sqrt{\beta^2\!+\!2\left(1\!+\!\frac{1}{w^2}\right)}  \\
\nn&&\hspace*{30pt}\times\left[(1+w^2)^4+20w^2(1+w^2)^3\beta^2+60w^4(1+w^2)^2\beta^4+56w^6(1+w^2)\beta^6+16w^8\beta^8\right]\!\Bigg{\}}  \\
\nn&&\hspace*{30pt}\times\Bigg{\{}384w^4\left[1+w^4+w^2(1+w^2)\beta^2\right]\left[2+w^2(2+\beta^2)\right]\hat{a}^2+(1+w^2+w^2\beta^2)\Big{\{}768+384w^2(20+\beta^2)   \\
\nn&&\hspace*{30pt}-81\pi^2w^4(4\!+\!w^2)^2\!+\!384w^4\!\!\left[56\!+\!9\beta^2\!+\!w^2(44\!+\!19\beta^2)\!+\!3w^4(2\!+\!\beta^2)\right]\!-\!1536Dw^2(1\!+\!w^2)^2\beta^2(2\!+\!2w^2\!+\!w^2\beta^2)     \\
\nn&&\hspace*{30pt}+\,256D^2(1+w^2)\left[-2(1+w^2)^3+17w^2(1+w^2)^2\beta^2+13w^4(1+w^2)\beta^4+2w^6\beta^6\right]     \\
   &&\hspace*{30pt}-\,6w^2\!\left[128\!-\!9\pi^2w^2(8\!+\!6w^2\!+\!w^4)\!+\!64w^2\!\left(14\!+\!\beta^2\!+\!w^4(2\!+\!\beta^2)\!+\!2w^2(7\!+\!3\beta^2)\right)\right]\!\hat{Q}^2
\!-\!9\pi^2w^4(2\!+\!w^2)^2\hat{Q}^4\!\Big{\}}\!\Bigg{\}}~.  \label{centroid-Series-beta-3}
\end{eqnarray}
The coefficients of the zeroth-, first-, and second-order contributions to the scaled centroid are plotted in Fig.\,\ref{Figure4}. In the limit $w\rightarrow1$, Eq.~\eqref{centroid-Series-2} is reduced to
\begin{eqnarray}
\nn&&{\it\Theta}_{\text{cent}}=\frac{|\beta|(3+\beta^2)}{2+\beta^2}-\frac{2(1+\beta^2)\hat{a}s^{+}}{(2+\beta^2)^2}\varepsilon+|\beta|\Bigg{\{}\frac{2(1+\beta^2)\hat{a}^2}{(2+\beta^2)^3}
-\frac{1}{384(4+\beta^2)(2+\beta^2)^2}\Big{\{}2025\pi^2    \\
&&\hspace*{34pt}-1024\!\left[6(4\!+\!\beta^2)(2\!-\!D\beta^2)\!-\!D^2(8\!-\!34\beta^2\!-\!13\beta^4\!-\!\beta^6)\right]
\!-\!6\left[135\pi^2\!-\!512(4\!+\!\beta^2)\right]\!\hat{Q}^2\!+\!81\pi^2\hat{Q}^4\Big{\}}\Bigg{\}}\varepsilon^2+\mathcal{O}(\varepsilon^3)~,~~~~~~~~  \label{centroid-w}
\end{eqnarray}
which is consistent with the result for the case of the Schwarzschild lensing of light~\cite{KP2005} when the electrical charge and angular momentum of the black hole are dropped.

\subsection{Differential time delay}
The difference between the time delays of the primary and secondary images is another traditional lensing observable. In order to obtain its analytical form, we have to derive the Shapiro time delay of a test particle propagating from the source to the observer in the equatorial plane of the KN black hole firstly.

To our knowledge, the calculations of the Shapiro time delay of light were performed via various approaches~\cite{LJ2020b,We1972,RM1983,Moyer2003,Sereno2004,PLT2004,TP2008,HBP2014,AB2010,KZ2010,HL2016b}, such as the classical one given in Ref.~\cite{We1972}, the Richter-Matzner method~\cite{RM1983}, the Fermat's principle method~\cite{Sereno2004}, and the approach based on the time transfer functions~\cite{PLT2004,TP2008,HBP2014}. However, it has been found that the result of the second-order contributions to the gravitational time delay takes diverse ways in different approaches, and further work is thus needed with respect to this issue to get a perfect agreement. In this work we adopt the classical method~\cite{We1972,KP2005} to perform our derivation.

According to Eqs.~\eqref{C1}, \eqref{C2}, \eqref{dot-t}, and \eqref{dot-r}, we have
\begin{eqnarray}
\left|\frac{dt}{dr}\right|=\frac{1+\frac{a^2}{r^2}\left(1+\frac{2M}{r}-\frac{Q^2}{r^2}\right)-\frac{swab}{r^2}\left(\frac{2M}{r}-\frac{Q^2}{r^2}\right) }
{b\left(1-\frac{2M}{r}+\frac{a^2+Q^2}{r^2}\right)\sqrt{\frac{1}{b^2}\left(1+\frac{a^2}{r^2}-\frac{swab}{r^2}\right)^2
-\frac{1}{r^2}\left(1-\frac{2M}{r}+\frac{a^2+Q^2}{r^2}\right)\left[\left(w-\frac{sa}{b}\right)^2+\frac{(1-w^2)r^2}{b^2}\right]}}~. \label{dt-dr}
\end{eqnarray}
The travelling time of a massive particle propagating from the point (with the radial coordinate $r_0$) of the closest approach to the black hole to an arbitrary but finite point (with a radial coordinate $R\geq r_0$) of its trajectory is then written as
\begin{eqnarray}
\nn&&T(R)=\int_{r_0}^R\left|\frac{dt}{dr}\right|dr  \\
\nn&&\hspace*{25.5pt}=r_0\int_{\frac{r_0}{R}}^{1}\frac{1+\hat{a}^2h^2x^2+\hat{a}\left(\hat{a}h-\frac{swb}{r_0}\right)\left(2-\hat{Q}^2hx\right)h^2x^3}{x^2\left[1-2hx+\left(\hat{a}^2+\hat{Q}^2\right)h^2x^2\right]}  \\
&&\hspace*{34pt}\times\left\{\left[1+\hat{a}h\left(\hat{a}h-\frac{swb}{r_0}\right)x^2\right]^2-\left[1-2hx+\left(\hat{a}^2+\hat{Q}^2\right)h^2x^2\right]
\left[1-w^2+\left(\frac{wb}{r_0}-s\hat{a}h\right)^2x^2\right]\right\}^{-\frac{1}{2}}dx~,   \label{T-Define}
\end{eqnarray}
where $h,~\hat{a}$, and $\hat{Q}$ have been defined in Eq.~\eqref{haQ}, and $x$ and $b/r_0$ have been given in Eqs.~\eqref{alpha-exact-2} and \eqref{b-r0-4}, respectively. By performing the series expansion of the integrand of Eq.~\eqref{T-Define} in $h$ and then integrating it over $x$, we obtain via defining $\xi\equiv r_0/R$
\begin{eqnarray}
\nn&&T(R)=\frac{\sqrt{R^2-r_0^2}}{w}+\frac{hr_0}{w^3}\!\left[\frac{\sqrt{1-\xi^2}}{1+\xi}+(3w^2\!-\!1)\ln\!\left(\frac{1+\sqrt{1-\xi^2}}{\xi}\right)\!\right]
\!+\!\frac{h^2r_0}{w}\Bigg{\{}\!\frac{3\left(5-\hat{Q}^2\right)}{2}\!\left(\frac{\pi}{2}-\arcsin{\xi}\right)+\frac{\sqrt{1-\xi^2}}{2w(1+\xi)^2}   \\
\nn&&\hspace*{36pt}\times\!\left[\frac{1\!+\!(1\!-\!6w^2)(1\!+\!\xi)}{w^3}\!-\!4s\hat{a}(1\!+\!\xi)[1\!+\!w^2(1\!+\!\xi)]\right]\!\!\Bigg{\}}\!+\!\frac{h^3r_0}{w}\!\Bigg{\{}\!w\!\!\left[\frac{15}{2w^3}
\!+\!3\left(1\!+\!\frac{4}{w^2}\right)s\hat{a}\!-\!\frac{3\hat{Q}^2}{2w^3}\!-\!\frac{1}{2}\left(1\!+\!\frac{2}{w^2}\right)s\hat{a}\hat{Q}^2\right]      \\
\nn&&\hspace*{36pt}\times\left(\arcsin{\xi}-\frac{\pi}{2}\right)+\sqrt{1-\xi^2}\,\Bigg{[}\frac{35-15\hat{Q}^2}{2}+\frac{6w^2-1+w^4\left[23-7\hat{Q}^2+2(1+w^2)\hat{a}^2\right]}{2w^6(1+\xi)}
+\frac{1-(1+3w^2)(1+\xi)}{2w^6(1+\xi)^3}   \\
&&\hspace*{36pt}+\,\frac{s\hat{a}\left\{4[(1+4w^2)(1+\xi)-1]-w^4(6-\hat{Q}^2)(1+\xi)^2\xi\right\} }{2w^3(1+\xi)^2}\Bigg{]}\!\Bigg{\}}+\mathcal{O}(h^4)~,   \label{T-Series}
\end{eqnarray}
in agreement with the result presented in Ref.~\cite{KP2005} for the case of $w=1$ and $a=Q=0$.

Therefore, the weak-field gravitational time delay for a massive particle propagating in the equatorial plane of the KN black hole from the source $S$ to the observer $O$ can be given by
\end{widetext}
\begin{equation}
\tau=T(R_S)+T(R_O)-\frac{d_S}{\cos\mathcal{B}} ~,   \label{T-total}
\end{equation}
where the radial coordinates of the source and observer are, respectively,
\begin{equation}
R_S=\sqrt{d_{LS}^2+d_{S}^2\tan^2\mathcal{B}}~,~~~~~~~R_O=d_L~.~~  \label{RS-RO}
\end{equation}

For the sake of comparison with the actual astronomical observations, it is more convenient to express Eq.~\eqref{T-total} in terms of the angular variables through Eqs.~\eqref{b-r0-4} and \eqref{variables}, and the relations $b=d_L\sin\vartheta$ and $M_{\bullet}=d_L\tan\vartheta_{\bullet}$. For this purpose, we first need to evaluate the magnitudes of the quantities $M/b$, $b/R_S$, and $b/R_O$. With the consideration that~\cite{KP2005}
\begin{eqnarray}
&&\frac{M}{b}\sim\varepsilon~,~~  \label{Magnitudes-1}  \\
&&\frac{b}{R_S}\sim\frac{D(1-D)}{\sqrt{D^2+\tan^2\mathcal{B}}}\varepsilon ~,~~~~  \label{Magnitudes-2}  \\
&&\frac{b}{R_O}\sim D\varepsilon ~,~~  \label{Magnitudes-3}
\end{eqnarray}
Eq.~\eqref{T-Series} can be expanded as power series in the small parameter $\varepsilon$
\begin{widetext}
\begin{equation}
\frac{T(R)}{R}=\frac{1}{w}-\frac{1}{2w}\frac{b}{R}\!\left[\frac{b}{R}-\frac{2}{w^2}\frac{M}{b}+2\left(\!3\!-\!\frac{1}{w^2}\!\right)\frac{M}{b}\ln\!\left(\frac{b}{2R}\right)\right]
+\frac{b}{R}\frac{M^2}{b^2}\!\left[\frac{3\pi(5-\hat{Q}^2)}{4w}\!-\!2\left(1+\frac{1}{w^2}\right)s\,\hat{a}\right]+\mathcal{O}(\varepsilon^4)~,   \label{T-Series-2}
\end{equation}
which yields immediately the power-series expansion of Eq.~\eqref{T-total} as
\begin{eqnarray}
\nn&&\tau=\left(\frac{1}{w}-1\right)d_S+\frac{8d_Ld_{LS}}{w^3d_S}\Bigg{\{}\!\left[1-w^2\theta_0^2+w^2\left(1+\frac{(1-w)d_{LS}}{d_L}\right)\beta^2
+\frac{1-3w^2}{2}\ln\left(\frac{d_L\theta_0^2\vartheta_E^2}{4d_{LS}}\right) \right]\!\varepsilon^2  \\
&&\hspace*{21pt}+\frac{3\pi w^2(5-\hat{Q}^2)-8ws\hat{a}(1+w^2)+4(1-3w^2-2w^2\theta_0^2)\,\theta_1}{4\theta_0} \varepsilon^3+\mathcal{O}(\varepsilon^4)\Bigg{\}}~.~~~~~~~  \label{T-Series-3}
\end{eqnarray}
\end{widetext}
The leading term on the right-hand side of Eq.~\eqref{T-Series-3} is a geometrical contribution induced by the velocity effect. From Eq.~\eqref{T-Series-3}, it is obvious that the leading-order contribution induced by the lens' electrical charge to the timelike gravitational time delay is always negative. In addition, the travelling time for a massive particle in prograde motion ($s=+1$) relative to the lens' rotation is less than that for the particle's retrograde motion ($s=-1$). These two conclusions are similar to the lightlike counterparts~\cite{Sereno2004,AKP2011}. Furthermore, by substituting Eq.~\eqref{theta1} into Eq.~\eqref{T-Series-3} and using the natural lensing time scale $\tau_E=d_L\vartheta_E^2/D=4M$, we obtain the desired scaled gravitational time delay
\begin{widetext}
\begin{eqnarray}
\nn&&\hat{\tau}\equiv\frac{\tau}{\tau_E}=\left(\frac{1}{w}-1\right)\frac{d_{LS}}{d_L\vartheta_E^2}
+\frac{1}{2w^3}\left\{1-w^2\theta_0^2+w^2\left[1+\frac{(1-w)d_{LS}}{d_{L}}\right]\beta^2+\frac{1-3w^2}{2}\ln\left(\frac{d_L\theta_0^2\vartheta_E^2}{4d_{LS}}\right)\right\}+\frac{1}{8w^3\theta_0}    \\
&&\hspace*{47pt}\times\!\left\{3\pi w^2(5\!-\!\hat{Q}^2)\!-\!8w(1\!+\!w^2)s\hat{a}\!+\!\frac{1\!-\!3w^2\!-\!2w^2\theta_0^2}{2(1\!+\!w^2\!+\!2w^2\theta_0^2)}
\!\left[3\pi(4\!+\!w^2)\!-\!16ws\hat{a}\!-\!\pi(2+w^2)\hat{Q}^2\right]\!\right\}\varepsilon+\mathcal{O}(\varepsilon^2)~.~~~~~~   \label{T-Series-4}
\end{eqnarray}
Here, there are two aspects which are worth emphasizing. The first one is that if the electrical charge of the lens is dropped and $w$ is equal to $1$, Eq.~\eqref{T-Series-4} will be reduced to the Kerr lensing result of light~\cite{AKP2011}
\begin{equation}
\hat{\tau}=\frac{1}{2}\left[1+\beta^2-\theta_0^2-\ln\left(\frac{d_L\theta_0^2\vartheta_E^2}{4d_{LS}}\right)\right]+\frac{15\pi-16s\hat{a}}{16\theta_0}\varepsilon+\,\mathcal{O}(\varepsilon^2) ~.    \label{T-Series-5}
\end{equation}
\begin{figure}
\centering
\begin{minipage}[b]{8.1cm}
\includegraphics[width=8.1cm]{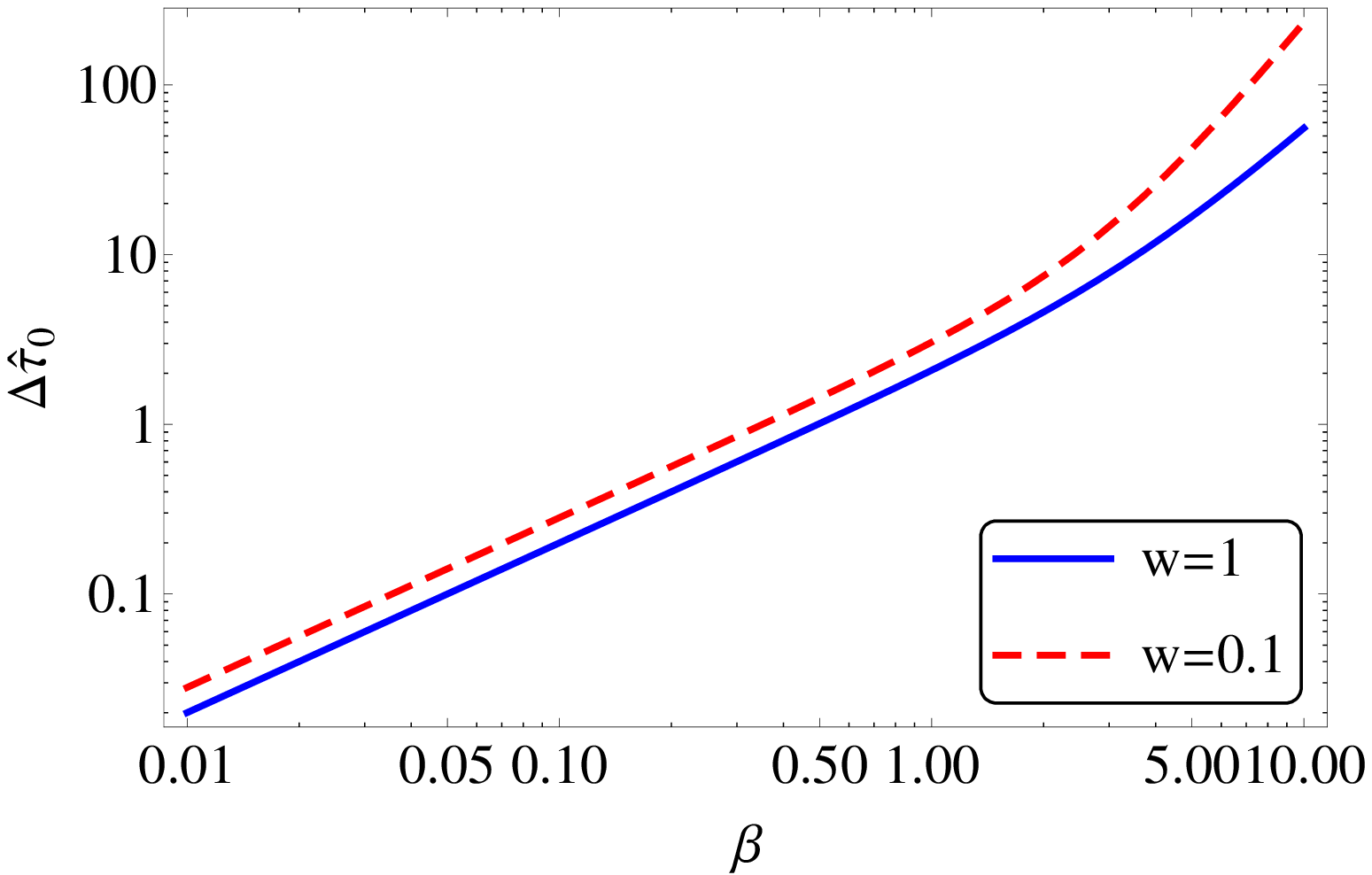} \vspace{-5pt}
  \centerline{(a) $\Delta\hat{\tau}_0(\beta)$}
\end{minipage} \hspace*{15pt}
\begin{minipage}[b]{8.1cm}
\includegraphics[width=8.1cm]{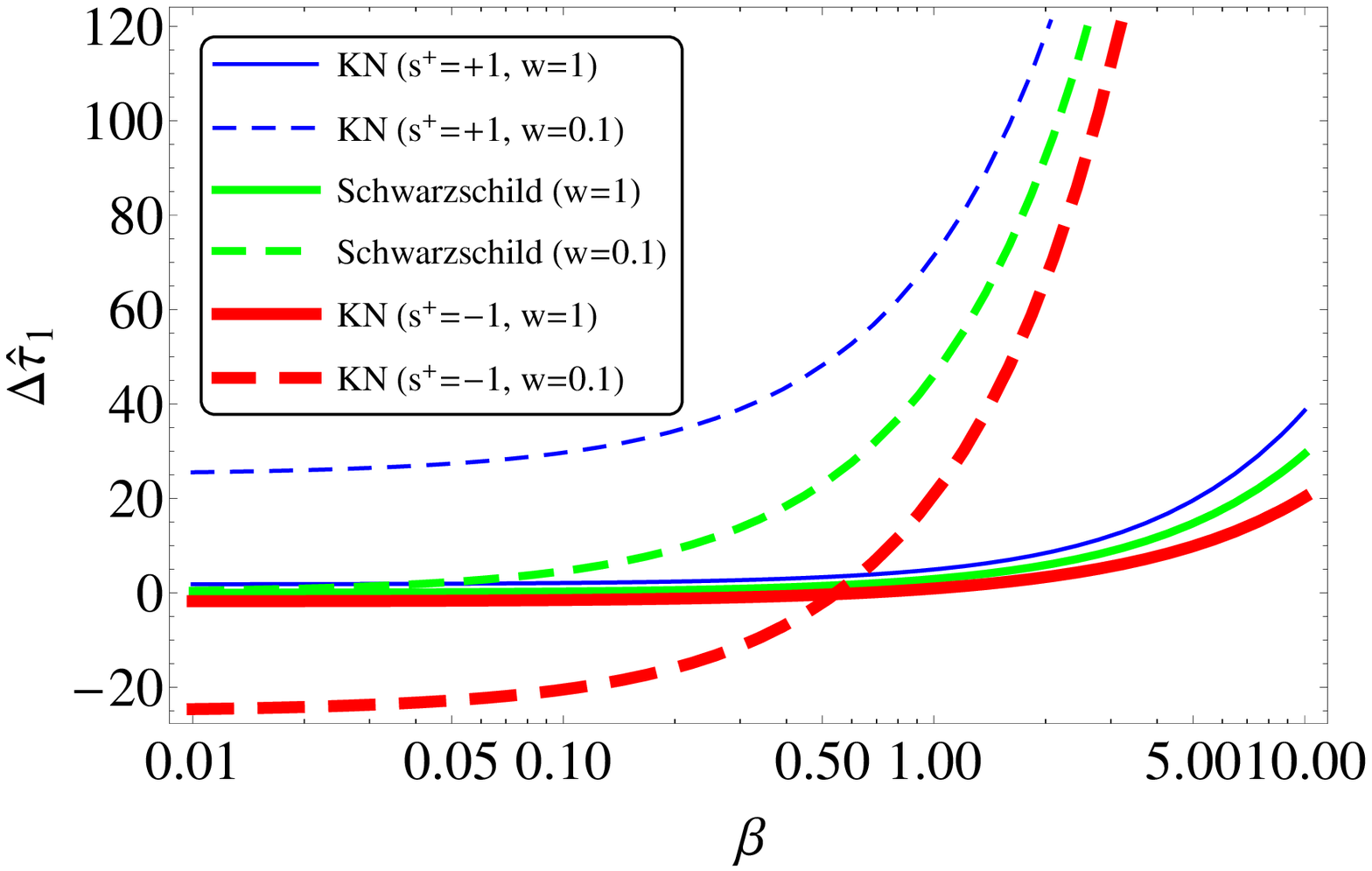}
  \centerline{(b) $\Delta\hat{\tau}_1(\beta)$}
\end{minipage}
\caption{$\Delta\hat{\tau}_0$ and $\Delta\hat{\tau}_1$ plotted as the functions of $\beta$ for the particle's prograde or retrograde motion. } \label{Figure5}
\end{figure}
Secondly, based on Eqs.~\eqref{T-Series-4} and \eqref{NewR-1}, as well as the relation $s^{+}=-s^{-}$, we finally achieve the scaled differential time delay between the positive- and negative-parity images as follows:
\begin{equation}
\Delta\hat{\tau}=\hat{\tau}_{-}-\hat{\tau}_{+}=\Delta\hat{\tau}_0+\Delta\hat{\tau}_1\varepsilon+\mathcal{O}(\varepsilon^2)~, \label{Differential-TD}
\end{equation}
with
\begin{eqnarray}
&&\Delta\hat{\tau}_0=\frac{(\theta_0^{+})^2-(\theta_0^{-})^2}{2w}+\frac{1-3w^2}{2w^3}\ln\left(\frac{\theta_0^{-}}{\theta_0^{+}}\right)~,~~  \label{Delta-1} \\
\nn&&\Delta\hat{\tau}_1=\frac{3\pi(5-\hat{Q}^2)}{8w}\left(\frac{1}{\theta_0^{-}}-\frac{1}{\theta_0^{+}}\right)+\frac{3\pi(4+w^2)-\pi(2+w^2)\hat{Q}^2}{16w^3}
\left\{\frac{1-3w^2-2w^2(\theta_0^{-})^2}{[1+w^2+2w^2(\theta_0^{-})^2]\theta_0^{-}}-\frac{1-3w^2-2w^2(\theta_0^{+})^2}{[1+w^2+2w^2(\theta_0^{+})^2]\theta_0^{+}}\right\}   \\
&&\hspace*{31pt}+\frac{s^{+}\hat{a}}{w^2}\left\{\frac{2-w^2+w^4+2w^4(\theta_0^{-})^2}{[1+w^2+2w^2(\theta_0^{-})^2]\theta_0^{-}}
+\frac{2-w^2+w^4+2w^4(\theta_0^{+})^2}{[1+w^2+2w^2(\theta_0^{+})^2]\theta_0^{+}}\right\} ~.  \label{Delta-2}
\end{eqnarray}
In terms of the angular source position $\beta$, Eqs.~\eqref{Delta-1} and \eqref{Delta-2} can also be expressed as
\begin{eqnarray}
&&\Delta\hat{\tau}_0=\frac{|\beta|}{2w}\sqrt{\beta^2+2\left(1+\frac{1}{w^2}\right)}
+\frac{1-3w^2}{2w^3}\ln\!\left[\frac{\sqrt{\beta^2+2\left(1+\frac{1}{w^2}\right)}-|\beta|}{\sqrt{\beta^2+2\left(1+\frac{1}{w^2}\right)}+|\beta|}\right]~, \label{Delta-1-beta}   \\
\nn&&\Delta\hat{\tau}_1=\frac{3\pi w(5\!-\!\hat{Q}^2)|\beta|}{4(1+w^2)}\!+\!\frac{1}{8w(1\!+\!w^2)^2(2\!+\!2w^2\!+\!w^2\beta^2)}
\Bigg{\{}\pi\!\left[3(4\!+\!w^2)\!-\!(2\!+\!w^2)\hat{Q}^2\right]\!\left[2(1\!-\!2w^2\!-\!3w^4)\!+\!w^2(1\!-\!3w^2)\beta^2\right]\!|\beta|   \\
&&\hspace*{31pt}+\,16ws^{+}\hat{a}\left[2(1+w^2)(1+w^4)+w^2(2-w^2+w^4)\beta^2\right]\sqrt{\beta^2+2\left(1+\frac{1}{w^2}\right)}\Bigg{\}}~. ~~~~~~   \label{Delta-2-beta}
\end{eqnarray}
Notice that Eq.~\eqref{Differential-TD} is consistent with the null result of the Schwarzschild lensing~\cite{KP2005} for the case of $w=1$ and $a=Q=0$, and with the result of the Kerr lensing of light~\cite{AKP2011} for the case of $w=1$ and $Q=0$. Fig.\,\ref{Figure5} gives the magnitudes of the coefficients of the zeroth- and first-order contributions to the scaled differential time delay.

\section{Velocity-induced effects on the lensing observables} \label{sect5}
The deviation of the initial velocity $w$ of a massive particle from the speed of light affects the geodesic motions and thus the related lensing observables. Considering their importance for discussing the gravitational lensing phenomena of massive particles, we present the explicit forms of the velocity-induced effects on the observables of the lensed images beyond the weak-deflection limit.

\subsection{Velocity effects on scaled angular image positions}
Based on Eqs.~\eqref{theta0-PN} - \eqref{theta2-PN}, the explicit forms of the velocity effects on the zeroth-, first-, and second-order coefficients of the scaled angular positions of the positive- and negative-parity images can be written respectively as follows:
\begin{eqnarray}
&&\delta\theta_0^{\pm}=\frac{1}{2}\left[\sqrt{\beta^2+2\left(1+\frac{1}{w^2}\right)}-\sqrt{\beta^2+4}\,\right]~,    \label{theta0-PN-VC}  \\
&&\delta\theta_1^{\pm}=\frac{3\pi\left(4+w^2\right)-16ws^{\pm}\hat{a}-\pi\left(2+w^2\right)\hat{Q}^2}{16\left(1+w^2\right)}\!\!\left[\!1\mp\frac{|\beta|}{\sqrt{\beta^2+2\left(1+\frac{1}{w^2}\right)}}\!\right]
\!-\!\frac{15\pi-16s^{\pm}\hat{a}-3\pi\hat{Q}^2}{32}\!\left(\!1\mp\frac{|\beta|}{\sqrt{\beta^2+4}}\!\right)~,~~~~~~~    \label{theta1-PN-VC}  \\
\nn&&\delta\theta_2^{\pm}=\frac{\left[2(1+6w^2+w^4)-\pi w^3s^{\pm}\hat{a}\right]\left(1+3w^2-w^2\hat{Q}^2\right)+w^3(1+w^2)\left(2w\hat{a}^2-3\pi s^{\pm}\hat{a}\right)-\frac{8D^2}{3}(1+w^2)^3}
{w^6\sqrt{\beta^2+2\left(1+\frac{1}{w^2}\right)}\left[\sqrt{\beta^2+2\left(1+\frac{1}{w^2}\right)}\pm|\beta|\right]^2}  \\
\nn&&\hspace*{26pt}-\frac{\left[3\pi(4+w^2)-16ws^{\pm}\hat{a}-\pi(2+w^2)\hat{Q}^2\right]^2}
{128w^2(1\!+\!w^2)\sqrt{\beta^2\!+\!2\left(1\!+\!\frac{1}{w^2}\right)}\left[\sqrt{\beta^2\!+\!2\left(1\!+\!\frac{1}{w^2}\right)}\pm|\beta|\right]^2}
\!\left[1\mp\frac{|\beta|}{\sqrt{\beta^2+2\left(1+\frac{1}{w^2}\right)}}\right]\!\!\left[3\pm\frac{|\beta|}{\sqrt{\beta^2+2\left(1+\frac{1}{w^2}\right)}}\right]   \\
\nn&&\hspace*{26pt}-\frac{4D}{w^4\sqrt{\beta^2+2\left(1+\frac{1}{w^2}\right)}}
\left\{(1+w^2)^2(1-D)-\frac{(1+w^2)w^2D}{12}\left[\sqrt{\beta^2+2\left(1+\frac{1}{w^2}\right)}\pm|\beta|\right]^2\right\}+\frac{16(1-D)D}{\sqrt{\beta^2+4}}  \\
\nn&&\hspace*{26pt}-\frac{1}{768\sqrt{\beta^2\!+\!4}\left(\sqrt{\beta^2\!+\!4}\pm|\beta|\right)^2}\Bigg{\{}\!512\!\left[8(12\!-\!4D^2\!-\!3\hat{Q}^2)\!+\!D^2\!\left(\sqrt{\beta^2\!+\!4}\pm|\beta|\right)^4\right]
+768\!\left[4\hat{a}^2-\pi s^{\pm}\hat{a}\,(10-\hat{Q}^2)\right]  \\
&&\hspace*{26pt}-\,3\!\left[3\pi(5-\hat{Q}^2)-16s^{\pm}\hat{a}\right]^2\left(1\mp\frac{|\beta|}{\sqrt{\beta^2+4}}\right)\left(3\pm\frac{|\beta|}{\sqrt{\beta^2+4}}\right)\!\!\Bigg{\}}~.  \label{theta2-PN-VC}
\end{eqnarray}

From Eqs.~\eqref{NewR-2} and \eqref{NewR-4} - \eqref{NewR-7}, the velocity effects on the zeroth-order sum, first- and second-order sum and difference relations for the coefficients of the scaled angular source positions are given respectively by
\begin{eqnarray}
&&\delta(\theta_0^{+}+\theta_0^{-})=\sqrt{\beta^2+2\left(1+\frac{1}{w^2}\right)}-\sqrt{\beta^2+4}~,~~~~ \label{NewR-2-VC}  \\
&&\delta(\theta_1^{+}+\theta_1^{-})=\frac{3\pi(4+w^2)-\pi(2+w^2)\hat{Q}^2}{8(1+w^2)}+\frac{2w}{1+w^2}\frac{s^{+}\hat{a}\,|\beta|}{\sqrt{\beta^2+2\left(1+\frac{1}{w^2}\right)}}
-\left[\frac{3\pi(5-\hat{Q}^2)}{16}+\frac{s^{+}\hat{a}\,|\beta|}{\sqrt{\beta^2+4}} \right]~, \label{NewR-4-VC}  \\
&&\delta(\theta_1^{+}-\theta_1^{-})=\frac{3\pi(5-\hat{Q}^2)|\beta|}{16\sqrt{\beta^2+4}}-\frac{3\pi(4+w^2)-\pi(2+w^2)\hat{Q}^2}{8(1+w^2)}\frac{|\beta|}{\sqrt{\beta^2+2\left(1+\frac{1}{w^2}\right)}}
+\frac{(1-w)^2s^{+}\hat{a}}{1+w^2}~, \label{NewR-5-VC}  \\
&&\nn\delta(\theta_2^{+}+\theta_2^{-})=\!\frac{1}{192\,w^2\,(1\!+\!w^2)^3}\Bigg{\{}\!\frac{384w^3\!\left\{\!2\!+\!2w^2(1\!+\!w^4\!+\!w^6)
\!+\!w^2(1\!-\!w^2)^2\!\left[3\!+\!w^2(3\!+\!\beta^2)\right]\!\beta^2\!\right\}\!\hat{a}^2}{(2+2w^2+w^2\beta^2)^{3/2}}\!+\!\frac{1}{w\,(2\!+\!2w^2\!+\!w^2\beta^2)^{3/2}}     \\
&&\nn\hspace*{1.75cm}\times\Big{\{}\!(1\!+\!w^2)^2\Big{[}768(1\!+\!10w^2)\!+\!48(448\!-\!27\pi^2)w^4\!+\!24(704\!-\!27\pi^2)w^6\!+\!9(256\!-\!9\pi^2)w^8\!-\!512(6\!-\!5D)D(1\!+\!w^2)^4     \\
&&\nn\hspace*{1.75cm}-\,6w^2\!\left(128+8(112-9\pi^2)w^2+(896-54\pi^2)w^4+(128-9\pi^2)w^6\right)\hat{Q}^2\!-\!9\pi^2w^4(2+w^2)^2\hat{Q}^4\Big{]}\!-6w^2(1+w^2)     \\
&&\nn\hspace*{1.75cm}\times\Big{[}128(2-D)D(1+w^2)^4-3\left(64(1+10w^2)+16(112-9\pi^2)w^4+8(176-9\pi^2)w^6+3(64-3\pi^2)w^8\right)  \\
&&\nn\hspace*{1.75cm}+\,6w^2\left(32+8(28-3\pi^2)w^2+2(112-9\pi^2)w^4+(32-3\pi^2)w^6\right)\hat{Q}^2+3\pi^2w^4(2+w^2)^2\hat{Q}^4\Big{]}\beta^2  \\
&&\nn\hspace*{1.75cm}-\,2w^4\Big{[}27\pi^2w^4(4+w^2)^2-192(1+10w^2+28w^4+22w^6+3w^8)+128D^2(1+w^2)^4     \\
&&\nn\hspace*{1.75cm}+\,w^2\left(192(1+7w^2+7w^4+w^6)-18\pi^2w^2(8+6w^2+w^4)\right)\hat{Q}^2+3\pi^2w^4(2+w^2)^2\hat{Q}^4\Big{]}\beta^4\Big{\}}
\end{eqnarray}
\begin{eqnarray}
&&\nn\hspace*{1.75cm}+\,192\pi w^3\,|\beta|\,s^{+}\hat{a}\left(4-2w^2+3w^4+w^2\hat{Q}^2\right)\!\!\Bigg{\}}+\frac{1}{768(\beta^2+4)^{3/2}}\Big{\{}2048D[12-D(10-\beta^2)](\beta^2+4)\\
&&\hspace*{1.75cm}+\,27\pi^2(5\!-\!\hat{Q}^2)^2(6\!+\!6\beta^2\!+\!\beta^4)\!-\!3072(4\!-\!\hat{Q}^2)(8\!+\!6\beta^2\!+\!\beta^4)
\!-\!96\!\left[16\hat{a}^2\!+\!\pi s^{+}\hat{a}(5\!+\!\hat{Q}^2)(\beta^2\!+\!4)^{3/2}|\beta|\right]\!\Big{\}}~,  \label{NewR-6-VC}  \\
&&\nn\delta(\theta_2^{+}-\theta_2^{-})=-\frac{1}{32w^2(1+w^2)^3}\Bigg{\{}\!\Big{\{}64(1\!+\!10w^2)\!+\!16(112\!-\!9\pi^2)w^4\!+\!8(176\!-\!9\pi^2)w^6\!+\!3(64-3\pi^2)w^8\!-\!128D^2(1+w^2)^4        \\
&&\nn\hspace*{1.75cm}+\,64w^4(1-w^2)^2\hat{a}^2-2w^2\!\left[32-w^2\!\left(3\pi^2(8+6w^2+w^4)-32(7+7w^2+w^4)\right)\right]\!\hat{Q}^2-\pi^2w^4(2+w^2)^2\hat{Q}^4\Big{\}}|\beta|        \\
&&\nn\hspace*{1.75cm}+\frac{16\pi w^2s^{+}\hat{a}}{(2+2w^2+w^2\beta^2)^{3/2}}\Big{\{}(16\!+\!4w^2\!+\!15w^4)(1+w^2)^2\!+\!6w^2(4\!+\!2w^2\!+\!w^4\!+\!3w^6)\beta^2\!+\!2w^4(4-2w^2+3w^4)\beta^4        \\
&&\nn\hspace*{1.75cm}+\,w^2\!\left[2\!+\!w^2\!\left(3\!-\!w^4\!+\!6(1\!+\!w^2)\beta^2\!+\!2w^2\beta^4\right)\right]\!\hat{Q}^2\!\Big{\}}\!\Bigg{\}}
\!-\!\frac{|\beta|\!\left[225\pi^2\!-\!4096\!+\!2048D^2\!+\!2(512\!-\!45\pi^2)\hat{Q}^2\!+\!9\pi^2\hat{Q}^4\right]}{256}  \\
&&   \hspace*{1.75cm}+\frac{\pi s^{+}\hat{a}\!\left[5(14+6\beta^2+\beta^4)+(2+6\beta^2+\beta^4)\hat{Q}^2\right]}{8(\beta^2+4)^{3/2}} ~.    \label{NewR-7-VC}
\end{eqnarray}

\subsection{Velocity effects on magnification relations}
The velocity effects on the coefficients of the zeroth-, first-, and second-order contributions to the magnifications of the positive- and negative-parity images shown in Eqs.~\eqref{mu0-beta} - \eqref{mu2-beta} are presented respectively as follows:
\begin{eqnarray}
&&\delta\mu_0^{\pm}=\pm\frac{1+\frac{1}{w^2}+\beta^2}{2|\beta|\sqrt{\beta^2+2\left(1+\frac{1}{w^2}\right)}}\mp\frac{\beta^2+2}{2|\beta|\sqrt{\beta^2+4}}~,  \label{mu0-beta-VC}  \\
&&\delta\mu_1^{\pm}=-\frac{3\pi\left(1+\frac{4}{w^2}\right)-\frac{16s^{\pm}\hat{a}}{w}-\pi\left(1+\frac{2}{w^2}\right)\hat{Q}^2}{16\left[\beta^2+2\left(1+\frac{1}{w^2}\right)\right]^{3/2}}
+\frac{3\pi(5-\hat{Q}^2)-16s^{\pm}\hat{a}}{16(\beta^2+4)^{3/2}} ~,~~~~ \label{mu1-beta-VC}  \\
\nn&&\delta\mu_2^{\pm}=\frac{\left(1+\frac{1}{w^2}\right)^3+\beta^2\left[2\beta^2+3\left(1+\frac{1}{w^2}\right)\right]^2
\pm\left[3\left(1+\frac{1}{w^2}\right)^2+8\left(1+\frac{1}{w^2}\right)\beta^2+4\beta^4\right]|\beta|\sqrt{\beta^2+2\left(1+\frac{1}{w^2}\right)}}
{3\left[\beta^2\pm|\beta|\sqrt{\beta^2+2\left(1+\frac{1}{w^2}\right)}\,\right]\left[\beta^2+2\left(1+\frac{1}{w^2}\right)\pm|\beta|\sqrt{\beta^2+2\left(1+\frac{1}{w^2}\right)}\,\right]^5}   \\
\nn&&\hspace*{26pt}\times\Bigg{\{}\!9\!\left[\frac{3\pi}{4}\!\left(\!1\!+\!\frac{4}{w^2}\!\right)\!-\!\frac{4s^{\pm}\hat{a}}{w}\!-\!\frac{\pi}{4}\!\left(\!1\!+\!\frac{2}{w^2}\!\right)\!\hat{Q}^2\right]^2
\!+\!4\!\left[\beta^2\!+\!2\left(1\!+\!\frac{1}{w^2}\right)\!\right]\!\Bigg{\{}\!4D^2\beta^2\!\left(\!1\!+\!\frac{1}{w^2}\!\right)^2\!-\!4(2\!+\!6D\!-\!9D^2)\!\left(\!1\!+\!\frac{1}{w^2}\!\right)^3  \\
\nn&&\hspace*{26pt}-\,3\!\left[\frac{2}{3}\!\left(\!5+\frac{45}{w^2}+\frac{15}{w^4}-\frac{1}{w^6}\!\right)-2\pi\left(\!\frac{3}{w}+\frac{2}{w^3}\!\right)s^{\pm}\hat{a}+2\left(\!1+\frac{1}{w^2}\!\right)\hat{a}^2
-2\left(\!1+\frac{6}{w^2}+\frac{1}{w^4}\!\right)\hat{Q}^2+\frac{\pi s^{\pm}\hat{a}\hat{Q}^2}{w}\right]\!\!\Bigg{\}}\!\Bigg{\}} \\
\nn&&\hspace*{26pt}-\frac{8+4\beta^2(3+\beta^2)^2\pm 4(3+4\beta^2+\beta^4)|\beta|\sqrt{4+\beta^2}}{48\left(\beta^2\pm|\beta|\sqrt{4+\beta^2}\right)\left(\beta^2+4\pm|\beta|\sqrt{4+\beta^2}\right)^5}
\Big{\{}3[675\pi^2-4096(4+\beta^2)]-1024D(4+\beta^2)[12-D(18+\beta^2)]  \\
&&\hspace*{26pt} -768(1+\beta^2)\hat{a}^2-6[135\pi^2-512(4+\beta^2)]\hat{Q}^2+81\pi^2\hat{Q}^4+96\pi s^{\pm}\hat{a}\!\left[5(7+4\beta^2)+(1-2\beta^2)\hat{Q}^2\right]\!\Big{\}}~.~~~~ \label{mu2-beta-VC}
\end{eqnarray}
Moreover, base on Eqs.~\eqref{mu0-PMN} - \eqref{mu2-PNN}, the velocity effects on the zeroth-order difference, first- and second-order sum and difference relations for the coefficients of the signed magnifications are presented respectively as
\begin{eqnarray}
&&\hspace*{-8.2cm}\delta(\mu_0^{+}-\mu_0^{-})=\frac{1}{|\beta|}\left[\frac{1+\frac{1}{w^2}+\beta^2}{\sqrt{\beta^2+2\left(1+\frac{1}{w^2}\right)}}-\frac{2+\beta^2}{\sqrt{4+\beta^2}}\right]~,  \label{mu0-PNN-VC}  \\
&&\hspace*{-8.2cm}\delta(\mu_1^{+}+\mu_1^{-})=\frac{\pi}{8}\left\{\frac{3(5-\hat{Q}^2)}{(4+\beta^2)^{3/2}}
-\frac{3\left(1+\frac{4}{w^2}\right)-\left(1+\frac{2}{w^2}\right)\hat{Q}^2}{\left[\beta^2+2\left(1+\frac{1}{w^2}\right)\right]^{3/2}}\right\}~,  \label{mu1-PPN-VC}  \\
&&\hspace*{-8.2cm}\delta(\mu_1^{+}-\mu_1^{-})=2s^{+}\hat{a}\left\{\frac{1}{w\!\left[\beta^2+2\left(1+\frac{1}{w^2}\right)\right]^{3/2}}-\frac{1}{(4+\beta^2)^{3/2}}\right\}~,   \label{mu1-PNN-VC}
\end{eqnarray}
\begin{eqnarray}
&&\delta(\mu_2^{+}+\mu_2^{-})=\frac{\pi s^{+}\hat{a}}{2|\beta|}\!
\left\{\!\frac{16\!+\!4w^2\!+\!15w^4\!+\!4w^2(2\!+\!3w^2)\beta^2\!+\!w^2(2\!-\!w^2\!-\!2w^2\beta^2)\hat{Q}^2}{\left[2+w^2(2+\beta^2)\right]^{5/2}}
\!-\!\frac{5(7\!+\!4\beta^2)\!+\!(1\!-\!2\beta^2)\hat{Q}^2}{(4+\beta^2)^{5/2}}\!\right\}~,    \label{mu2-PPN-VC}  \\
\nn&&\delta(\mu_2^{+}-\mu_2^{-})=\frac{1}{192w^3|\beta|\left(2\!+\!2w^2\!+\!w^2\beta^2\right)^{5/2}}
\Big{\{}\!\!-\!3\Big{[}256(1\!+\!10w^2)\!+\!16(448\!-\!27\pi^2)w^4\!+\!8(704\!-\!27\pi^2)w^6\!+\!3(256\!-\!9\pi^2)w^8 \\
\nn&&\hspace*{1.4cm}+\,512(2\!-\!3D)D(1\!+\!w^2)^4\Big{]}\!+\!128w^2\!\left[2(11D\!-\!6)D(1\!+\!w^2)^3\!-\!3(1\!+\!9w^2\!+\!19w^4\!+\!3w^6)\right]\beta^2\!+\!256D^2w^4(1\!+\!w^2)^2\beta^4   \\
\nn&&\hspace*{1.4cm}-\,384w^4\!\left[2(1\!-\!w^2\!+\!w^4)\!+\!w^2(1\!+\!w^2)\beta^2\right]\hat{a}^2\!+\!6w^2\left[128\!+\!8(112\!-\!9\pi^2)w^2+(896-54\pi^2)w^4+(128-9\pi^2)w^6  \right.  \\
\nn&&\hspace*{1.4cm}\left. +\,64w^2(1+6w^2+w^4)\beta^2\right]\hat{Q}^2+9\pi^2w^4(2+w^2)^2\hat{Q}^4\Big{\}}-\frac{1}{192|\beta|(4+\beta^2)^{5/2}}\Big{\{}3\left[675\pi^2-4096(4+\beta^2)\right]  \\
   &&\hspace*{1.4cm}-1024D(4+\beta^2)\left[12-D(18+\beta^2)\right]-768(1+\beta^2)\hat{a}^2-6\left[135\pi^2-512(4+\beta^2)\right]\hat{Q}^2+81\pi^2\hat{Q}^4\Big{\}}~.   \label{mu2-PNN-VC}
\end{eqnarray}
Note that the velocity effects on the total magnification can be indicated by Eqs.~\eqref{mu0-PNN-VC}, \eqref{mu1-PNN-VC}, and \eqref{mu2-PNN-VC}.

Finally, according to Eqs.~\eqref{centroid-Series-beta-1} - \eqref{centroid-Series-beta-3}, we also give respectively the velocity effects on the coefficients of the zeroth-, first-, and second-order contributions to the scaled magnification-weighted centroid
\begin{eqnarray}
&&\delta{\it\Theta}_{\text{cent,0}}=\frac{|\beta|\beta^2}{2}\left(\frac{1}{2+\beta^2}-\frac{w^2}{1+w^2+w^2\beta^2}\right)~,    \label{DeltaTheta-0-VC} \\
&&\delta{\it\Theta}_{\text{cent,1}}=s^{+}\hat{a}\left[\frac{2(1+\beta^2)}{(2+\beta^2)^2}-\frac{w(1+w^2+2w^2\beta^2)}{(1+w^2+w^2\beta^2)^2}\right]~,     \label{DeltaTheta-1-VC} \\
\nn&&\delta{\it\Theta}_{\text{cent,2}}=\frac{32|\beta|}{3w^{18}\!\left[\sqrt{\beta^2\!+\!2\left(1\!+\!\frac{1}{w^2}\right)}\!+\!|\beta|\right]\!\!
\left[2\left(1\!+\!\frac{1}{w^2}\right)\!+\!\left(\!\sqrt{\beta^2\!+\!2\left(1\!+\!\frac{1}{w^2}\right)}\!+\!|\beta|\!\right)^{\!2}\right]^2\!\!
\left[4\left(1\!+\!\frac{1}{w^2}\right)^2\!+\!\left(\!\sqrt{\beta^2\!+\!2\left(1\!+\!\frac{1}{w^2}\right)}\!+\!|\beta|\!\right)^{\!4}\right]^3}   \\
\nn&&\hspace*{44pt}\times\Bigg{\{}|\beta|\left[9(1\!+\!w^2)^4\!+\!60w^2(1\!+\!w^2)^3\beta^2\!+\!108w^4(1\!+\!w^2)^2\beta^4
\!+\!72w^6(1\!+\!w^2)\beta^6\!+\!16w^8\beta^8\right]\!+\!\sqrt{\beta^2\!+\!2\left(1\!+\!\frac{1}{w^2}\right)}  \\
\nn&&\hspace*{44pt}\times\left[(1+w^2)^4+20w^2(1+w^2)^3\beta^2+60w^4(1+w^2)^2\beta^4+56w^6(1+w^2)\beta^6+16w^8\beta^8\right]\!\Bigg{\}}  \\
\nn&&\hspace*{44pt}\times\Bigg{\{}384w^4\left[1+w^4+w^2(1+w^2)\beta^2\right]\left[2+w^2(2+\beta^2)\right]\hat{a}^2+(1+w^2+w^2\beta^2)\Big{\{}768+384w^2(20+\beta^2)   \\
\nn&&\hspace*{44pt}-\,81\pi^2w^4(4\!+\!w^2)^2\!+\!384w^4\!\!\left[56\!+\!9\beta^2\!+\!w^2(44\!+\!19\beta^2)\!+\!3w^4(2\!+\!\beta^2)\right]\!-\!1536Dw^2(1\!+\!w^2)^2\beta^2(2\!+\!2w^2\!+\!w^2\beta^2)     \\
\nn&&\hspace*{44pt}+\,256D^2(1+w^2)\left[-2(1+w^2)^3+17w^2(1+w^2)^2\beta^2+13w^4(1+w^2)\beta^4+2w^6\beta^6\right]     \\
\nn&&\hspace*{44pt}-\,6w^2\!\left[128\!-\!9\pi^2w^2(8\!+\!6w^2\!+\!w^4)\!+\!64w^2\!\left(14\!+\!\beta^2\!+\!w^4(2\!+\!\beta^2)\!+\!2w^2(7\!+\!3\beta^2)\right)\right]\!\hat{Q}^2
\!-\!9\pi^2w^4(2\!+\!w^2)^2\hat{Q}^4\!\Big{\}}\!\Bigg{\}}  \\
\nn&&\hspace*{44pt}-\frac{|\beta|\left\{(1\!+\!\beta^2)\!\left[1\!+\!\beta^2(3+\beta^2)^2\right]\!\sqrt{4+\beta^2}+|\beta|(3+\beta^2)\left[3+\beta^2(3+\beta^2)^2\right]\!\right\}}
{12(4+\beta^2)(2+\beta^2)^3\left(2+\beta^2+|\beta|\sqrt{4+\beta^2}\right)^3\left(\sqrt{4+\beta^2}+|\beta|\right)^3}\Big{\{}768(4\!+\!5\beta^2\!+\!\beta^4)\hat{a}^2\!-\!(2\!+\!\beta^2)    \\
&&\hspace*{44pt}\times\!\left[2025\pi^2\!+\!1024(4\!+\!\beta^2)\left((2\!-\!\beta^4)D^2\!+\!(6\!-\!9D)D\beta^2\!-\!12\right)
\!-\!6\left(135\pi^2\!-\!512(4\!+\!\beta^2)\right)\hat{Q}^2\!+\!81\pi^2\hat{Q}^4\right]\!\Big{\}} ~.  \label{DeltaTheta-2-VC}
\end{eqnarray}

\subsection{Velocity effects on scaled differential time delay}
Similarly, the velocity effects on the coefficients of the zeroth- and first-order contributions to the scaled differential time delay between the positive- and negative-parity images can be obtained from Eqs.~\eqref{Delta-1-beta} - \eqref{Delta-2-beta}
\begin{eqnarray}
&&\delta\Delta\hat{\tau}_0=\frac{|\beta|}{2}\!\!\left[\frac{1}{w}\sqrt{\beta^2+2\left(1+\frac{1}{w^2}\right)}-\sqrt{\beta^2+4}\,\right]
\!+\!\frac{1\!-\!3w^2}{2w^3}\ln\!\!\left[\frac{\sqrt{\beta^2+2\left(1+\frac{1}{w^2}\right)}-|\beta|}{\sqrt{\beta^2+2\left(1+\frac{1}{w^2}\right)}+|\beta|}\right]
\!-\ln\!\left(\frac{\sqrt{\beta^2+4}+|\beta|}{\sqrt{\beta^2+4}-|\beta|}\right)~,~~~~~~~ \label{Delta-1-beta-VC}
\end{eqnarray}
\begin{eqnarray}
\nn&&\hspace*{-55pt}\delta\Delta\hat{\tau}_1=\frac{3\pi(5-\hat{Q}^2)|\beta|}{16}\left(\frac{4w}{1+w^2}-1\right)+\frac{1}{8w(1+w^2)^2(2+2w^2+w^2\beta^2)}\Bigg{\{}\pi\!\left[3(4+w^2)-(2+w^2)\hat{Q}^2\right]\!   \\
\nn&&\hspace*{-55pt}\hspace*{35pt}\times\!\left[2(1-2w^2-3w^4)+w^2(1-3w^2)\beta^2\right]\!|\beta|+16ws^{+}\hat{a}\!\left[2(1+w^2)(1+w^4)+w^2(2-w^2+w^4)\beta^2\right]  \\
   &&\hspace*{-55pt}\hspace*{35pt}\times\sqrt{\beta^2+2\left(1+\frac{1}{w^2}\right)}\Bigg{\}}-s^{+}\hat{a}\sqrt{4+\beta^2}~. ~~~~~~~   \label{Delta-2-beta-VC}
\end{eqnarray}

Finally, it is recognized that the terms on the right-hand side of Eqs.~\eqref{theta0-PN-VC} - \eqref{Delta-2-beta-VC} will vanish in the limit $w\rightarrow 1$.
\end{widetext}

\section{Lensing by the galactic supermassive black hole} \label{sect6}
As an application of the analytical results given above, we model the supermassive black hole at the galactic center as a KN lens. Since the null lensing observables for the scenario where Sagittarius A$^{\ast}$ acts as a Schwarzschild lens have been studied in detail in the previous works (see, e.g.,~\cite{VE2000,KP2005}), in this section we concentrate on the analysis of the velocity-induced correctional effects on the practical observables of the lensed images. The possibilities of their astronomical detection will also be discussed.

\subsection{Basics} \label{Basics}
The basic parameters under consideration are given as follows. The mass of Sagittarius A$^{\ast}$ and the distance to it are $M=4.2\times10^6M_{\odot}$~\cite{BG2016,Parsa2017} and $d_L=8.2$\,kpc~\cite{BG2016}, respectively, with $M_{\odot}$ ($=1.475$\,km) being the mass of the Sun. The special angular gravitational radius is $\vartheta_{\bullet}\!=\!5.06\,\mu$as. The natural lensing time scale is $\tau_E=82.6\,$s.
Since the distance of the source from the lens is much smaller than $d_L$ in general, we may assume $d_{LS}=0.01$\,kpc. Hence, $D=1.22\times10^{-3}$, the angular Einstein radius is $\vartheta_E=0.071$\,as, and the small dimensionless parameter is $\varepsilon=7.12\times10^{-5}$. For the convenience of discussion, the initial velocity of the relativistic massive particle is assumed to have a rough range of $0.05\lesssim w<1$. Moreover, we know the spin and electrical charge of a massive black hole are determined by the competition between many physical processes. Since the observational evidences indicate that the galactic supermassive black hole may have high spin parameter~\cite{GSM2004,Shapiro2005,VMQR2005,ZLY2015,Reynolds2019} and very weak electrical charge~\cite{Wald1974,KLL2001,ZTEB2018,Turs2020}, we adopt $\hat{a}\!=\!0.9$~\cite{GSM2004} and then $\hat{Q}=7.56\times10^{-13}$ (the equilibrium Wald charge)~\cite{KLL2001,ZTEB2018} for Sagittarius A$^{\ast}$ in our scenario.

Considering the complexity resulted from the motion direction of the particle relative to the rotating lens (indicated by the sign parameter $s$), we take the positive-parity image with $|\beta|=\beta$ and a sign parameter $s^{+}\in\{+1,\,-1\}$ as an example to perform our discussions of the image properties. We follow the idea of Ref.~\cite{CX2021} to take the domain $[0.01,~10]$ for the scaled angular source position $\beta$. Notice that the sum and difference relations for the coefficients of the signed positions or magnifications, as well as the centroid and differential time delay given above, have been formulated in terms of the quantities including $s^{+}$.

It should be pointed out that the magnification is related to the image flux $F\,(=F_0+F_1\,\varepsilon+F_2\,\varepsilon^2+\mathcal{O}(\varepsilon^3))$, which is one of the practical lensing observables, via $F_i=|\mu_i|F_{\text{s}}$~\cite{KP2006a}. Here, $i\in N$, and $F_{\text{s}}~(>0)$ denotes the intrinsic flux of the particle's source without experiencing the lensing effect. To relate with the practical observations, we use the old lensing quantities $(\vartheta,~\mathcal{B},~F,~{\it\Xi}_{\text{cent}},~\tau)$ rather than the scaled quantities ($\theta,~\beta,~\mu,~{\it\Theta}_{\text{cent}},~\hat{\tau}$) in this section, with ${\it\Xi}_{\text{cent}}=\vartheta_E{\it\Theta}_{\text{cent}}$.

\subsection{Result: velocity-induced effects on the observables}
The velocity effects on the zeroth-, first-, or second-order contribution to the lensing observables (including the sum and difference relations for the positions and fluxes) of the primary and secondary images can be written in terms of the quantities $(\vartheta,~\mathcal{B},~F,~{\it\Xi}_{\text{cent}},~\tau)$ as follows:
\begin{eqnarray}
&&\delta\vartheta_i^{+}\varepsilon^i=\vartheta_E\,\delta\theta_i^{+}\varepsilon^i~,   \label{NS-1} \\
&&\delta F_i^{+}\varepsilon^i=F_{\text{s}}\,\delta\mu_i^{+}\varepsilon^i~,  \label{NS-2}  \\
&&\delta{\it\Xi}_{\text{cent,}i}\,\varepsilon^i=\vartheta_E\,\delta{\it\Theta}_{\text{cent,}i}\,\varepsilon^i~,  \label{NS-3}  \\
&&\delta\Delta\tau_j\,\varepsilon^j=\tau_E\,\delta\Delta\hat{\tau}_j\,\varepsilon^j~,   \label{NS-4}  \\
&&\delta(\vartheta_i^{+}\pm\vartheta_i^{-})\,\varepsilon^i=\vartheta_E\,\delta(\theta_i^{+}\pm\theta_i^{-})\,\varepsilon^i~,  \label{NS-5} \\
&&\delta(F_i^{+}\pm F_i^{-})\,\varepsilon^i=F_{\text{s}}\,\delta(\mu_i^{+}\mp\mu_i^{-})\,\varepsilon^i~, ~~~~  \label{NS-6}
\end{eqnarray}
where $i\in\{0,~1,~2\}$ and $j\in\{0,~1\}$. Note that $\delta\Delta\tau_j\varepsilon^j$ is roughly of the order of $\varepsilon^{j+2}$, since $\tau_E$ is of the order of $d_LD\varepsilon^2$. In order to analyze the image flux more conveniently, three auxiliary differential apparent magnitudes resulted from the deviation of $w$ from $c$ are defined as
\begin{widetext}
{\small
\begin{eqnarray}
&&\delta m_1\equiv-2.5\lg\!\left[1+\frac{\delta F^{+}}{F^{+}\!\!\left.\right|_{w=1}}\right]
=-2.5\lg\!\left[1+\frac{\delta F_0^{+}+\delta F_1^{+}\varepsilon+\delta F_2^{+}\varepsilon^2}
{F_0^{+}\!\!\left.\right|_{w=1}+F_1^{+}\!\!\left.\right|_{w=1}\varepsilon+F_2^{+}\!\!\left.\right|_{w=1}\varepsilon^2}+\mathcal{O}(\varepsilon^3)\right] ~,   \label{M-1} \\
&&\delta m_2\equiv-2.5\lg\!\left[1\!+\!\frac{\delta\left(F^{+}+F^{-}\right)}{\left(F^{+}\!+\!F^{-}\right)\!\left.\right|_{w=1}}\right]
=-2.5\lg\!\left[1\!+\!\frac{\delta(F_0^{+}+F_0^{-})+\delta(F_1^{+}+F_1^{-})\varepsilon+\delta(F_2^{+}+F_2^{-})\varepsilon^2}
{(F_0^{+}\!+\!F_0^{-})\!\!\left.\right|_{w=1}\!+\!(F_1^{+}\!+\!F_1^{-})\!\!\left.\right|_{w=1}\varepsilon
\!+\!(F_2^{+}\!+\!F_2^{-})\!\!\left.\right|_{w=1}\varepsilon^2}+\mathcal{O}(\varepsilon^3)\right]~,~~~~~~~  \label{M-2}
\end{eqnarray}
\begin{eqnarray}
\hspace*{-1cm}&&\delta m_3\equiv-2.5\lg\!\left[1+\frac{\delta \left(F^{+}-F^{-}\right)}{\left(F^{+}-F^{-}\right)\!\left.\right|_{w=1}}\right]
=-2.5\lg\!\left[1+\frac{\delta(F_1^{+}-F_1^{-})\varepsilon+\delta(F_2^{+}-F_2^{-})\varepsilon^2}
{F_{\text{s}}+(F_1^{+}-F_1^{-})\!\!\left.\right|_{w=1}\varepsilon+(F_2^{+}-F_2^{-})\!\!\left.\right|_{w=1}\varepsilon^2}+\mathcal{O}(\varepsilon^3)\right]~.~~~~~~~~~  \label{M-3}
\end{eqnarray}}
\end{widetext}

\begin{figure*}
\centering
\begin{minipage}[b]{5.35cm}
\includegraphics[width=5.35cm]{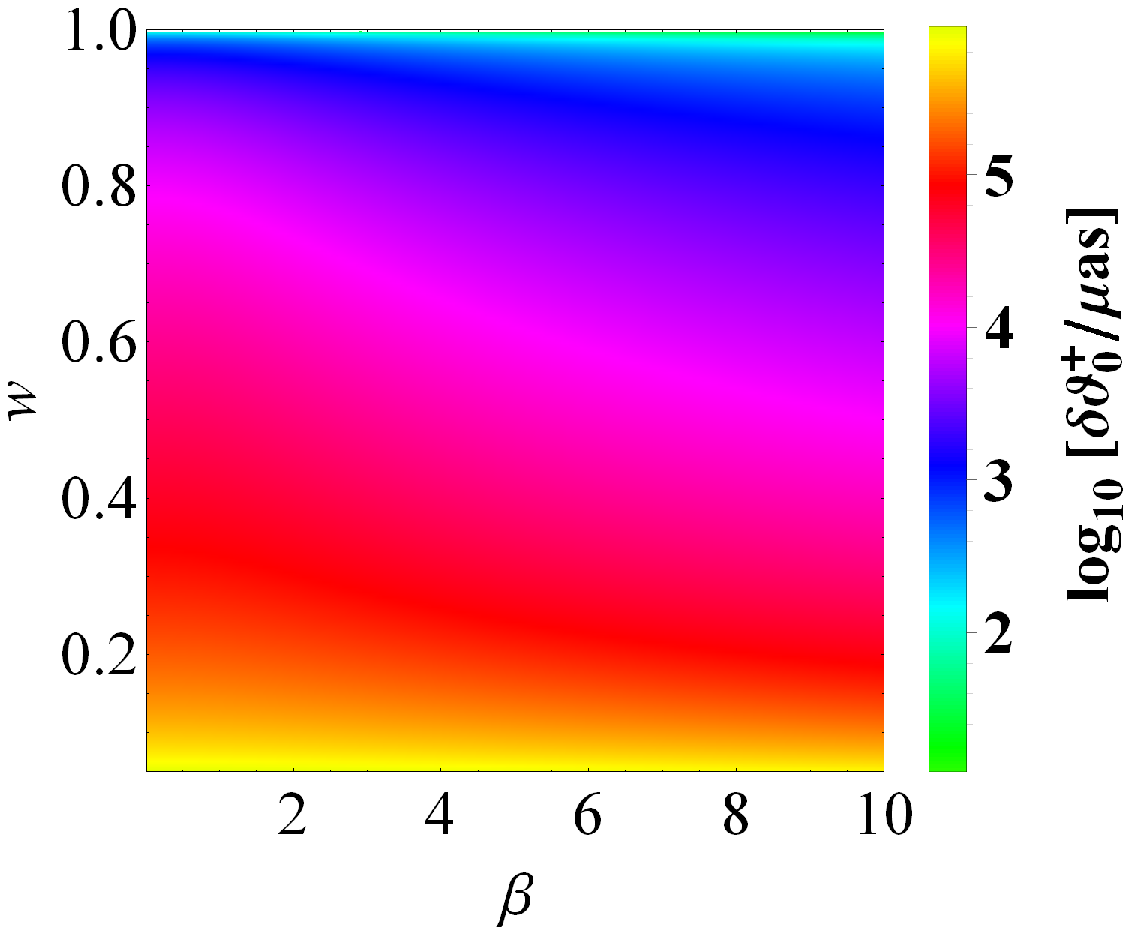} \vspace*{-3pt}
  \centerline{(a) $\delta\vartheta_0^{+}$ }
\end{minipage} \hspace*{13pt}
\begin{minipage}[b]{5.55cm}
\includegraphics[width=5.55cm]{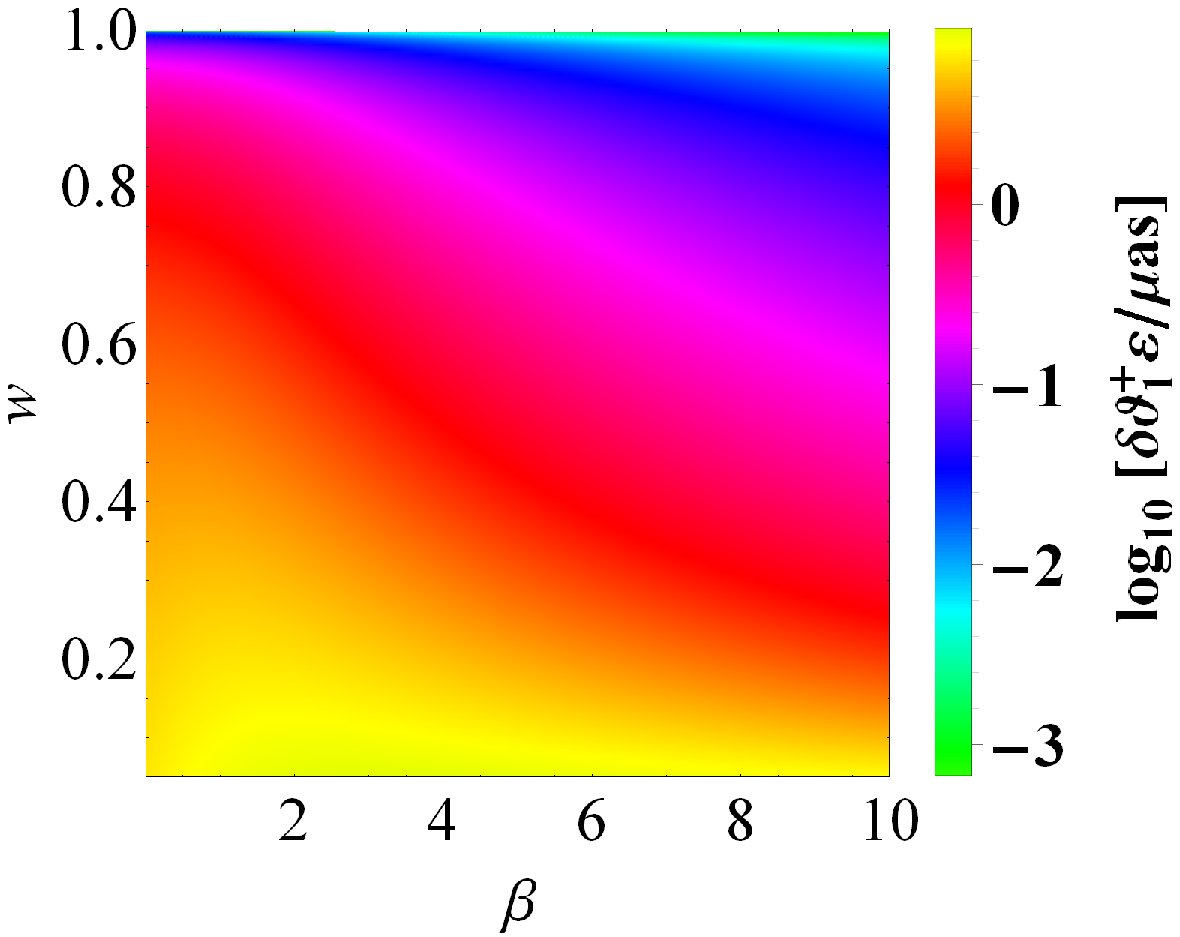}
  \centerline{(b$_1$) $\delta\vartheta_1^{+}\varepsilon$ for $s^{+}=+1$}
\end{minipage} \hspace*{8pt}
\begin{minipage}[b]{5.48cm}
\includegraphics[width=5.48cm]{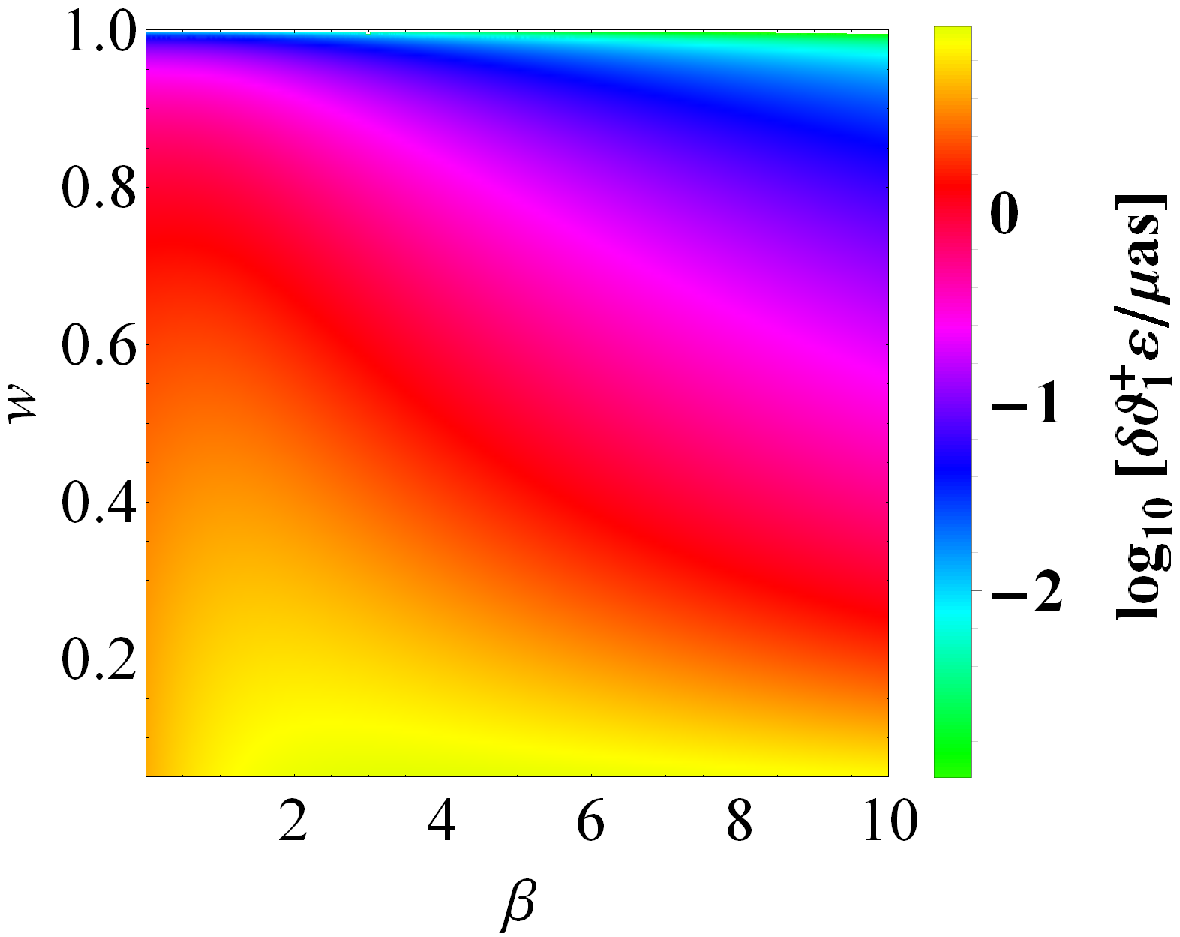}
  \centerline{(b$_2$) $\delta\vartheta_1^{+}\varepsilon$ for $s^{+}=-1$}
\end{minipage} \vspace*{2pt} \\ \hspace*{3pt}
\begin{minipage}[b]{5.55cm}
\includegraphics[width=5.55cm]{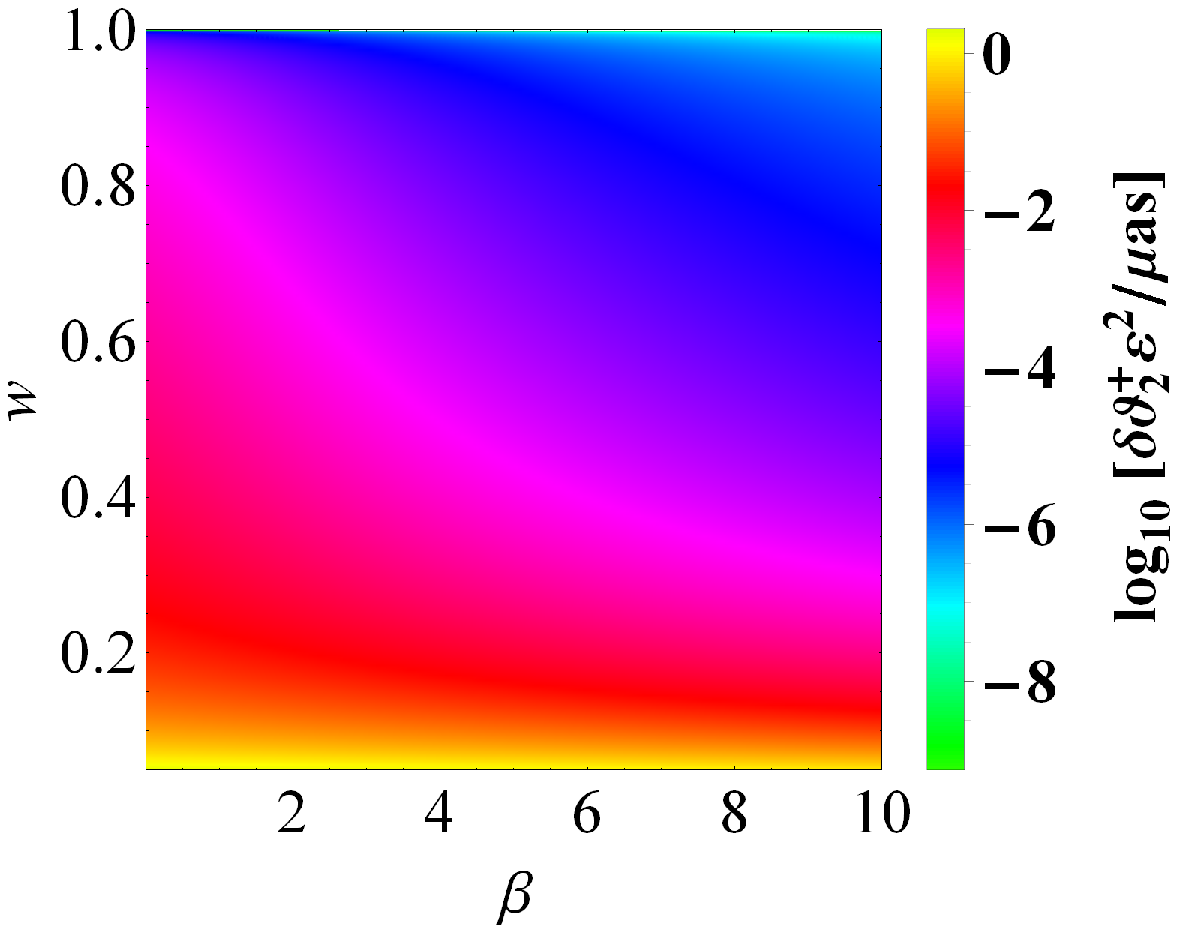} \vspace*{-7pt}
  \centerline{(c) $\delta\vartheta_2^{+}\varepsilon^2$ for $s^{+}=+1$ }
\end{minipage} \hspace*{6pt}
\begin{minipage}[b]{5.35cm}
\includegraphics[width=5.35cm]{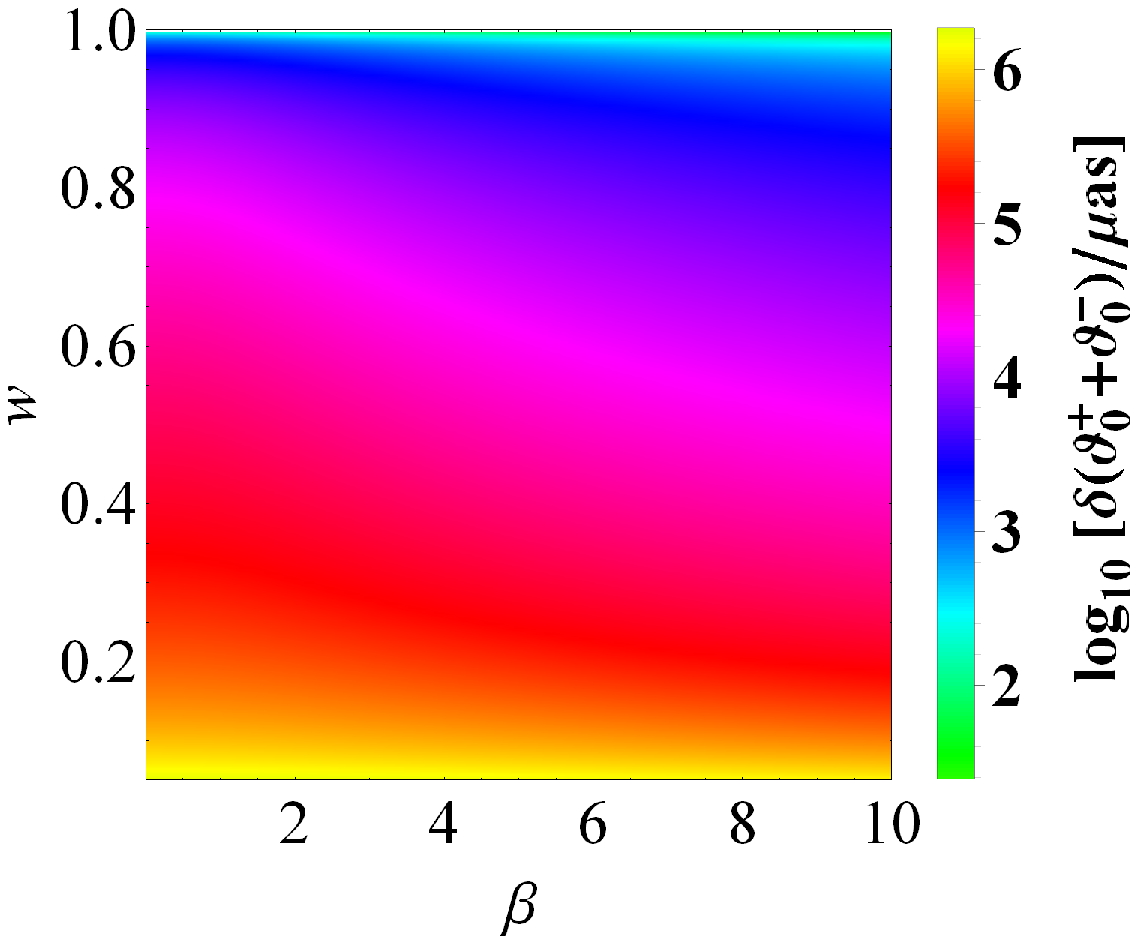} \vspace*{-7pt}
  \centerline{(d) $\delta(\vartheta_0^{+}+\vartheta_0^{-})$ }
\end{minipage} \hspace*{15pt}
\begin{minipage}[b]{5.65cm}
\includegraphics[width=5.65cm]{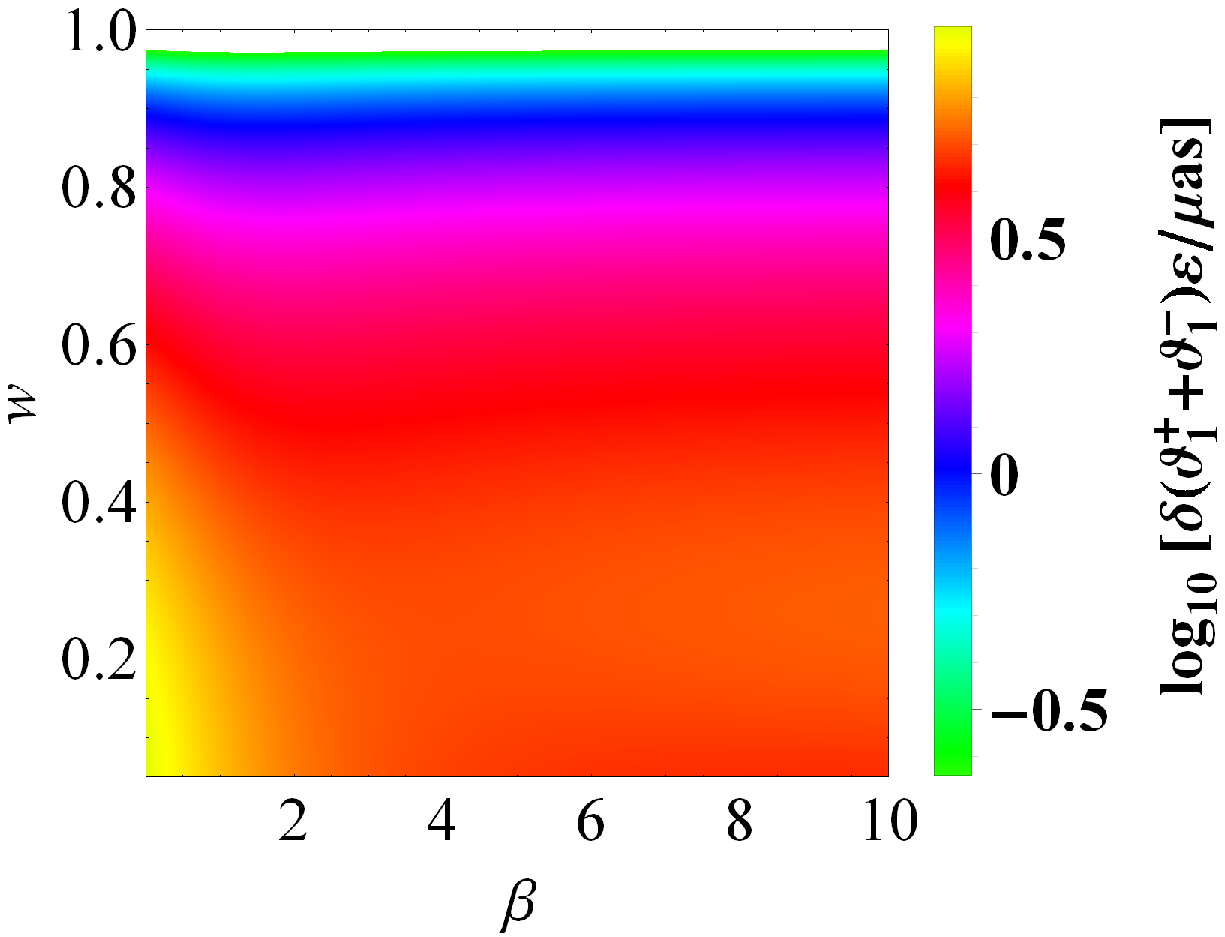}  \vspace*{-4pt}
  \centerline{(e$_1$) $\delta(\vartheta_1^{+}+\vartheta_1^{-})\,\varepsilon$ for $s^{+}=+1$ }
\end{minipage} \vspace*{10pt}  \\  \hspace*{-2pt}
\begin{minipage}[b]{5.73cm}
\includegraphics[width=5.73cm]{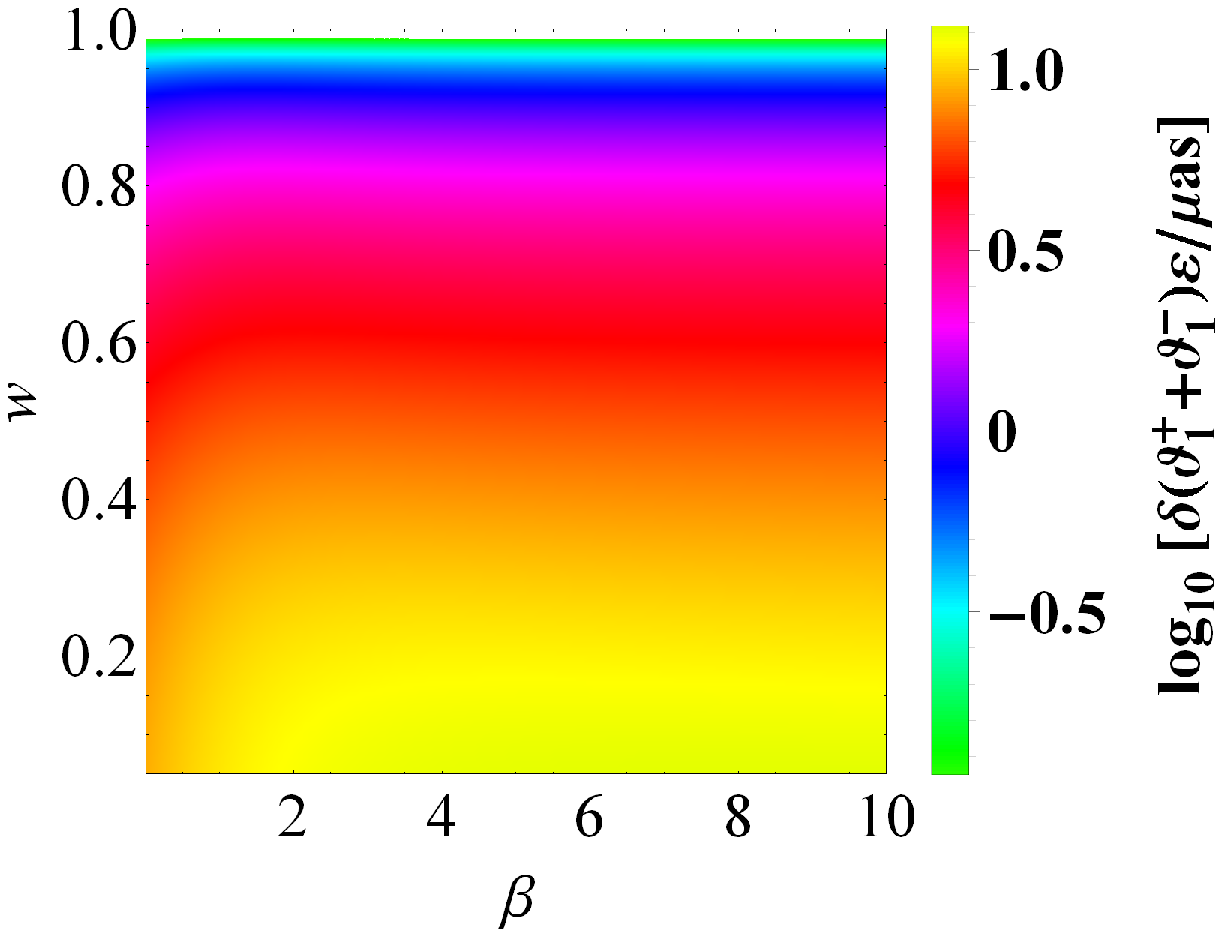}   \vspace*{-5pt}
  \centerline{(e$_2$) $\delta(\vartheta_1^{+}+\vartheta_1^{-})\,\varepsilon$ for $s^{+}=-1$ }
\end{minipage} \vspace*{5pt}  \hspace*{4pt}
\begin{minipage}[b]{5.4cm}
\includegraphics[width=5.4cm]{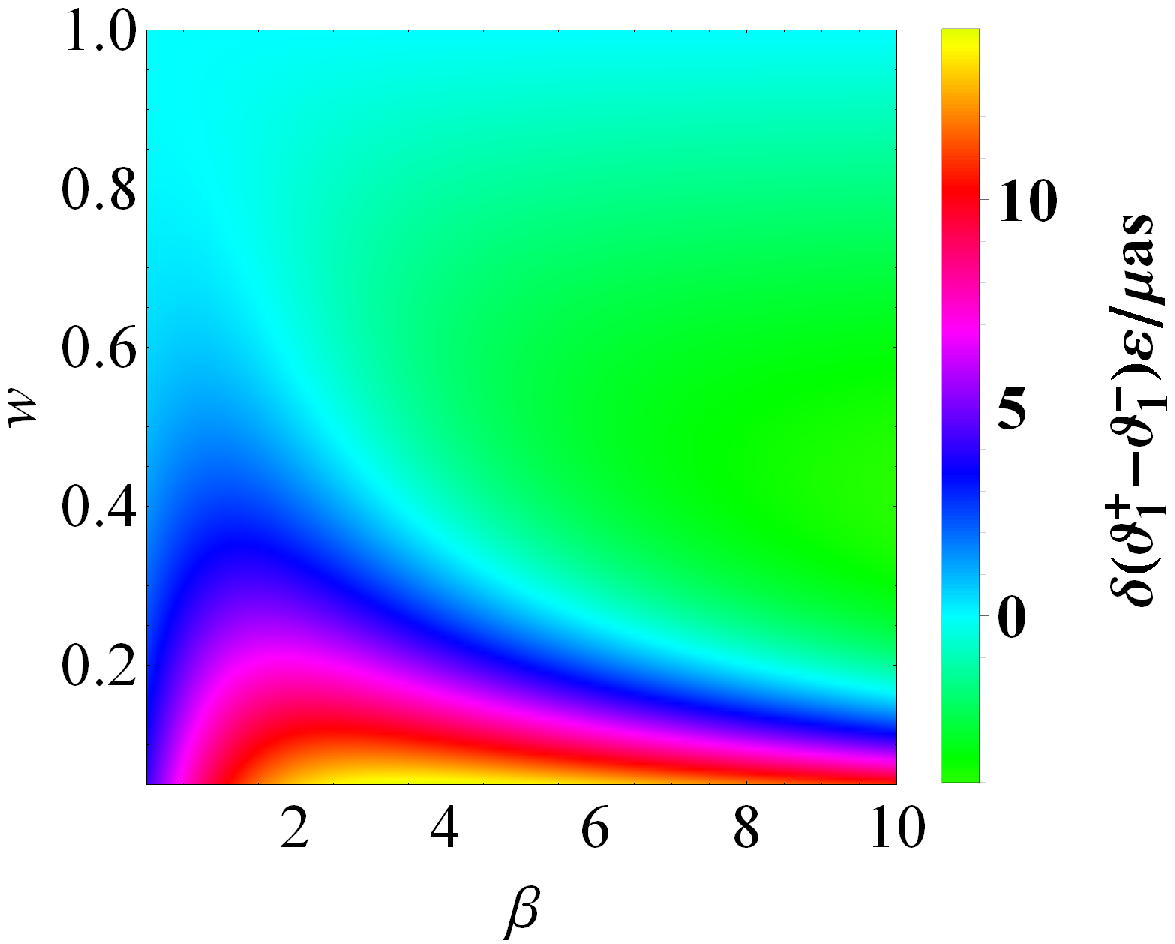}  \vspace*{-4pt}
  \centerline{(f$_1$) $\delta(\vartheta_1^{+}-\vartheta_1^{-})\,\varepsilon$ for $s^{+}=+1$ }
\end{minipage} \hspace*{12pt}
\begin{minipage}[b]{5.62cm}
\includegraphics[width=5.62cm]{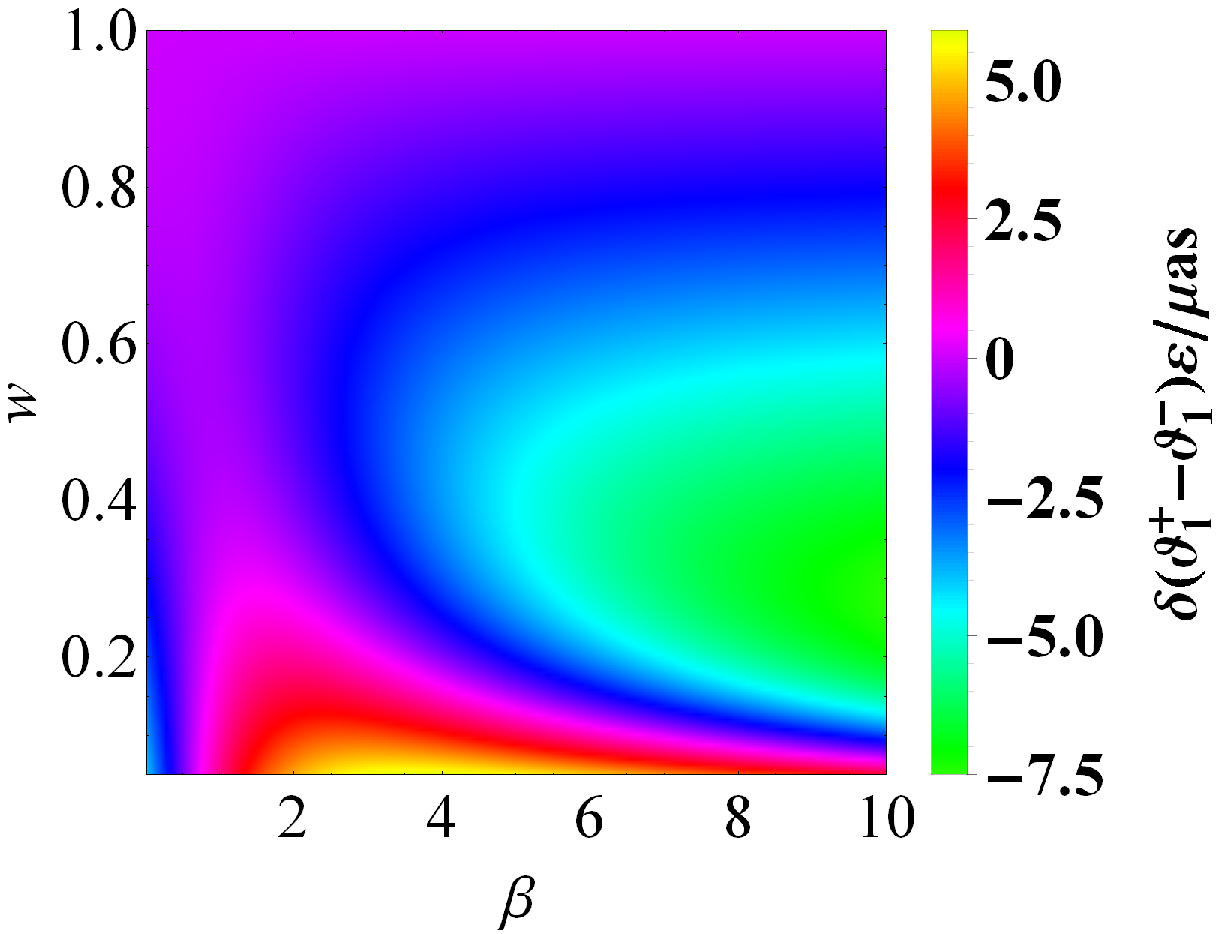}
  \centerline{(f$_2$) $\delta(\vartheta_1^{+}-\vartheta_1^{-})\,\varepsilon$ for $s^{+}=-1$ }
\end{minipage} \\    \hspace*{-1pt}
\begin{minipage}[b]{5.55cm}
\includegraphics[width=5.55cm]{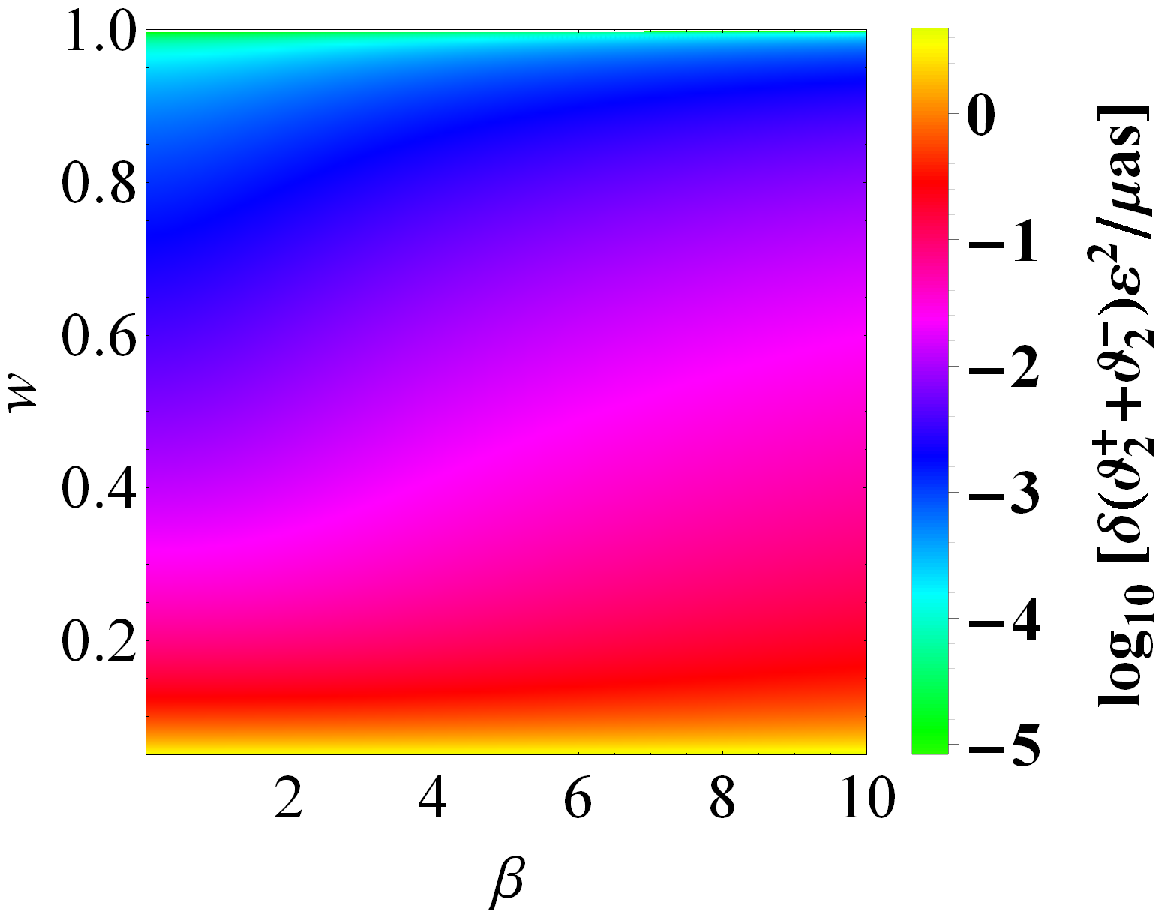} \vspace*{-5pt}
  \centerline{(g) $\delta(\vartheta_2^{+}+\vartheta_2^{-})\,\varepsilon^2$ for $s^{+}=+1$ }
\end{minipage}  \hspace*{9pt}
\begin{minipage}[b]{5.78cm}
\includegraphics[width=5.78cm]{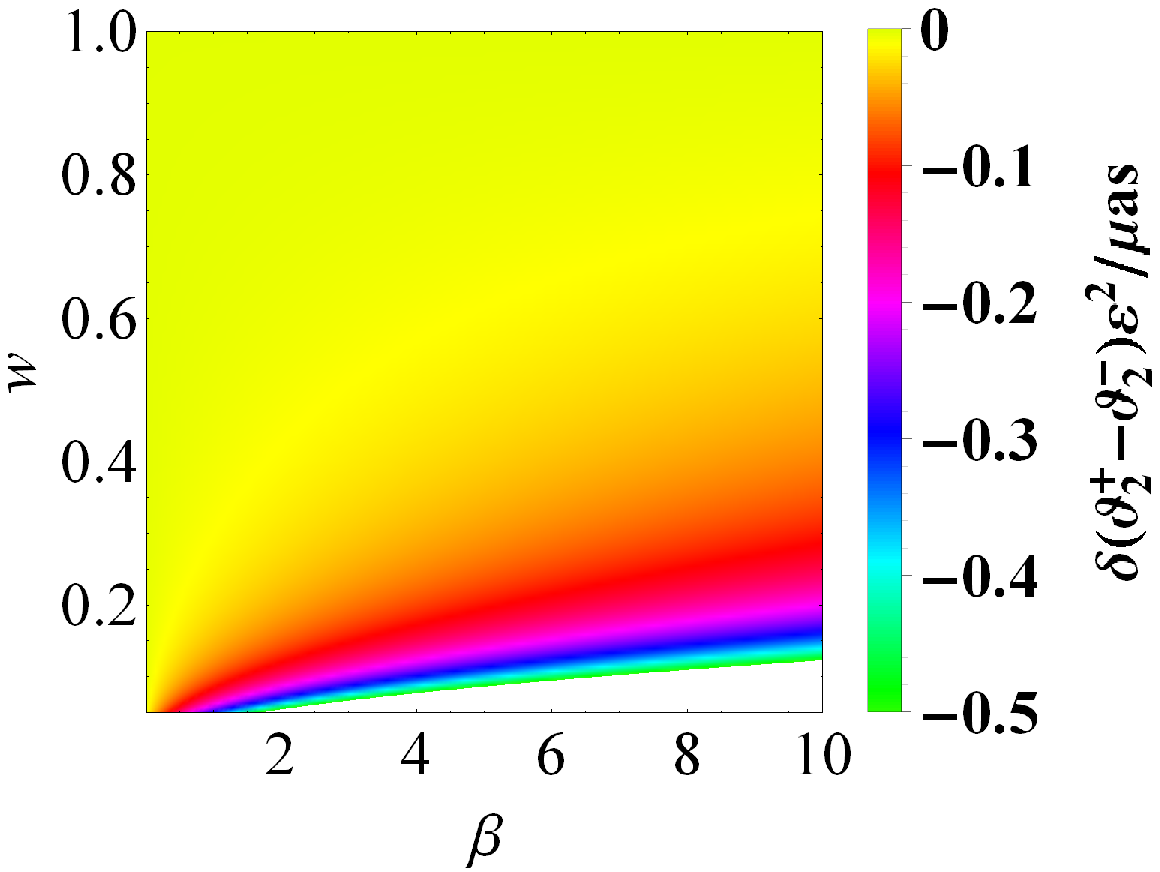}  \vspace*{-4pt}
  \centerline{(h$_1$) $\delta(\vartheta_2^{+}-\vartheta_2^{-})\,\varepsilon^2$ for $s^{+}=+1$ }
\end{minipage} \hspace*{1pt}
\begin{minipage}[b]{5.7cm}
\includegraphics[width=5.7cm]{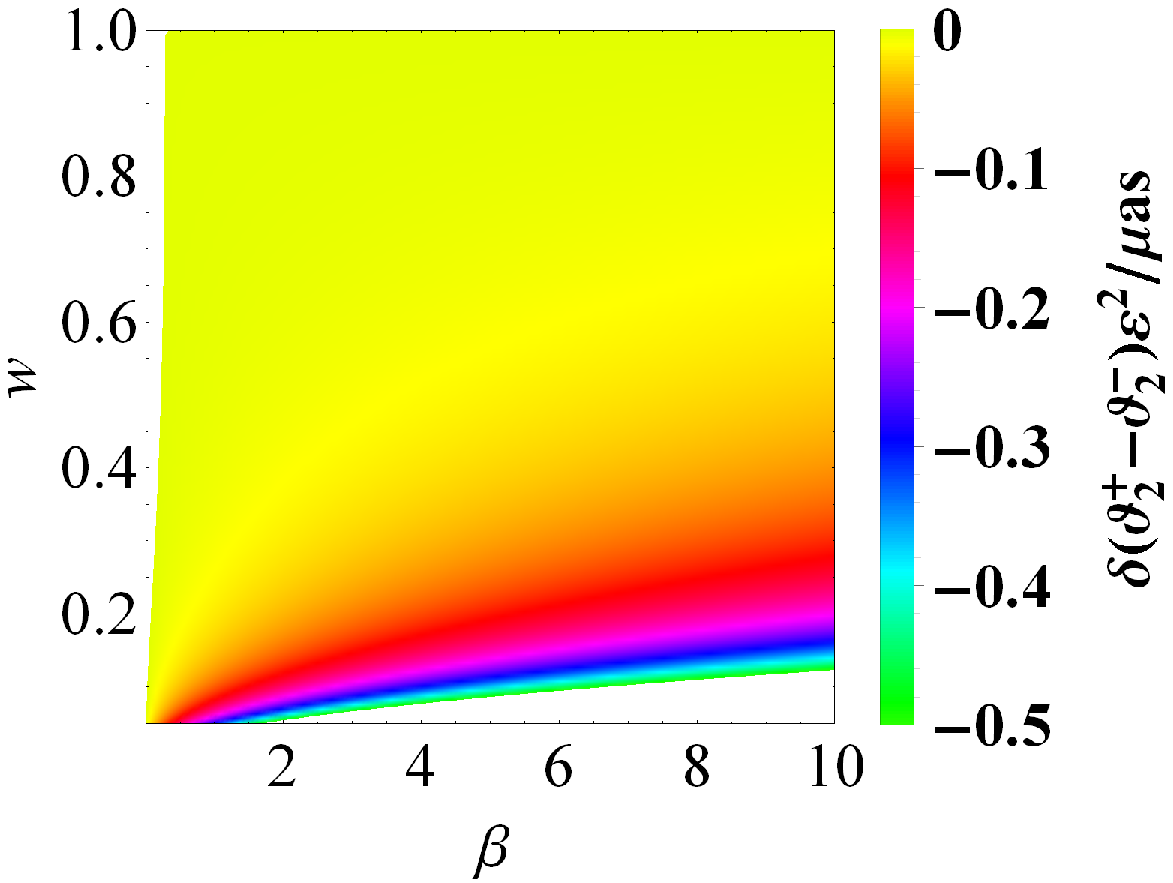}
  \centerline{(h$_2$) $\delta(\vartheta_2^{+}-\vartheta_2^{-})\,\varepsilon^2$ for $s^{+}=-1$ }
\end{minipage}

\caption{$\delta\vartheta_0^{+}$, $\delta\vartheta_1^{+}\varepsilon$, $\delta\vartheta_2^{+}\varepsilon^2$, $\delta(\vartheta_0^{+}+\vartheta_0^{-})$, $\delta(\vartheta_1^{+}\pm\vartheta_1^{-})\,\varepsilon$, and $\delta(\vartheta_2^{+}\pm\vartheta_2^{-})\,\varepsilon^2$ displayed in color-indexed form as the bivariate functions of $w$ and $\beta$, for the particle's prograde ($s^{+}=+1$) or retrograde ($s^{+}=-1$) motion.
The values of the related parameters are given in Sect.~\ref{Basics}. Note that we don't show $\delta\vartheta_2^{+}\varepsilon^2$ and $\delta(\vartheta_2^{+}+\vartheta_2^{-})\,\varepsilon^2$ for the case of $s^{+}=-1$, since it is hard to distinguish them from the corresponding results for the case of $s^{+}=+1$, respectively. Note also that $\delta(\vartheta_1^{+}-\vartheta_1^{-})\,\varepsilon$ and $\delta(\vartheta_2^{+}-\vartheta_2^{-})\,\varepsilon^2$ may take negative values for $0.05\lesssim w<1$ and $0.01\leq\beta\leq10$. Here and thereafter, a white region of a figure indicates the value domain where the magnitude of the velocity effect is too large or too small to be shown properly, and we don't fill them by adjusting the value range for the convenience of display.  } \label{Figure6}
\end{figure*}

\begin{widetext}

\begin{table}
\begin{minipage}[t]{0.49\textwidth}
  \centering
\begin{tabular}{cccccccc} \toprule[1.2px]
      $\beta\:\backslash\:w$    &    ~~~~$ 0.05 $ ~~~~   &     ~~~~$0.1$~~~~       &     ~~~~$0.5$~~~~        &    ~~~~$0.9$~~~~    &      $0.999999$         \\   \midrule[0.5pt] \vspace*{-8pt}  \\
                   0.01         &   9.34$\times10^{5}$   &   4.34$\times10^{5}$    &    4.13$\times10^{4}$    &  4.05$\times10^{3}$ &       $\star$           \\
                    0.1         &   9.34$\times10^{5}$   &   4.33$\times10^{5}$    &    4.12$\times10^{4}$    &  4.04$\times10^{3}$ &       $\star$           \\
                    0.5         &   9.32$\times10^{5}$   &   4.32$\times10^{5}$    &    4.05$\times10^{4}$    &  3.93$\times10^{3}$ &       $\star$           \\
                      1         &   9.27$\times10^{5}$   &   4.26$\times10^{5}$    &    3.84$\times10^{4}$    &  3.64$\times10^{3}$ &       $\star$           \\
                      5         &   8.30$\times10^{5}$   &   3.44$\times10^{5}$    &    1.88$\times10^{4}$    &  1.54$\times10^{3}$ &       $\star$           \\
                     10         &   7.04$\times10^{5}$   &   2.55$\times10^{5}$    &    1.03$\times10^{4}$    &  8.16$\times10^{2}$ &       $\star$           \\   \bottomrule[1.2px]
\end{tabular} \par  \vspace*{3pt}
     \centerline{(a) $\delta\vartheta_0^{+}$ } \vspace*{10pt}
\end{minipage}
\begin{minipage}[t]{0.49\textwidth}
  \centering
\begin{tabular}{cccccccc} \toprule[1.2px]
      $\beta\:\backslash\:w$    &    ~~~~$ 0.05 $ ~~~~   &     ~~~~$0.1$~~~~       &     ~~~~$0.5$~~~~        &    ~~~~$0.9$~~~~    &      $0.999999$         \\   \midrule[0.5pt] \vspace*{-8pt}  \\
                   0.01         &          6.51          &          6.22           &          3.13            &         0.48        &       $\star$           \\
                    0.1         &          6.71          &          6.38           &          3.13            &         0.47        &       $\star$           \\
                    0.5         &          7.54          &          7.06           &          3.09            &         0.44        &       $\star$           \\
                      1         &          8.39          &          7.72           &          2.94            &         0.38        &       $\star$           \\
                      5         &          9.26          &          7.23           &          0.92            &         0.08        &       $\star$           \\
                     10         &          7.68          &          4.73           &          0.29            &       $\star$       &       $\star$           \\   \bottomrule[1.2px]
\end{tabular} \par  \vspace*{3pt}
     \centerline{(b) $\delta\vartheta_1^{+}\varepsilon$ for $s^{+}=+1$} \vspace*{10pt}
\end{minipage}

\begin{minipage}[t]{0.49\textwidth}
  \centering
\begin{tabular}{cccccccc} \toprule[1.2px]
      $\beta\:\backslash\:w$    &    ~~~~$ 0.05 $ ~~~~   &     ~~~~$0.1$~~~~       &     ~~~~$0.5$~~~~        &    ~~~~$0.9$~~~~    &      $0.999999$         \\   \midrule[0.5pt] \vspace*{-8pt}  \\
                   0.01         &          2.06          &          0.27           &        $\star$           &       $\star$       &       $\star$           \\
                    0.1         &          2.05          &          0.27           &        $\star$           &       $\star$       &       $\star$           \\
                    0.5         &          1.99          &          0.25           &        $\star$           &       $\star$       &       $\star$           \\
                      1         &          1.92          &          0.23           &        $\star$           &       $\star$       &       $\star$           \\
                      5         &          1.43          &          0.13           &        $\star$           &       $\star$       &       $\star$           \\
                     10         &          0.97          &          0.06           &        $\star$           &       $\star$       &       $\star$           \\   \bottomrule[1.2px]
\end{tabular} \par  \vspace*{3pt}
     \centerline{(c) $\delta\vartheta_2^{+}\varepsilon^2$ for $s^{+}=+1$ }  \vspace*{10pt}
\end{minipage}
\begin{minipage}[t]{0.49\textwidth}
  \centering
\begin{tabular}{cccccccc} \toprule[1.2px]
      $\beta\:\backslash\:w$    &    ~~~~$ 0.05 $ ~~~~   &     ~~~~$0.1$~~~~       &     ~~~~$0.5$~~~~        &    ~~~~$0.9$~~~~    &     $0.999999$          \\   \midrule[0.5pt] \vspace*{-8pt}  \\
                   0.01         &   1.87$\times10^{6}$   &   8.67$\times10^{5}$    &    8.25$\times10^{4}$    &  8.10$\times10^{3}$ &       0.07              \\
                    0.1         &   1.87$\times10^{6}$   &   8.67$\times10^{5}$    &    8.25$\times10^{4}$    &  8.09$\times10^{3}$ &       0.07              \\
                    0.5         &   1.86$\times10^{6}$   &   8.63$\times10^{5}$    &    8.09$\times10^{4}$    &  7.87$\times10^{3}$ &       0.07              \\
                      1         &   1.85$\times10^{6}$   &   8.53$\times10^{5}$    &    7.67$\times10^{4}$    &  7.28$\times10^{3}$ &       0.06              \\
                      5         &   1.66$\times10^{6}$   &   6.87$\times10^{5}$    &    3.77$\times10^{4}$    &  3.08$\times10^{3}$ &       $\star$           \\
                     10         &   1.41$\times10^{6}$   &   5.10$\times10^{5}$    &    2.06$\times10^{4}$    &  1.63$\times10^{3}$ &       $\star$           \\   \bottomrule[1.2px]
\end{tabular} \par  \vspace*{3pt}
     \centerline{(d) $\delta(\vartheta_0^{+}+\vartheta_0^{-})$ } \vspace*{10pt}
\end{minipage}

\begin{minipage}[t]{0.49\textwidth}
  \centering
\begin{tabular}{cccccccc} \toprule[1.2px]
      $\beta\:\backslash\:w$    &    ~~~~$ 0.05 $ ~~~~   &     ~~~~$0.1$~~~~       &     ~~~~$0.5$~~~~        &    ~~~~$0.9$~~~~    &     $0.999999$          \\   \midrule[0.5pt] \vspace*{-8pt}  \\
                   0.01         &          8.87          &          8.73           &          5.35            &         0.94        &       $\star$           \\
                    0.1         &          8.66          &          8.54           &          5.25            &         0.92        &       $\star$           \\
                    0.5         &          7.79          &          7.68           &          4.82            &         0.88        &       $\star$           \\
                      1         &          6.87          &          6.78           &          4.42            &         0.84        &       $\star$           \\
                      5         &          4.74          &          4.83           &          4.21            &         0.88        &       $\star$           \\
                     10         &          4.58          &          4.81           &          4.37            &         0.90        &       $\star$           \\   \bottomrule[1.2px]
\end{tabular} \par  \vspace*{3pt}
     \centerline{(e) $\delta(\vartheta_1^{+}+\vartheta_1^{-})\,\varepsilon$ for $s^{+}=+1$}
\end{minipage}
\begin{minipage}[t]{0.49\textwidth}
  \centering
\begin{tabular}{cccccccc} \toprule[1.2px]
      $\beta\:\backslash\:w$    &    ~~~~$ 0.05 $ ~~~~   &     ~~~~$0.1$~~~~       &     ~~~~$0.5$~~~~        &    ~~~~$0.9$~~~~    &      $0.999999$         \\   \midrule[0.5pt] \vspace*{-8pt}  \\
                   0.01         &          4.16          &          3.71           &          0.92            &       $\star$       &       $\star$           \\
                    0.1         &          4.76          &          4.23           &          1.01            &       $\star$       &       $\star$           \\
                    0.5         &          7.29          &          6.43           &          1.36            &       $\star$       &       $\star$           \\
                      1         &          9.92          &          8.65           &          1.46            &       $-$0.08       &       $\star$           \\
                      5         &          13.79         &          9.63           &       $-$2.38            &       $-$0.73       &       $\star$           \\
                     10         &          10.78         &          4.64           &       $-$3.80            &       $-$0.86       &       $\star$           \\   \bottomrule[1.2px]
\end{tabular} \par  \vspace*{3pt}
     \centerline{(f) $\delta(\vartheta_1^{+}-\vartheta_1^{-})\,\varepsilon$ for $s^{+}=+1$ } \vspace*{10pt}
\end{minipage}

\begin{minipage}[t]{0.49\textwidth}
  \centering
\begin{tabular}{cccccccc} \toprule[1.2px]
      $\beta\:\backslash\:w$    &    ~~~~$ 0.05 $ ~~~~   &     ~~~~$0.1$~~~~       &     ~~~~$0.5$~~~~        &    ~~~~$0.9$~~~~    &      $0.999999$         \\  \midrule[0.5pt] \vspace*{-8pt}  \\
                   0.01         &          4.12          &          0.54           &        $\star$           &       $\star$       &       $\star$           \\
                    0.1         &          4.12          &          0.54           &        $\star$           &       $\star$       &       $\star$           \\
                    0.5         &          4.13          &          0.54           &        $\star$           &       $\star$       &       $\star$           \\
                      1         &          4.13          &          0.54           &        $\star$           &       $\star$       &       $\star$           \\
                      5         &          4.30          &          0.63           &        $\star$           &       $\star$       &       $\star$           \\
                     10         &          4.84          &          0.86           &        $\star$           &       $\star$       &       $\star$           \\   \bottomrule[1.2px]
\end{tabular} \par  \vspace*{3pt}
     \centerline{(g) $\delta(\vartheta_2^{+}+\vartheta_2^{-})\,\varepsilon^2$ for $s^{+}=+1$}
\end{minipage}
\begin{minipage}[t]{0.49\textwidth}
  \centering
\begin{tabular}{cccccccc} \toprule[1.2px]
      $\beta\:\backslash\:w$    &    ~~~~$ 0.05 $ ~~~~   &     ~~~~$0.1$~~~~       &     ~~~~$0.5$~~~~        &    ~~~~$0.9$~~~~    &      $0.999999$         \\   \midrule[0.5pt] \vspace*{-8pt}  \\
                   0.01         &         $\star$        &        $\star$          &        $\star$           &       $\star$       &       $\star$           \\
                    0.1         &         $\star$        &        $\star$          &        $\star$           &       $\star$       &       $\star$           \\
                    0.5         &         $-$0.15        &        $\star$          &        $\star$           &       $\star$       &       $\star$           \\
                      1         &         $-$0.29        &        $-$0.08          &        $\star$           &       $\star$       &       $\star$           \\
                      5         &         $-$1.45        &        $-$3.71          &        $\star$           &       $\star$       &       $\star$           \\
                     10         &         $-$2.90        &        $-$7.41          &        $\star$           &       $\star$       &       $\star$           \\   \bottomrule[1.2px]
\end{tabular} \par  \vspace*{3pt}
     \centerline{(h) $\delta(\vartheta_2^{+}-\vartheta_2^{-})\,\varepsilon^2$ for $s^{+}=+1$ }
\end{minipage}

\caption{The magnitudes (in units of $\mu$as) of $\delta\vartheta_0^{+}$, $\delta\vartheta_1^{+}\varepsilon$, $\delta\vartheta_2^{+}\varepsilon^2$, $\delta(\vartheta_0^{+}+\vartheta_0^{-})$, $\delta(\vartheta_1^{+}\pm\vartheta_1^{-})\,\varepsilon$, and $\delta(\vartheta_2^{+}\pm\vartheta_2^{-})\,\varepsilon^2$ for various $w$ and $\beta$. Hereafter, our attention is focused on the absolute value of magnitudes of the velocity effects when analyzing their measurability. The star ``$\star$" denotes the magnitude whose absolute value is less than $0.05\mu$as (the capability of NEAT).  }   \label{Table1}
\end{table}
\end{widetext}

Figure~\ref{Figure6} shows the color-indexed velocity effects on the zeroth-, first-, and second-order contributions to the positive-parity image position, as well as on the sum and difference relations for the positive- and negative-parity image positions, as the bivariate functions of $w$ and $\beta$ for prograde $(s^{+}=+1)$ or retrograde $(s^{+}=-1)$ motion of the massive particle. For the readers' convenience, the magnitudes of these velocity effects for particle's prograde motion are presented in Tab.~\ref{Table1}. According to the results given in Fig.~\ref{Figure6} and Tab.~\ref{Table1}, three aspects are summarized. Firstly, for a given angular source position $\beta$ in its domain, it is found that the velocity effects on the zeroth-, first-, and second-order contributions to positive-parity image position, as well as on the zeroth- and second-order positional sum relations, increase monotonically with decreasing $w$. It also applies to the velocity effect on the first-order positional sum relation when $s^{+}=-1$. Contrary to this trend, the velocity effect on the second-order positional difference relation decreases when decreasing $w$ from $1$ to $0.05$, for a given $\beta$. Compared with them, the velocity effect on the first-order positional sum for $s^{+}=+1$ and a given $\beta$ within the domain $3.3\lesssim\beta\leq10$ firstly increases to a maximum value and then decreases with the decrease of $w$, although the value of $w$ for the peak value of the velocity effect varies with $\beta$. Moreover, we find the velocity effect on the first-order positional difference relation first decreases to a minimum value and then increases to some value with decreasing $w$ for a given $\beta$. The magnitude of it can be positive, negative, or zero. It is interesting to find that its zero-value region has an approximate $C$ sharp with a short tail. Secondly, we consider the possibilities to detect the velocity-induced effects qualitatively. One can see from Fig.~\ref{Figure6} and Tab.~\ref{Table1} that the magnitude of the velocity effect on the zeroth-order contribution to the positive-parity image position or the positional sum relation for almost all relativistic massive particles is much larger than current observational accuracy ($\sim\!\mu$as). For instance, the magnitude of $\delta(\vartheta_0^{+}+\vartheta_0^{-})$ with $\beta=0.5$ still exceeds the NEAT's accuracy ($0.05\mu$as) for an ultrarelativistic massive particle with an initial velocity $w=0.999999$ (such as a common neutrino~\cite{KBF1979,Adam2012}) as the test particle. We also notice that there is a large possibility to detect the velocity effect on the first-order contribution of the primary image position or the positional sum relation, since their magnitudes are much larger than $0.05\mu$as for most relativistic massive particles (with a rough range of $0.05\lesssim w\lesssim0.8$) and a given $\beta\in\left[0.01,~10\right]$. The smaller the source position $\beta$ is, the higher upper limit the rough range of $w$ will have for $\delta\vartheta_1^{+}\varepsilon$. The possibility to detect the velocity effect (focusing on the absolute value) on the first-order positional difference relation is relatively large, which requires a proper combination of $w$ and $\beta$. With respect to the velocity effect on the second-order contribution to the image position or to the positional sum or difference relation, it is likely to detect them only when the massive particle has a relatively small relativistic initial velocity (e.g., $w\lesssim 0.2$ for $\delta(\vartheta_2^{+}+\vartheta_2^{-})\,\varepsilon^2$). Thirdly, it should be mentioned that the direction of the orbital angular momentum of the particle's motion relative to the lens' rotation may make a difference to the magnitudes of these velocity effects and their detection. For example, the difference between the magnitudes of $\delta(\vartheta_1^{+}-\vartheta_1^{-})\,\varepsilon$ for $s^{+}=+1$ and $s^{+}=-1$ is considerably in excess of $0.05\mu$as for a fixed source position $\beta\in[0.01,~10]$, provided $w$ is relatively small (e.g., $w\lesssim0.5$).
\begin{figure*}
\centering
\begin{minipage}[b]{5.25cm}
\includegraphics[width=5.25cm]{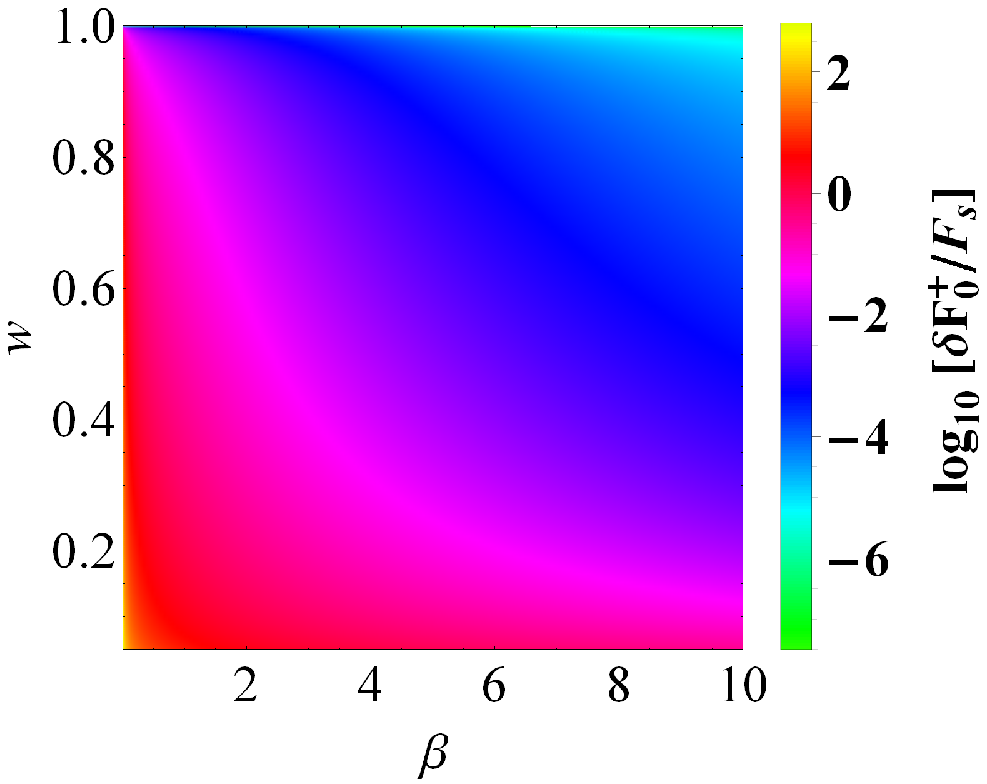}
  \centerline{ (a) $\delta F_0^{+}/F_{\text{s}}$ }
\end{minipage} \hspace*{2pt}
\begin{minipage}[b]{6.1cm}
\includegraphics[width=6.1cm]{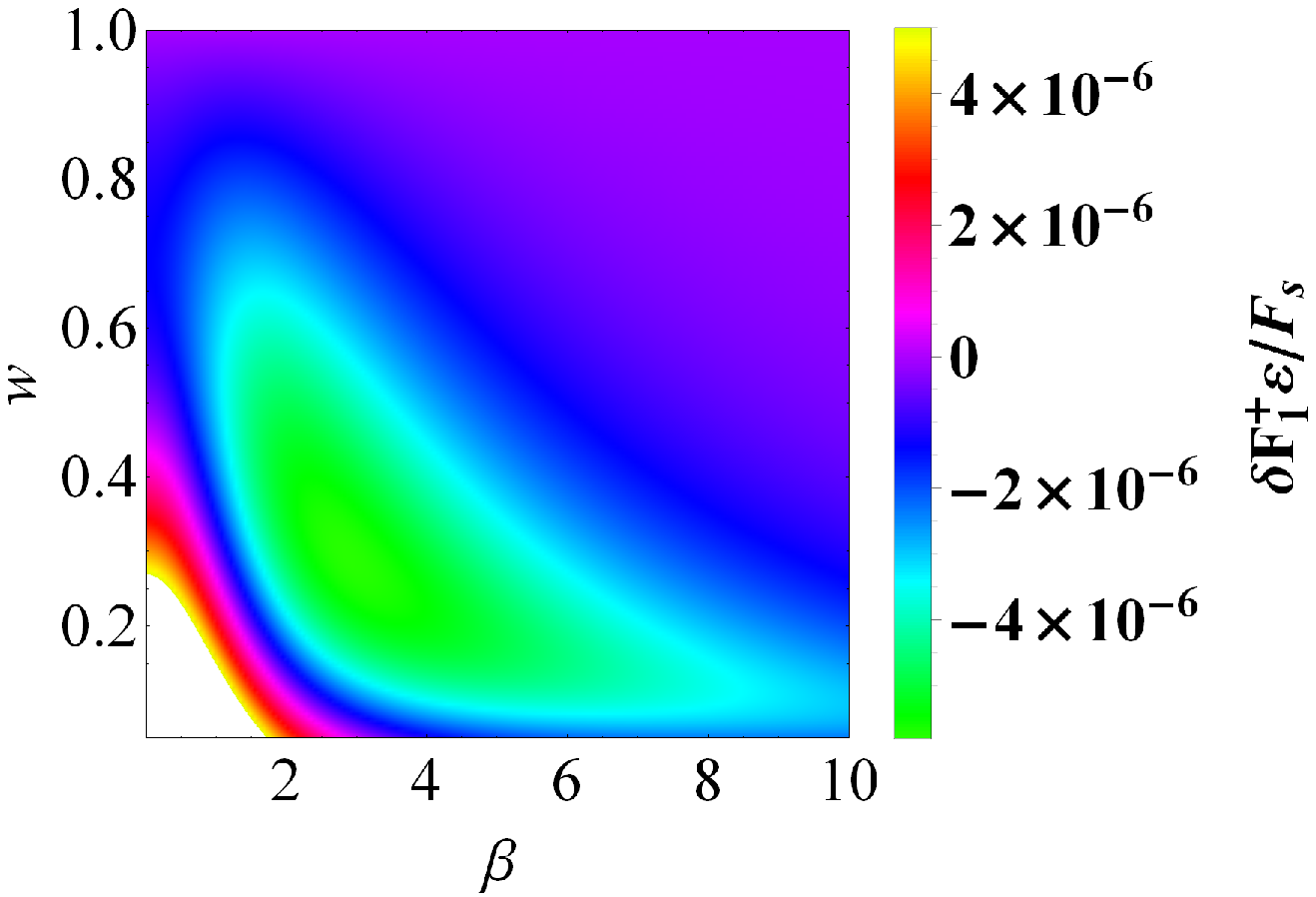}
  \centerline{(b$_1$) $\delta F_1^{+}\varepsilon/F_{\text{s}}$ for $s^{+}=+1$}
\end{minipage} \hspace*{2pt} \vspace*{-2pt}
\begin{minipage}[b]{6.1cm}
\includegraphics[width=6.1cm]{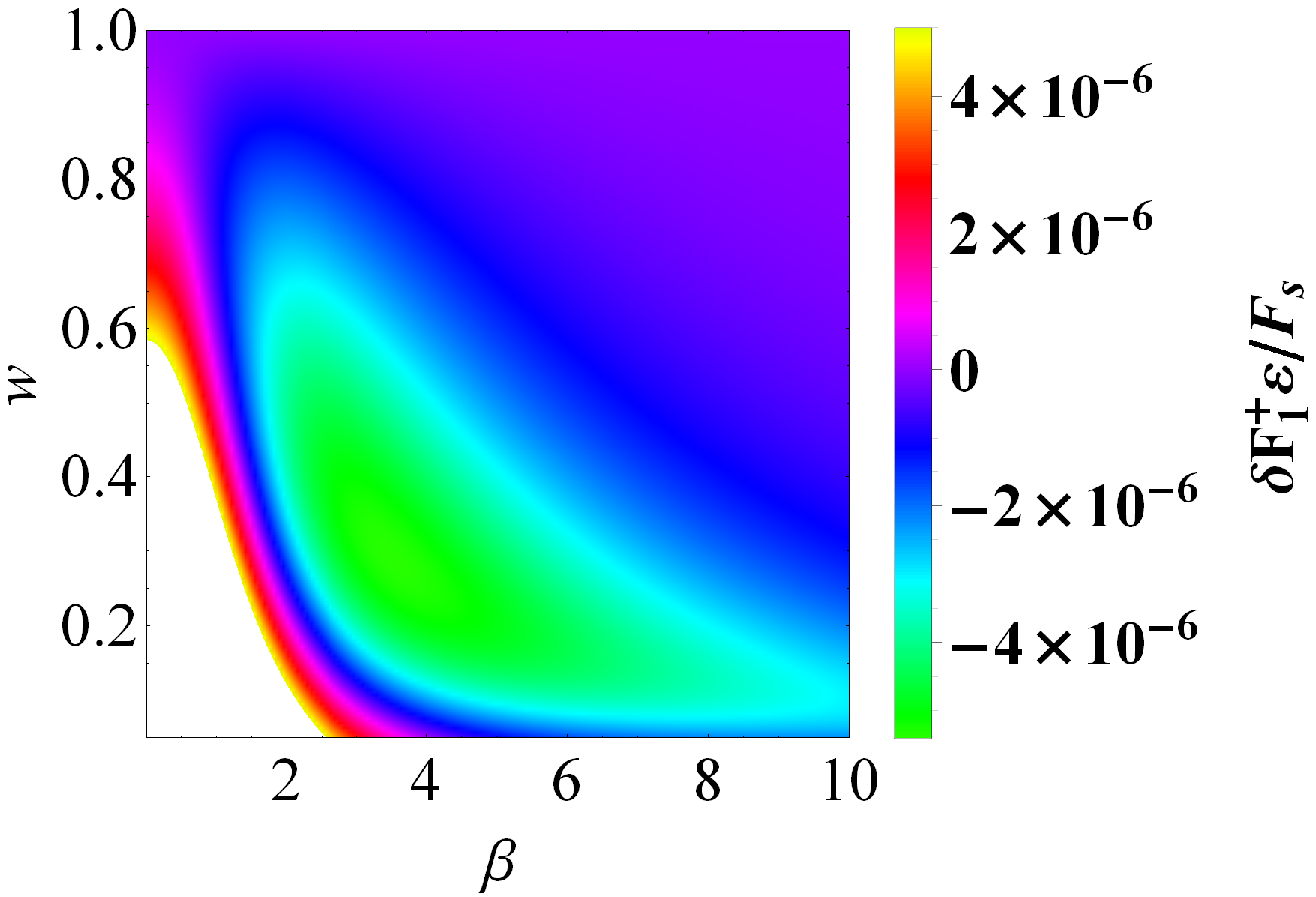}
  \centerline{(b$_2$) $\delta F_1^{+}\varepsilon/F_{\text{s}}$ for $s^{+}=-1$}
\end{minipage}    \vspace*{0.1pt} \\
\begin{minipage}[b]{6.1cm}
\includegraphics[width=6.1cm]{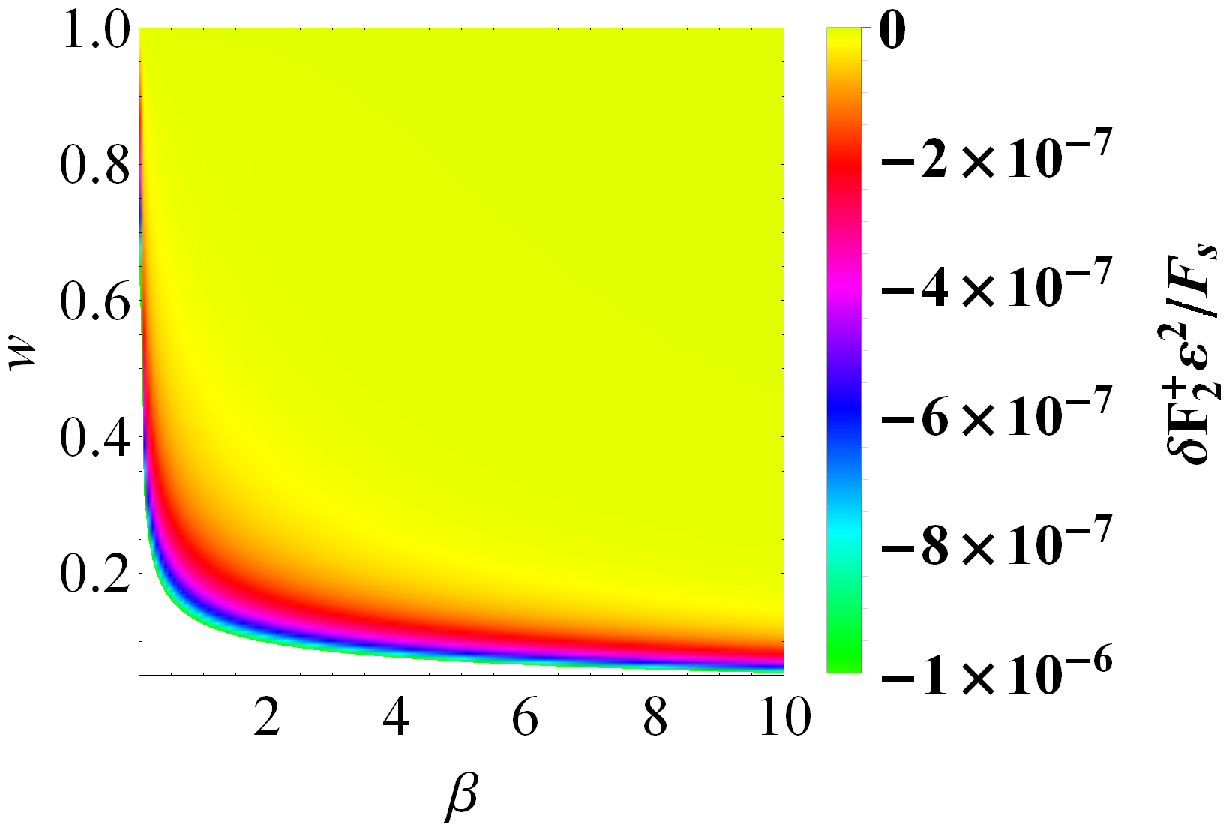}  \vspace*{-4pt}
  \centerline{(c) $\delta F_2^{+}\varepsilon^2/F_{\text{s}}$ for $s^{+}=+1$ }
\end{minipage} \hspace*{20pt}
\begin{minipage}[b]{5.17cm}
\includegraphics[width=5.17cm]{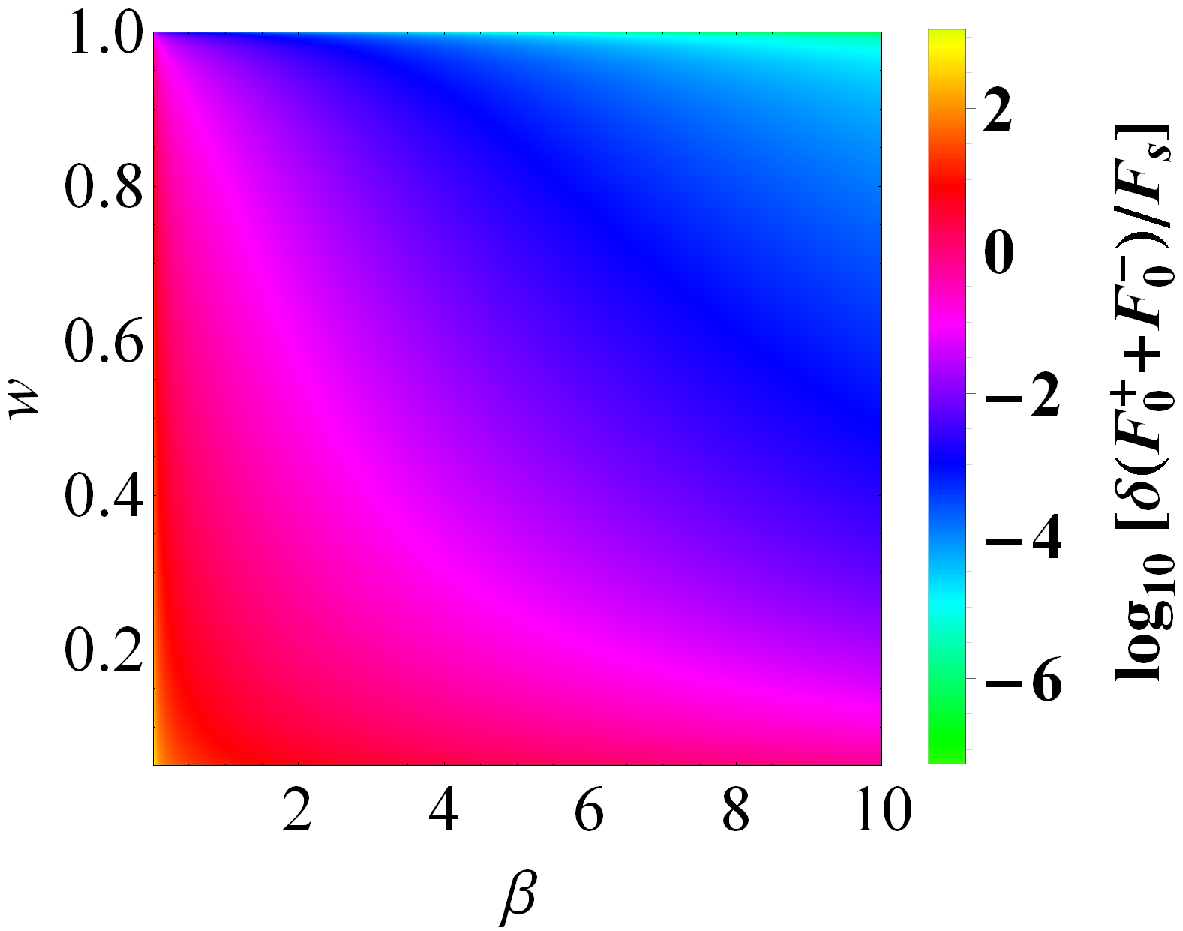}
  \centerline{(d) $\delta(F_0^{+}+F_0^{-})/F_{\text{s}}$ }
\end{minipage}   \vspace*{4pt} \\   \hspace*{21pt}
\begin{minipage}[b]{5.85cm}
\includegraphics[width=5.85cm]{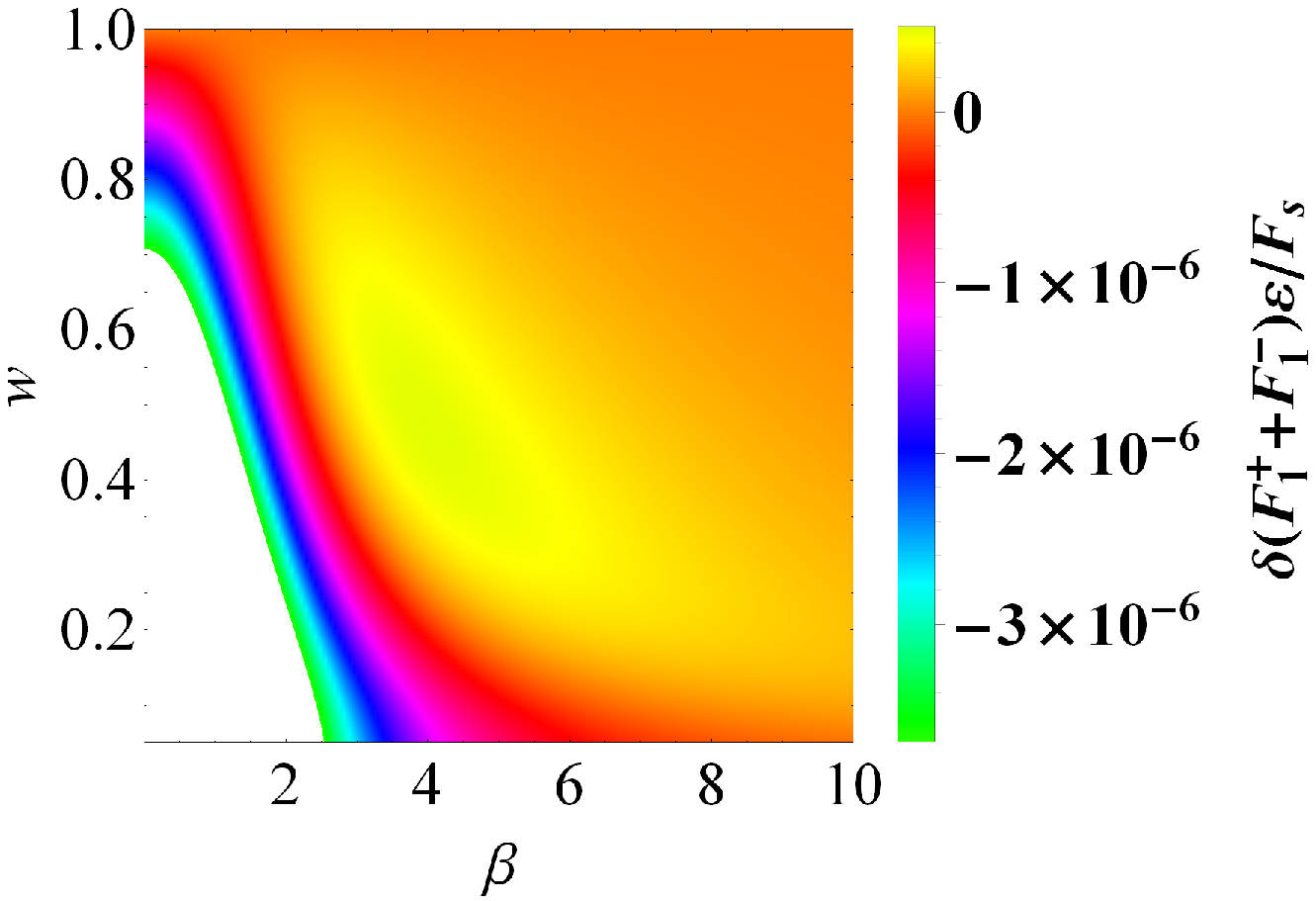}
  \centerline{(e) $\delta(F_1^{+}+F_1^{-})\,\varepsilon/F_{\text{s}}$ for $s^{+}=+1$ }
\end{minipage}    \hspace*{25pt}
\begin{minipage}[b]{5.9cm}
\includegraphics[width=5.9cm]{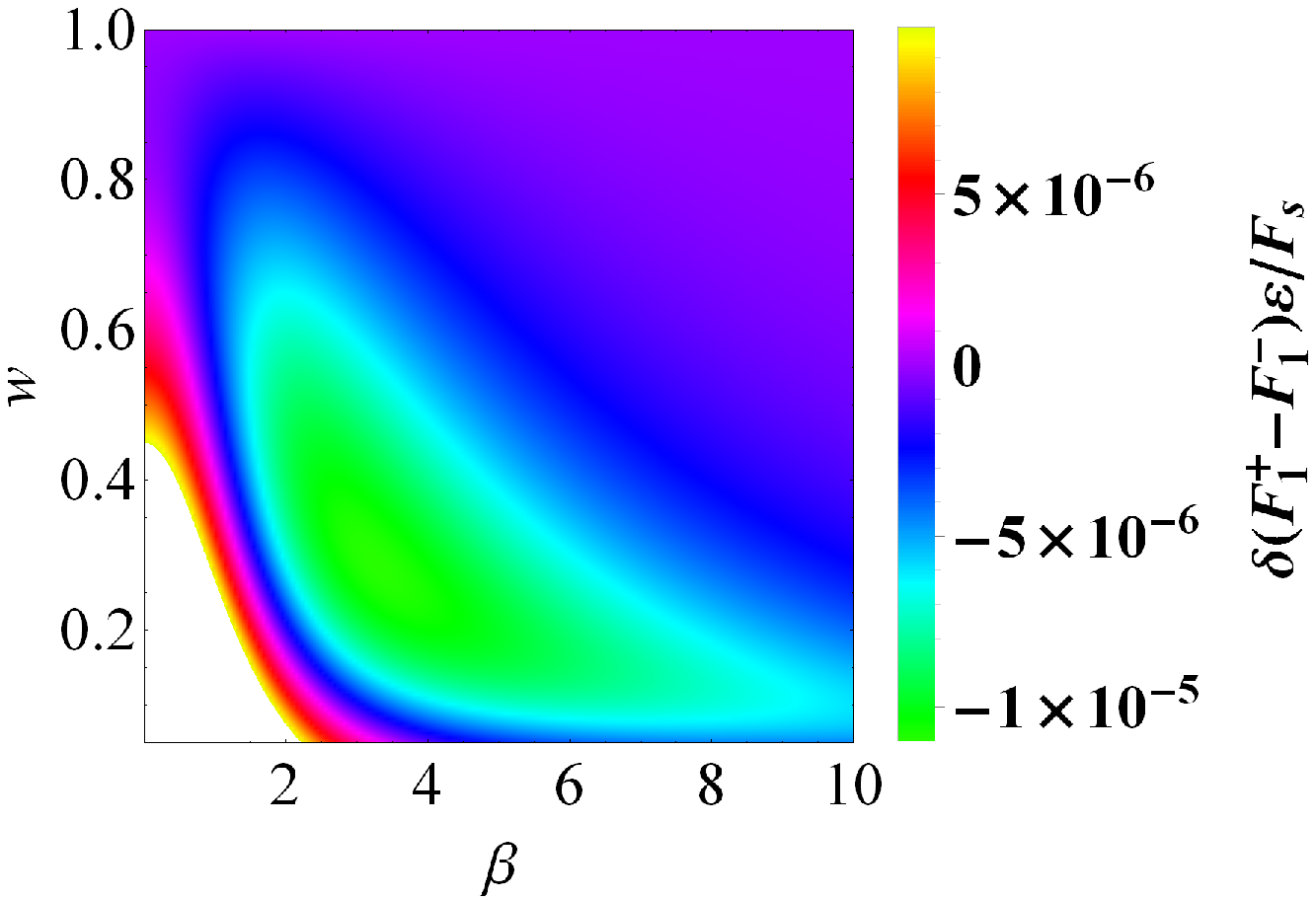}     \vspace*{-3pt}
  \centerline{(f) $\delta(F_1^{+}-F_1^{-})\,\varepsilon/F_{\text{s}}$ }
\end{minipage}   \vspace*{3pt} \\    \hspace*{-3pt}
\begin{minipage}[b]{6.05cm}
\includegraphics[width=6.05cm]{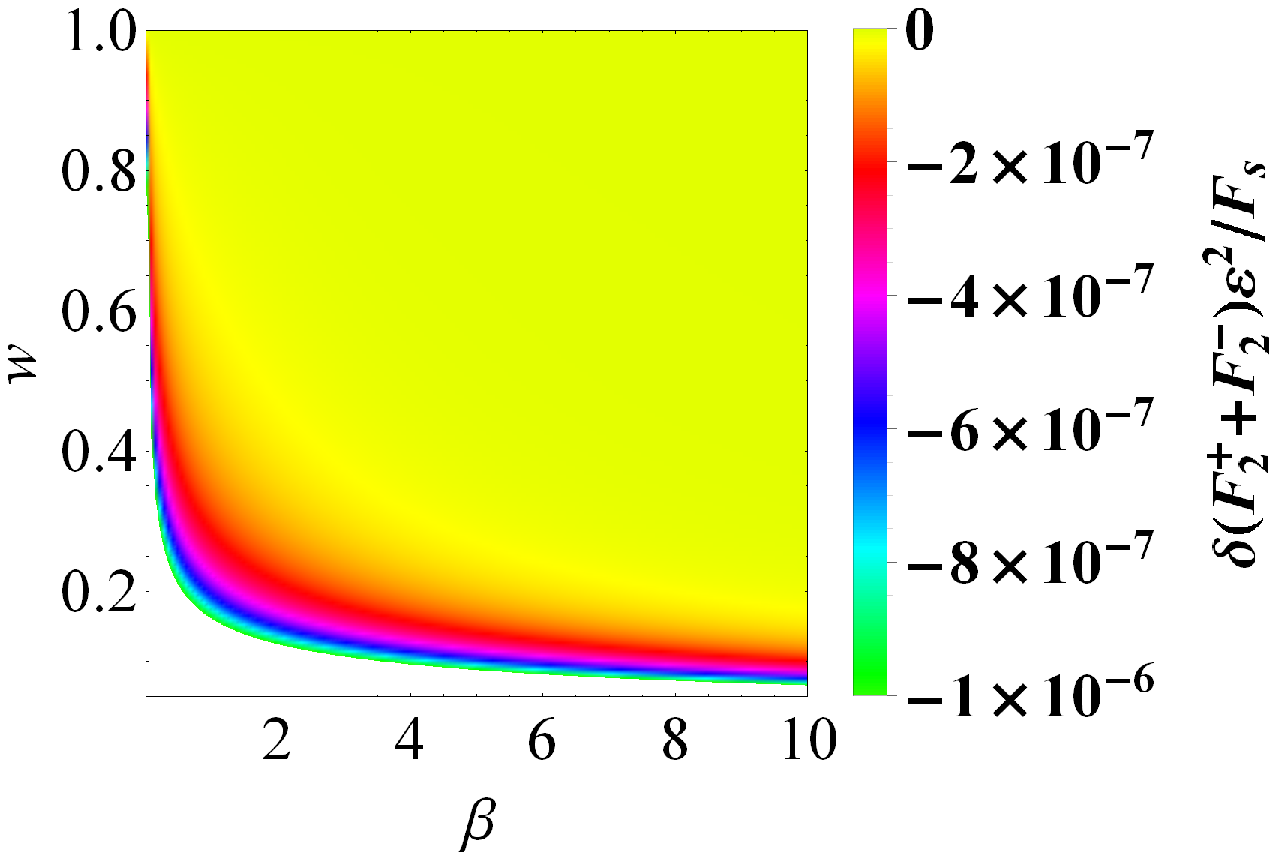}  \vspace*{-4pt}
  \centerline{(g) $\delta(F_2^{+}+F_2^{-})\,\varepsilon^2/F_{\text{s}}$ }
\end{minipage} \hspace*{23.pt}
\begin{minipage}[b]{5.2cm}
\includegraphics[width=5.2cm]{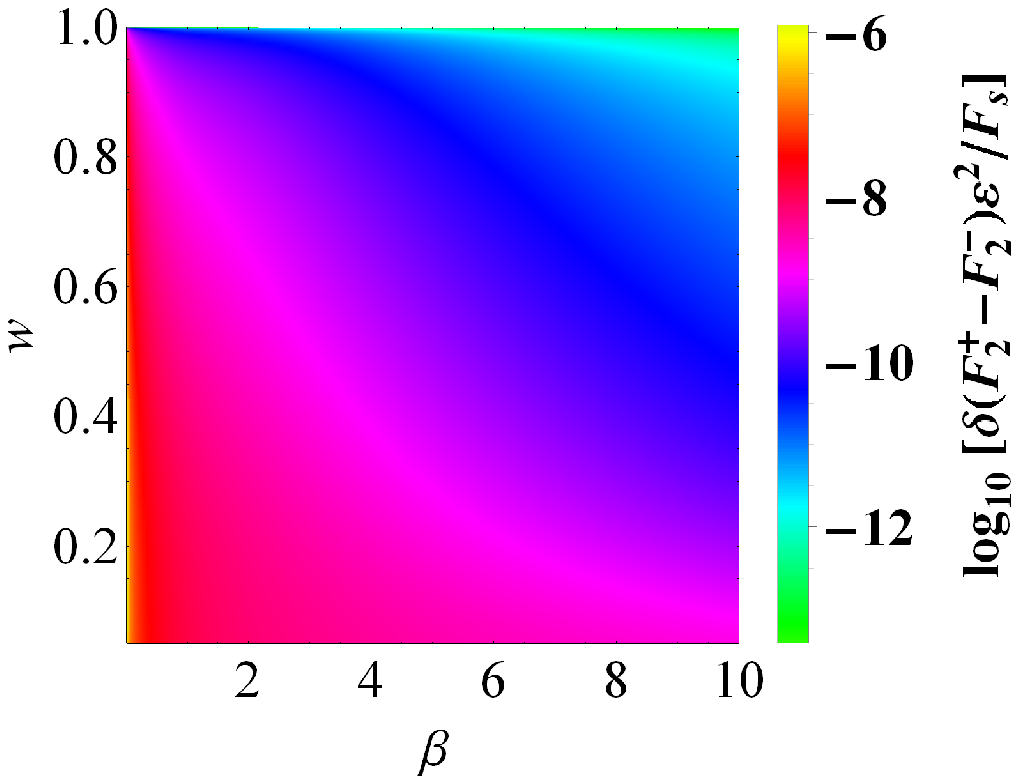}
  \centerline{(h) $\delta(F_2^{+}-F_2^{-})\,\varepsilon^2/F_{\text{s}}$ for $s^{+}=+1$ }
\end{minipage}
\caption{$\delta F_0^{+}/F_{\text{s}}$, $\delta F_1^{+}\varepsilon/F_{\text{s}}$, $\delta F_2^{+}\varepsilon^2/F_{\text{s}}$, $\delta(F_0^{+}+F_0^{-})/F_{\text{s}}$, $\delta(F_1^{+}\pm F_1^{-})\,\varepsilon/F_{\text{s}}$, and $\delta(F_2^{+}\pm F_2^{-})\,\varepsilon^2/F_{\text{s}}$ plotted as the color-indexed functions of $w$ and $\beta$ for the particle's prograde or retrograde motion.
We don't show $\delta F_2^{+}\varepsilon^2/F_{\text{s}}$, $\delta(F_1^{+}+F_1^{-})\,\varepsilon/F_{\text{s}}$, and $\delta(F_2^{+}-F_2^{-})\,\varepsilon^2/F_{\text{s}}$ for the case of $s^{+}=-1$ due to the same reason given in Fig.~\ref{Figure6} or the symmetry. } \label{Figure7}
\end{figure*}

We then consider the velocity effects on the image flux relations. The velocity effects on the zeroth-, first-, and second-order contributions to the normalized flux of the positive-parity image and to the normalized-flux sum and difference relations are plotted as the functions of $w$ and $\beta$ in Fig.~\ref{Figure7}. The magnitudes of these velocity effects are given in Tab.~\ref{Table2}. Similarly, three aspects with respect to these results should be pointed out. First, Figure~\ref{Figure7} and Table~\ref{Table2} show that for a given $\beta$ the velocity effect on the zeroth-order term of the normalized primary-image flux increases monotonically with the decrease of $w$, which holds for the velocity effects on the zeroth-order sum and second-order difference of the normalized fluxes. In contrast, the velocity effect on the second-order contribution to the normalized primary-image flux or the normalized-flux sum relation decreases monotonically when decreasing $w$. Differently, the velocity effect on the first-order contribution to the normalized image flux or the normalized-flux difference relation experiences a trend of first decrease and then increase with decreasing $w$ for a given $\beta~(1\lesssim\beta\leq10)$. Contrary to this trend, the velocity effect on the first-order contribution to the normalized-flux sum relation first increases and then decreases when decreasing $w$, with a fixed $\beta~(2\lesssim\beta\leq10)$. Second, we discuss the possibilities to detect these velocity effects. The results presented in Fig.~\ref{Figure7} and Tab.~\ref{Table2} indicate that there is a relatively large possibility to detect the velocity effect on the zeroth-order contribution to the normalized flux of the positive-parity image or the normalized-flux sum. For instance, for the case of $w=0.9$ and $\beta=1$, $\delta F_0^{+}/F_{\text{s}}$ and $\delta(F_0^{+}+F_0^{-})/F_{\text{s}}$ can reach approximately $0.02$ and $0.04$, respectively. The resulting differential apparent magnitudes $(\delta m_1)_{\text{zeroth}}$ and $(\delta m_2)_{\text{zeroth}}$ are about $-0.019\,$mag and $-0.033\,$mag, respectively, whose absolute values are much larger than the photometric precision of the \emph{Kepler} Mission. Interestingly, we notice that it is likely to detect the velocity effect on the second-order contribution to the normalized image flux or the normalized-flux sum relation in current resolution, provided both $w$ and $\beta$ take small values. For example, if $w=0.06$ and $\beta=0.025$ are preset, the differential apparent magnitude $(\delta m_2)_{\text{second}}$ resulted from the velocity effect $\delta(F_2^{+}+F_2^{-})\,\varepsilon^2/F_{\text{s}}$ will have a value of $18.56\,\mu$mag, which is larger than \emph{Kepler}'s precision evidently. Moreover, there is a small possibility to detect the velocity effect on the first-order contribution to the normalized-flux difference relation for very limited values of $w$ and $\beta$. For instance, the differential apparent magnitude $(\delta m_3)_{\text{first}}$ caused by $\delta(F_1^{+}-F_1^{-})\,\varepsilon/F_{\text{s}}$ is about $11.95\,\mu$mag for the case of $w=0.27$ and $\beta=3.5$. Compared with them, it is not possible to measure in current precision the velocity effect on the first-order contribution to the normalized image flux or the normalized-flux sum relation, or on the second-order normalized-flux difference relation. Finally, we stress that the influence of the sign of $\hat{L}$ on the velocity effects on the image flux relations is limited, and the qualitative conclusions are not changed when $s^{+}$ takes a different value.

\begin{widetext}

\begin{table}
{\footnotesize
\begin{minipage}[t]{0.49\textwidth}
  \centering
\begin{tabular}{cccccccc} \toprule[1.2px]
      $\beta\:\backslash\:w$    &    ~~~$ 0.05 $ ~~~     &       ~~~$0.1$~~~       &       ~~~$0.5$~~~        &     ~~~$0.9$~~~     &          $0.999999$             \\   \midrule[0.5pt] \vspace*{-8pt}  \\
                   0.01         &         657.99         &         305.32          &         29.06            &         2.85        &      2.50$\times10^{-5}$         \\
                    0.1         &         65.78          &         30.52           &         2.90             &         0.28        &      2.49$\times10^{-6}$         \\
                    0.5         &         13.08          &         6.03            &         0.55             &         0.05        &       $\blacktriangle$           \\
                      1         &         6.42           &         2.91            &         0.23             &         0.02        &       $\blacktriangle$           \\
                      5         &         0.98           &         0.33            &   5.72$\times10^{-3}$    & 3.14$\times10^{-4}$ &       $\blacktriangle$           \\
                     10         &         0.33           &         0.08            &   4.72$\times10^{-4}$    & 2.33$\times10^{-5}$ &       $\blacktriangle$           \\   \bottomrule[1.2px]
\end{tabular} \par  \vspace*{3pt}
     \centerline{(a) $\delta F_0^{+}/F_{\text{s}}$ } \vspace*{10pt}
\end{minipage}
\begin{minipage}[t]{0.49\textwidth}
  \centering
\begin{tabular}{cccccccc} \toprule[1.2px]
      $\beta\:\backslash\:w$    &    ~~~$ 0.05 $ ~~~     &       ~~~$0.1$~~~       &       ~~~$0.5$~~~        &     ~~~$0.9$~~~     &       $0.999999$        \\   \midrule[0.5pt] \vspace*{-8pt}  \\
                   0.01         &  1.53$\times10^{-5}$   &   1.26$\times10^{-5}$   &    $\blacktriangle$      &  $\blacktriangle$   &   $\blacktriangle$      \\
                    0.1         &  1.52$\times10^{-5}$   &   1.25$\times10^{-5}$   &    $\blacktriangle$      &  $\blacktriangle$   &   $\blacktriangle$      \\
                    0.5         &  1.37$\times10^{-5}$   &   1.10$\times10^{-5}$   &  $-$1.20$\times10^{-6}$  &  $\blacktriangle$   &   $\blacktriangle$      \\
                      1         &  1.01$\times10^{-5}$   &   7.43$\times10^{-6}$   &  $-$3.00$\times10^{-6}$  &  $\blacktriangle$   &   $\blacktriangle$      \\
                      5         & $-$1.84$\times10^{-6}$ & $-$3.80$\times10^{-6}$  &  $-$1.89$\times10^{-6}$  &  $\blacktriangle$   &   $\blacktriangle$      \\
                     10         & $-$2.29$\times10^{-6}$ & $-$2.95$\times10^{-6}$  &    $\blacktriangle$      &  $\blacktriangle$   &   $\blacktriangle$      \\   \bottomrule[1.2px]
\end{tabular} \par  \vspace*{3pt}
     \centerline{(b) $\delta F_1^{+}\varepsilon/F_{\text{s}}$ for $s^{+}=+1$} \vspace*{10pt}
\end{minipage}

\begin{minipage}[t]{0.49\textwidth}
  \centering
\begin{tabular}{cccccccc} \toprule[1.2px]
      $\beta\:\backslash\:w$    &      ~~$ 0.05 $ ~~     &        ~~$0.1$~~        &         ~~$0.5$~~        &      ~~$0.9$~~      &       $0.999999$       \\   \midrule[0.5pt] \vspace*{-8pt}  \\
                   0.01         & $-$1.47$\times10^{-3}$ & $-$1.92$\times10^{-4}$  &  $-$2.19$\times10^{-6}$  &  $\blacktriangle$   &    $\blacktriangle$    \\
                    0.1         & $-$1.47$\times10^{-4}$ & $-$1.92$\times10^{-5}$  &    $\blacktriangle$      &  $\blacktriangle$   &    $\blacktriangle$    \\
                    0.5         & $-$2.93$\times10^{-5}$ & $-$3.83$\times10^{-6}$  &    $\blacktriangle$      &  $\blacktriangle$   &    $\blacktriangle$    \\
                      1         & $-$1.46$\times10^{-5}$ & $-$1.90$\times10^{-6}$  &    $\blacktriangle$      &  $\blacktriangle$   &    $\blacktriangle$    \\
                      5         & $-$2.80$\times10^{-6}$ &    $\blacktriangle$     &    $\blacktriangle$      &  $\blacktriangle$   &    $\blacktriangle$    \\
                     10         & $-$1.23$\times10^{-6}$ &    $\blacktriangle$     &    $\blacktriangle$      &  $\blacktriangle$   &    $\blacktriangle$    \\   \bottomrule[1.2px]
\end{tabular} \par  \vspace*{3pt}
     \centerline{(c) $\delta F_2^{+}\varepsilon^2/F_{\text{s}}$ for $s^{+}=+1$ }  \vspace*{10pt}
\end{minipage}
\begin{minipage}[t]{0.49\textwidth}
  \centering
\begin{tabular}{cccccccc} \toprule[1.2px]
      $\beta\:\backslash\:w$    &    ~~~$ 0.05 $ ~~~     &       ~~~$0.1$~~~       &       ~~~$0.5$~~~        &     ~~~$0.9$~~~      &       $0.999999$        \\   \midrule[0.5pt] \vspace*{-8pt}  \\
                   0.01         &       1315.98          &          610.63         &          58.11           &         5.70         &   5.00$\times10^{-5}$   \\
                    0.1         &        131.56          &           61.03         &           5.80           &         0.57         &   4.98$\times10^{-6}$   \\
                    0.5         &         26.15          &           12.06         &           1.10           &         0.10         &    $\blacktriangle$     \\
                      1         &         12.84          &            5.82         &           0.47           &         0.04         &    $\blacktriangle$     \\
                      5         &          1.96          &            0.67         &           0.01           &  6.28$\times10^{-4}$ &    $\blacktriangle$     \\
                     10         &          0.67          &            0.16         &   9.43$\times10^{-4}$    &  4.67$\times10^{-5}$ &    $\blacktriangle$     \\   \bottomrule[1.2px]
\end{tabular} \par  \vspace*{3pt}
     \centerline{(d) $\delta(F_0^{+}+F_0^{-})/F_{\text{s}}$ } \vspace*{10pt}
\end{minipage}
\begin{minipage}[t]{0.49\textwidth}
  \centering
\begin{tabular}{cccccccc} \toprule[1.2px]
      $\beta\!\:\backslash\:\!w$    &    ~~$ 0.05 $ ~~     &       ~~$0.1$~~       &        ~~$0.5$~~         &      ~~$0.9$~~      &      \!\!$0.999999$         \\   \midrule[0.5pt] \vspace*{-8pt}  \\
                   0.01         & $-$1.59$\times10^{-5}$ & $-$1.56$\times10^{-5}$  &  $-$7.91$\times10^{-6}$  &  $\blacktriangle$   &   $\blacktriangle$      \\
                    0.1         & $-$1.58$\times10^{-5}$ & $-$1.55$\times10^{-5}$  &  $-$7.87$\times10^{-6}$  &  $\blacktriangle$   &   $\blacktriangle$      \\
                    0.5         & $-$1.45$\times10^{-5}$ & $-$1.42$\times10^{-5}$  &  $-$6.82$\times10^{-6}$  &  $\blacktriangle$   &   $\blacktriangle$      \\
                      1         & $-$1.14$\times10^{-5}$ & $-$1.10$\times10^{-5}$  &  $-$4.44$\times10^{-6}$  &  $\blacktriangle$   &   $\blacktriangle$      \\
                      5         &    $\blacktriangle$    &     $\blacktriangle$    &     $\blacktriangle$     &  $\blacktriangle$   &   $\blacktriangle$      \\
                     10         &    $\blacktriangle$    &     $\blacktriangle$    &     $\blacktriangle$     &  $\blacktriangle$   &   $\blacktriangle$      \\   \bottomrule[1.2px]
\end{tabular} \par  \vspace*{3pt}
     \centerline{(e) $\delta(F_1^{+}+F_1^{-})\,\varepsilon/F_{\text{s}}$ for $s^{+}=+1$ }
\end{minipage}
\begin{minipage}[t]{0.5\textwidth}
  \centering
\begin{tabular}{cccccccc} \toprule[1.2px]
  $\!\beta\:\backslash\:\!w$    &       ~~$ 0.05 $~~     &        ~~$0.1$~~        &         ~~$0.5$~~        &      ~~$0.9$~~         &  \!\!$0.999999$          \\   \midrule[0.5pt] \vspace*{-8pt}  \\
                   0.01         &   4.65$\times10^{-5}$  &   4.07$\times10^{-5}$   &   7.33$\times10^{-6}$    &   $\blacktriangle$     &  $\blacktriangle$       \\
                    0.1         &   4.63$\times10^{-5}$  &   4.05$\times10^{-5}$   &   7.20$\times10^{-6}$    &   $\blacktriangle$     &  $\blacktriangle$       \\
                    0.5         &   4.20$\times10^{-5}$  &   3.62$\times10^{-5}$   &   4.41$\times10^{-6}$    &   $\blacktriangle$     &  $\blacktriangle$       \\
                      1         &   3.16$\times10^{-5}$  &   2.59$\times10^{-5}$   &  $-$1.57$\times10^{-6}$  & $-$1.43$\times10^{-6}$ &  $\blacktriangle$       \\
                      5         & $-$2.96$\times10^{-6}$ & $-$7.15$\times10^{-6}$  &  $-$4.20$\times10^{-6}$  &   $\blacktriangle$     &  $\blacktriangle$       \\
                     10         & $-$4.56$\times10^{-6}$ & $-$6.01$\times10^{-6}$  &     $\blacktriangle$     &   $\blacktriangle$     &  $\blacktriangle$       \\   \bottomrule[1.2px]
\end{tabular} \par  \vspace*{3pt}
     \centerline{(f) $\delta(F_1^{+}-F_1^{-})\,\varepsilon/F_{\text{s}}$ }
\end{minipage} \\ \vspace*{10pt}

\begin{minipage}[t]{0.48\textwidth}
  \centering
\begin{tabular}{cccccccc} \toprule[1.2px]
      $\beta\:\backslash\:w$    &    ~~~$ 0.05 $ ~~~     &       ~~~$0.1$~~~       &       ~~~$0.5$~~~        &     ~~~$0.9$~~~     &      $0.999999$         \\   \midrule[0.5pt] \vspace*{-8pt}  \\
                   0.01         & $-$2.93$\times10^{-3}$ & $-$3.84$\times10^{-4}$  &  $-$4.90$\times10^{-6}$  &  $\blacktriangle$   &   $\blacktriangle$      \\
                    0.1         & $-$2.93$\times10^{-4}$ & $-$3.84$\times10^{-5}$  &    $\blacktriangle$      &  $\blacktriangle$   &   $\blacktriangle$      \\
                    0.5         & $-$5.86$\times10^{-5}$ & $-$7.68$\times10^{-6}$  &    $\blacktriangle$      &  $\blacktriangle$   &   $\blacktriangle$      \\
                      1         & $-$2.93$\times10^{-5}$ & $-$3.82$\times10^{-6}$  &    $\blacktriangle$      &  $\blacktriangle$   &   $\blacktriangle$      \\
                      5         & $-$5.61$\times10^{-6}$ &     $\blacktriangle$    &    $\blacktriangle$      &  $\blacktriangle$   &   $\blacktriangle$      \\
                     10         & $-$2.46$\times10^{-6}$ &     $\blacktriangle$    &    $\blacktriangle$      &  $\blacktriangle$   &   $\blacktriangle$      \\   \bottomrule[1.2px]
\end{tabular} \par  \vspace*{3pt}
     \centerline{(g) $\delta(F_2^{+}+F_2^{-})\,\varepsilon^2/F_{\text{s}}$ }
\end{minipage}  \hspace*{10pt}
\begin{minipage}[t]{0.48\textwidth}
  \centering
\begin{tabular}{cccccccc} \toprule[1.2px]
      $\beta\:\backslash\:w$    &    ~~~$ 0.05 $ ~~~     &       ~~~$0.1$~~~       &       ~~~$0.5$~~~        &     ~~~$0.9$~~~     &        $0.999999$       \\   \midrule[0.5pt] \vspace*{-8pt}  \\
                   0.01         &   1.23$\times10^{-6}$  &   1.20$\times10^{-6}$   &    $\blacktriangle$      &   $\blacktriangle$  &     $\blacktriangle$    \\
                    0.1         &    $\blacktriangle$    &     $\blacktriangle$    &    $\blacktriangle$      &   $\blacktriangle$  &     $\blacktriangle$    \\
                    0.5         &    $\blacktriangle$    &     $\blacktriangle$    &    $\blacktriangle$      &   $\blacktriangle$  &     $\blacktriangle$    \\
                      1         &    $\blacktriangle$    &     $\blacktriangle$    &    $\blacktriangle$      &   $\blacktriangle$  &     $\blacktriangle$    \\
                      5         &    $\blacktriangle$    &     $\blacktriangle$    &    $\blacktriangle$      &   $\blacktriangle$  &     $\blacktriangle$    \\
                     10         &    $\blacktriangle$    &     $\blacktriangle$    &    $\blacktriangle$      &   $\blacktriangle$  &     $\blacktriangle$    \\   \bottomrule[1.2px]
\end{tabular} \par  \vspace*{3pt}
     \centerline{(h) $\delta(F_2^{+}-F_2^{-})\,\varepsilon^2/F_{\text{s}}$ for $s^{+}=+1$}
\end{minipage}}

\caption{The magnitudes of $\delta F_0^{+}/F_{\text{s}}$, $\delta F_1^{+}\varepsilon/F_{\text{s}}$, $\delta F_2^{+}\varepsilon^2/F_{\text{s}}$, $\delta(F_0^{+}+F_0^{-})/F_{\text{s}}$,
$\delta(F_1^{+}\pm F_1^{-})\,\varepsilon/F_{\text{s}}$, and $\delta(F_2^{+}\pm F_2^{-})\,\varepsilon^2/F_{\text{s}}$ for various $w$ and $\beta$. Here, the black triangle ``$\blacktriangle$" denotes the magnitude
whose absolute value is less than $1.0\times10^{-6}$.  }   \label{Table2}
\end{table}
\end{widetext}

\begin{figure*}
\centering
\begin{minipage}[b]{5.55cm}
\includegraphics[width=5.55cm]{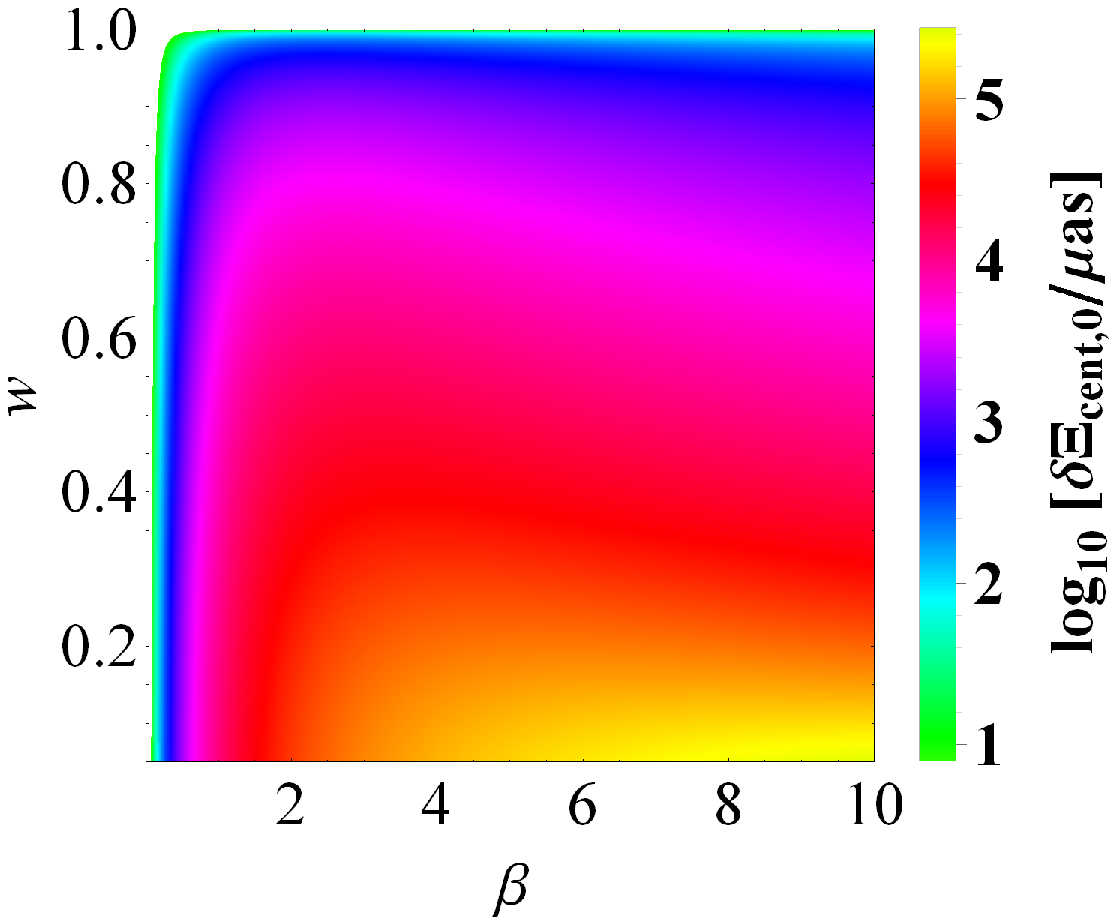} \vspace*{-4pt}
  \centerline{ (a) $\delta{\it\Xi}_{\text{cent,0}}$ }
\end{minipage} \hspace*{5pt}
\begin{minipage}[b]{5.9cm}
\includegraphics[width=5.9cm]{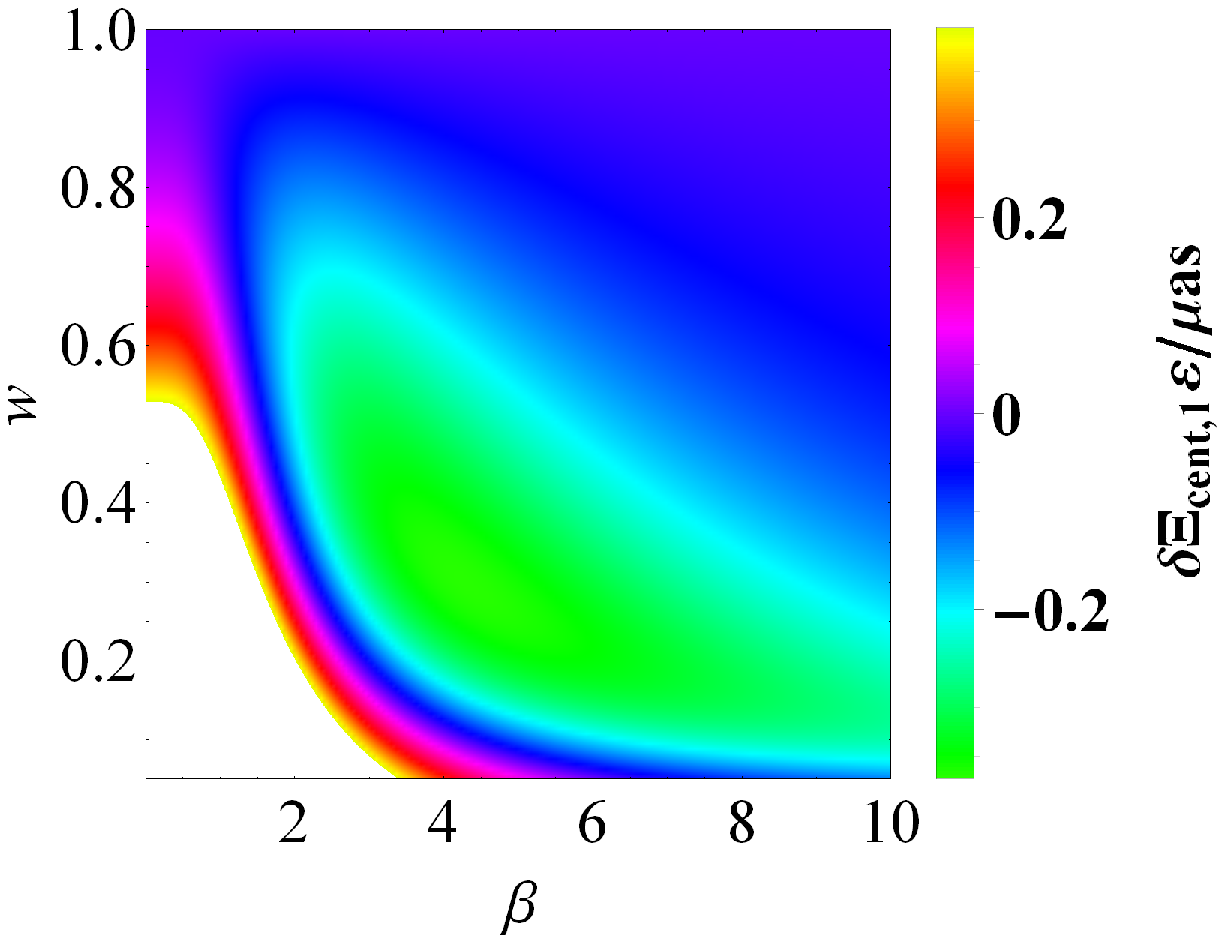} \vspace*{-2pt}
  \centerline{(b) $\delta{\it\Xi}_{\text{cent,1}}\,\varepsilon$ for $s^{+}=+1$}
\end{minipage} \hspace*{2pt}
\begin{minipage}[b]{5.7cm}
\includegraphics[width=5.7cm]{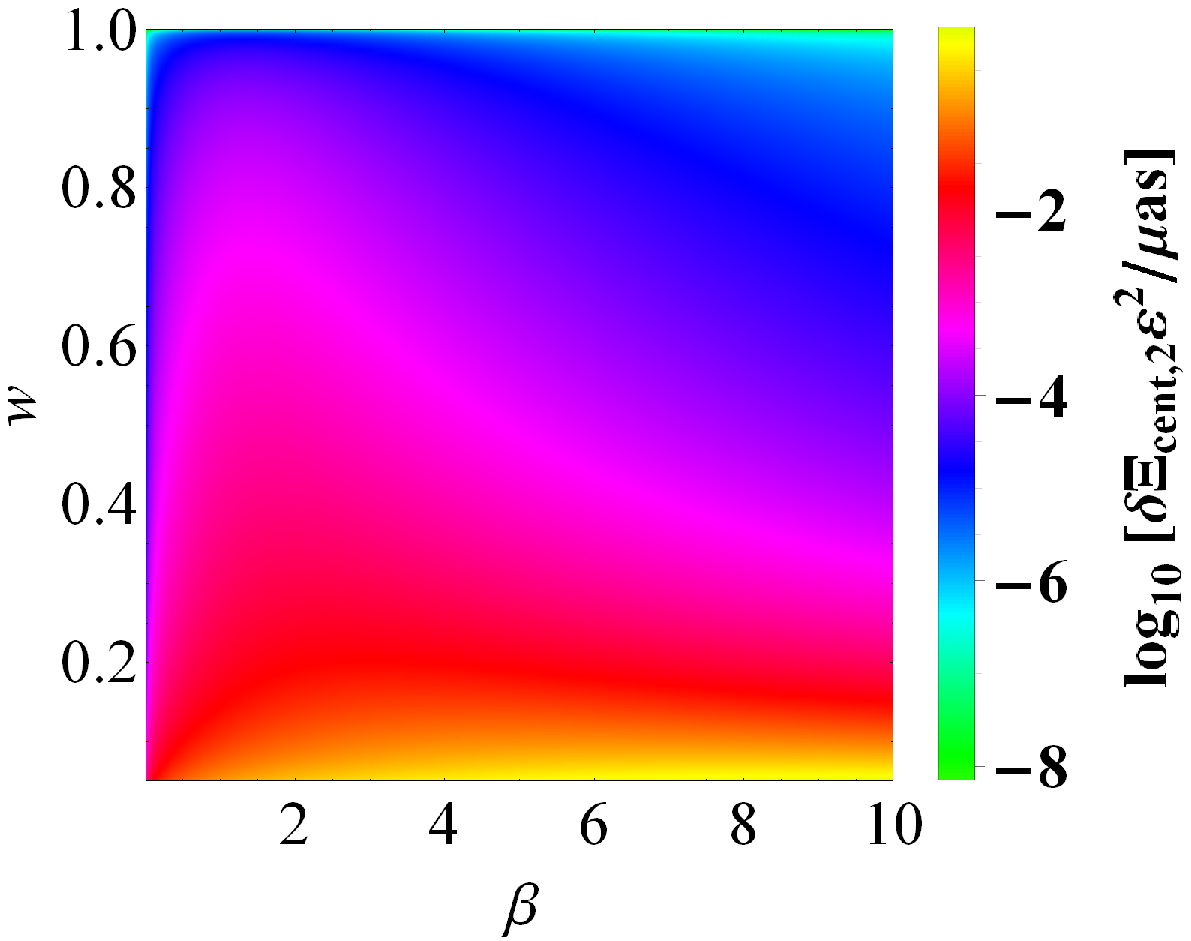}
  \centerline{(c) $\delta{\it\Xi}_{\text{cent,2}}\,\varepsilon^2$ }
\end{minipage}  \vspace*{6pt} \\   \hspace*{1.5pt}
\begin{minipage}[b]{5.6cm}
\includegraphics[width=5.6cm]{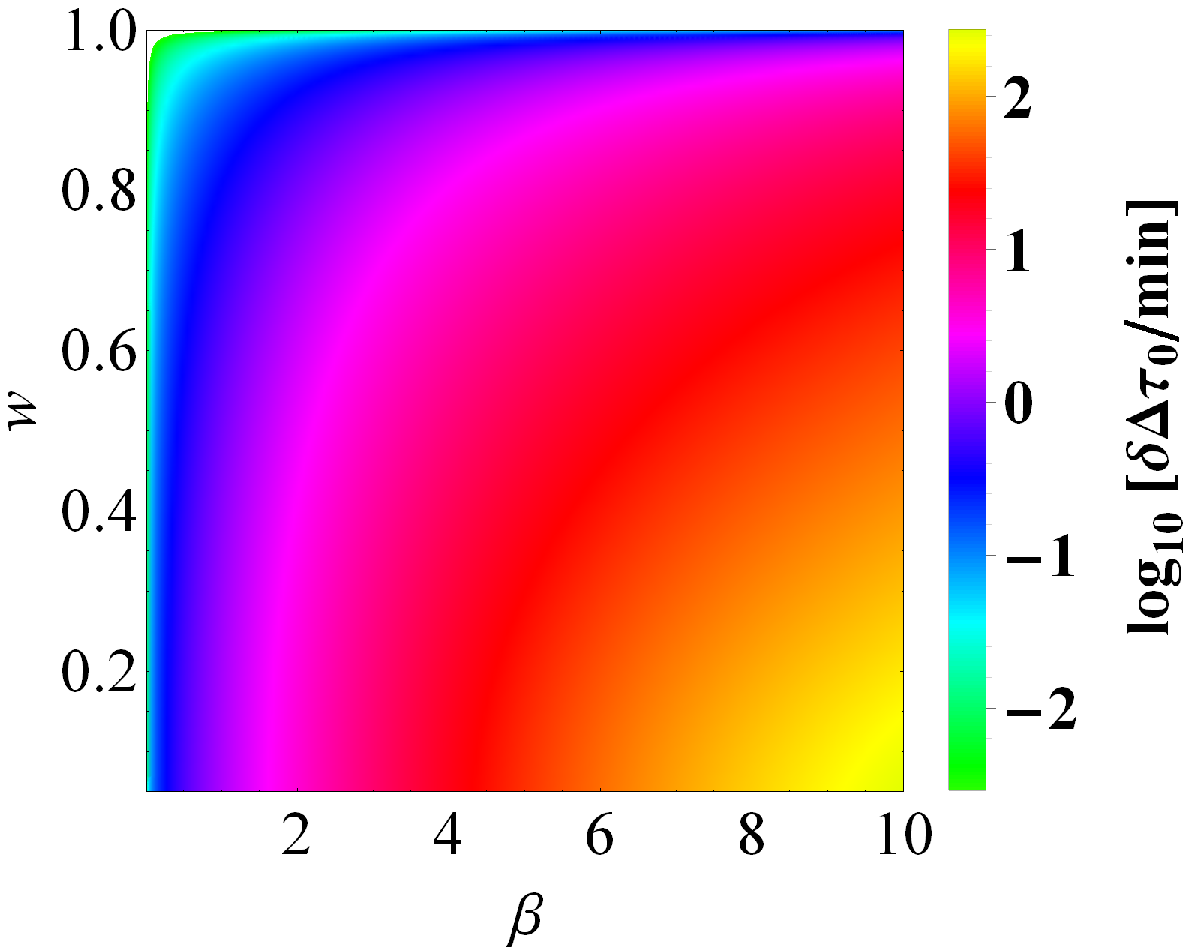}  \vspace*{-3pt}
  \centerline{ (d) $\delta\Delta\tau_0$ }
\end{minipage} \hspace*{0.5pt}
\begin{minipage}[b]{5.8cm}
\includegraphics[width=5.8cm]{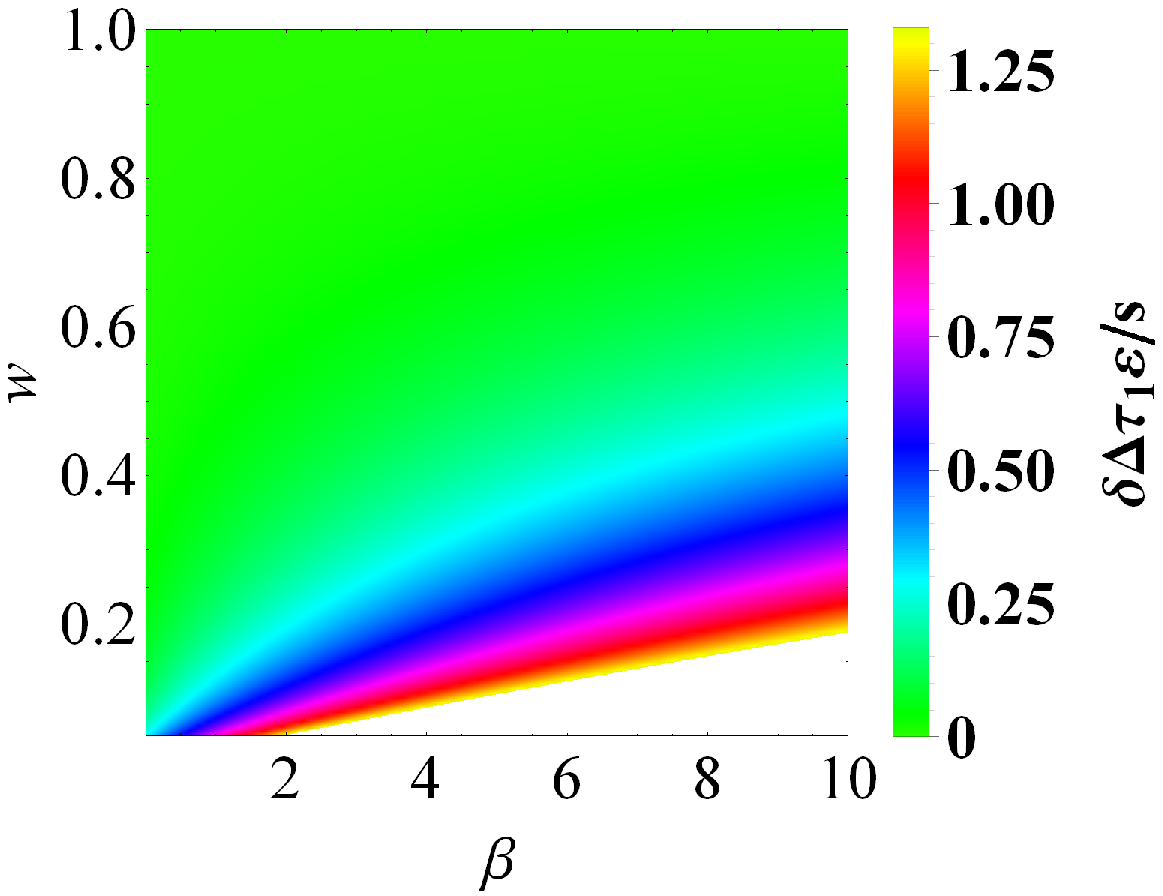}
  \centerline{(e$_1$) $\delta\Delta\tau_1\,\varepsilon$ for $s^{+}=+1$ } \vspace*{-11.5pt}
\end{minipage} \hspace*{6pt}
\begin{minipage}[b]{5.8cm}
\includegraphics[width=5.8cm]{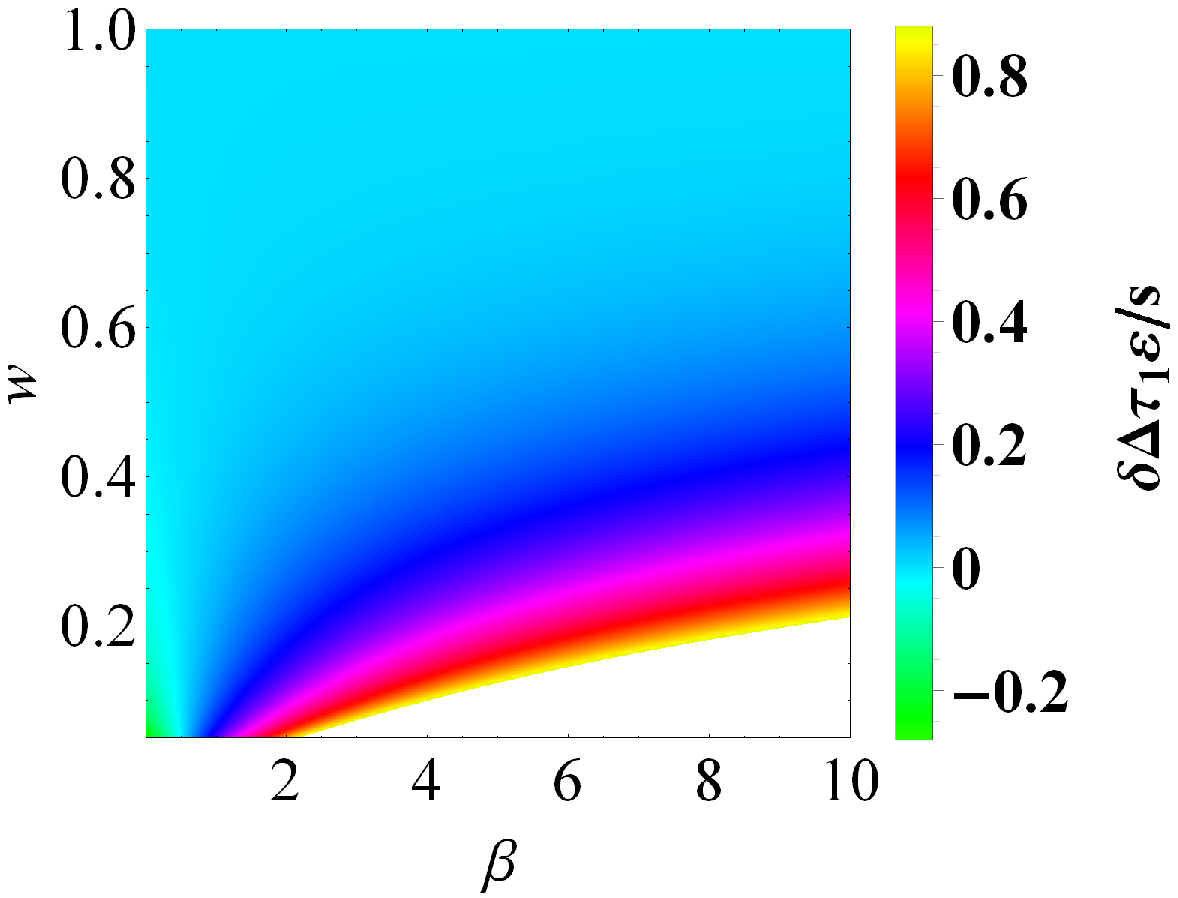}
  \centerline{(e$_2$) $\delta\Delta\tau_1\,\varepsilon$ for $s^{+}=-1$ }
\end{minipage}
\caption{$\delta{\it\Xi}_{\text{cent,0}}$, $\delta{\it\Xi}_{\text{cent,1}}\,\varepsilon$, and $\delta{\it\Xi}_{\text{cent,2}}\,\varepsilon^2$, $\delta\Delta\tau_0$, and $\delta\Delta\tau_1\,\varepsilon$ plotted as the color-indexed functions of $w$ and $\beta$. } \label{Figure8}
\end{figure*}

Now we discuss the velocity-induced effects on the zeroth-, first-, and second-order contributions to the centroid, which are shown on the top of Fig.~\ref{Figure8} in color-indexed form for the scenario of Sagittarius A$^{\ast}$. Their magnitudes for various $w$ and $\beta$ are listed in Tab.~\ref{Table3}. We can see from Fig.~\ref{Figure8} that the velocity effect on the zeroth- or second-order contribution to the centroid increases monotonically with the decrease of $w$ when the angular source position is fixed. This is not the case for $\delta{\it\Xi}_{\text{cent,1}}\,\varepsilon$ with $s^{+}=+1$. It decreases firstly to a minimum value and then increases with decreasing $w$ for $1\lesssim\beta\leq10$. However, it will monotonically increase with decreasing $w$ when $0.01\leq\beta\lesssim1$. For $s^{+}=-1$, the behavior of $\delta{\it\Xi}_{\text{cent,1}}\,\varepsilon$ is then reversed. As to the possibilities to detect them, Figure~\ref{Figure8} (a) indicates $\delta{\it\Xi}_{\text{cent,0}}$ is very likely to be detected, as long as $\beta$ and $w$ don't take very small and ultrarelativistic values, respectively. We argue that it is also possible to detect the velocity effect on the first-order contribution to the centroid position with a proper combination of $\beta$ and $w$. It is only when $w$ and $\beta$ take respectively small and relatively large values that a possibility to observe $\delta{\it\Xi}_{\text{cent,2}}\,\varepsilon^2$ in current resolution exists.

\begin{table*}[t]
\begin{minipage}[t]{0.45\textwidth}
  \centering
\begin{tabular}{cccccccc} \toprule[1.2px]
      $\beta\:\backslash\:w$    &    ~~~~$ 0.05 $ ~~~~   &     ~~~~$0.1$~~~~       &     ~~~~$0.5$~~~~        &    ~~~~$0.9$~~~~    &      $0.999999$         \\  \midrule[0.5pt] \vspace*{-8pt}  \\
                   0.01         &        $\star$         &        $\star$          &        $\star$           &       $\star$       &       $\star$           \\
                    0.1         &         17.57          &         17.31           &         10.58            &        1.85         &       $\star$           \\
                    0.5         &   1.96$\times10^{3}$   &   1.93$\times10^{3}$    &    1.13$\times10^{3}$    &       186.20        &       $\star$           \\
                      1         &   1.17$\times10^{4}$   &   1.15$\times10^{4}$    &    5.92$\times10^{3}$    &       858.14        &       $\star$           \\
                      5         &   1.54$\times10^{5}$   &   1.29$\times10^{5}$    &    1.64$\times10^{4}$    &  1.42$\times10^{3}$ &       $\star$           \\
                     10         &   2.77$\times10^{5}$   &   1.71$\times10^{5}$    &    9.94$\times10^{3}$    &       798.54        &       $\star$           \\   \bottomrule[1.2px]
\end{tabular} \par  \vspace*{3pt}
     \centerline{(a) $\delta{\it\Xi}_{\text{cent,0}}$ } \vspace*{10pt}
\end{minipage}
\begin{minipage}[t]{0.48\textwidth}
  \centering
\begin{tabular}{cccccccc} \toprule[1.2px]
      $\beta\:\backslash\:w$    &    ~~~~$ 0.05 $ ~~~~   &     ~~~~$0.1$~~~~       &     ~~~~$0.5$~~~~        &    ~~~~$0.9$~~~~    &     $0.999999$          \\   \midrule[0.5pt] \vspace*{-8pt}  \\
                   0.01         &          2.05          &          1.82           &          0.45            &       $\star$       &       $\star$           \\
                    0.1         &          2.05          &          1.82           &          0.45            &       $\star$       &       $\star$           \\
                    0.5         &          2.02          &          1.80           &          0.43            &       $\star$       &       $\star$           \\
                      1         &          1.80          &          1.57           &          0.25            &       $\star$       &       $\star$           \\
                      5         &          0.10          &       $-$0.11           &       $-$0.23            &       $\star$       &       $\star$           \\
                     10         &       $-$0.13          &       $-$0.25           &       $-$0.08            &       $\star$       &       $\star$           \\   \bottomrule[1.2px]
\end{tabular} \par  \vspace*{3pt}
     \centerline{(b) $\delta{\it\Xi}_{\text{cent,1}}\,\varepsilon$ for $s^{+}=+1$ } \vspace*{10pt}
\end{minipage}

\begin{minipage}[t]{0.48\textwidth}
  \centering
\begin{tabular}{cccccccc} \toprule[1.2px]
      $\beta\:\backslash\:w$    &    ~~~~$ 0.05 $ ~~~~   &     ~~~~$0.1$~~~~       &     ~~~~$0.5$~~~~        &    ~~~~$0.9$~~~~    &      $0.999999$         \\  \midrule[0.5pt] \vspace*{-8pt}  \\
                   0.01         &         $\star$        &        $\star$          &        $\star$           &       $\star$       &       $\star$           \\
                    0.1         &         $\star$        &        $\star$          &        $\star$           &       $\star$       &       $\star$           \\
                    0.5         &          0.07          &        $\star$          &        $\star$           &       $\star$       &       $\star$           \\
                      1         &          0.14          &        $\star$          &        $\star$           &       $\star$       &       $\star$           \\
                      5         &          0.65          &          0.12           &        $\star$           &       $\star$       &       $\star$           \\
                     10         &          0.94          &          0.10           &        $\star$           &       $\star$       &       $\star$           \\   \bottomrule[1.2px]
\end{tabular} \par  \vspace*{3pt}
     \centerline{(c) $\delta{\it\Xi}_{\text{cent,2}}\,\varepsilon^2$ }
\end{minipage}

\caption{The magnitudes (in units of $\mu$as) of $\delta{\it\Xi}_{\text{cent,0}}$, $\delta{\it\Xi}_{\text{cent,1}}\,\varepsilon$, and $\delta{\it\Xi}_{\text{cent,2}}\,\varepsilon^2$ for various $w$ and $\beta$. Here the star ``$\star$" denotes the magnitude whose absolute value is less than $0.05\mu$as.  }   \label{Table3}
\end{table*}

\begin{table*}[t]
\begin{minipage}[t]{0.49\textwidth}
  \centering
\begin{tabular}{cccccccc} \toprule[1.2px]
  $\beta\!\:\backslash\:\!w$    &    ~~~~$ 0.05 $ ~~~~   &     ~~~~$0.1$~~~~       &     ~~~~$0.5$~~~~        &    ~~~~$0.9$~~~~      &            $0.999999$            \\ \midrule[0.5pt] \vspace*{-8pt}  \\
                   0.01         &          0.01          &          0.01           &    7.29$\times10^{-3}$   &  1.41$\times10^{-3}$  &       1.38$\times10^{-8}$        \\
                    0.1         &          0.11          &          0.11           &          0.07            &         0.01          &       1.38$\times10^{-7}$        \\
                    0.5         &          0.59          &          0.59           &          0.38            &         0.07          &       7.10$\times10^{-7}$        \\
                      1         &          1.35          &          1.33           &          0.85            &         0.16          &       1.54$\times10^{-6}$        \\
                      5         &         36.41          &         34.88           &         14.24            &         2.00          &       1.85$\times10^{-5}$        \\
                     10         &         274.27         &        244.15           &         62.68            &         7.71          &       7.02$\times10^{-5}$        \\   \bottomrule[1.2px]
\end{tabular} \par  \vspace*{3pt}
     \centerline{(a) $\delta\Delta\tau_0$ } \vspace*{10pt}
\end{minipage}
\begin{minipage}[t]{0.49\textwidth}
  \centering
\begin{tabular}{cccccccc} \toprule[1.2px]
      $\beta\:\backslash\:w$    &    ~~~~$ 0.05 $ ~~~~   &     ~~~~$0.1$~~~~       &     ~~~~$0.5$~~~~        &    ~~~~$0.9$~~~~      &            $0.999999$     \\   \midrule[0.5pt] \vspace*{-8pt}  \\
                   0.01         &          0.29          &          0.14           &          0.01            &  7.37$\times10^{-4}$  &       5.36$\times10^{-9}$        \\
                    0.1         &          0.34          &          0.16           &          0.01            &  8.26$\times10^{-4}$  &       6.01$\times10^{-9}$        \\
                    0.5         &          0.55          &          0.26           &          0.02            &  1.27$\times10^{-3}$  &       9.24$\times10^{-9}$        \\
                      1         &          0.82          &          0.39           &          0.03            &  1.91$\times10^{-3}$  &       1.40$\times10^{-8}$        \\
                      5         &          2.95          &          1.41           &          0.14            &  8.21$\times10^{-3}$  &       6.12$\times10^{-8}$        \\
                     10         &          5.64          &          2.72           &          0.27            &         0.02          &       1.22$\times10^{-7}$        \\   \bottomrule[1.2px]
\end{tabular} \par  \vspace*{3pt}
     \centerline{(b$_1$) $\delta\Delta\tau_1\,\varepsilon$ for $s^{+}=+1$ } \vspace*{10pt}
\end{minipage}
\begin{minipage}[t]{0.49\textwidth}
  \centering
\begin{tabular}{cccccccc} \toprule[1.2px]
      $\beta\:\backslash\:w$    &    ~~~~$ 0.05 $ ~~~~   &     ~~~~$0.1$~~~~       &     ~~~~$0.5$~~~~        &    ~~~~~$0.9$~~~~~     &            $0.999999$          \\   \midrule[0.5pt] \vspace*{-8pt}  \\
                   0.01         &        $-$0.28         &        $-$0.13          &         $-$0.01          & $-$7.17$\times10^{-4}$ &    $-$5.22$\times10^{-9}$      \\
                    0.1         &        $-$0.23         &        $-$0.12          &         $-$0.01          & $-$6.33$\times10^{-4}$ &    $-$4.62$\times10^{-9}$      \\
                    0.5         &        $-$0.02         &        $-$0.01          &  $-$2.62$\times10^{-3}$  & $-$3.00$\times10^{-4}$ &    $-$2.31$\times10^{-9}$      \\
                      1         &          0.25          &           0.12          &    6.24$\times10^{-3}$   &   2.52$\times10^{-5}$  &    $-$1.73$\times10^{-10}$     \\
                      5         &          2.39          &           1.12          &           0.06           &   1.46$\times10^{-3}$  &       8.10$\times10^{-9}$      \\
                     10         &          5.04          &           2.35          &           0.13           &   2.94$\times10^{-3}$  &       1.63$\times10^{-8}$      \\   \bottomrule[1.2px]
\end{tabular} \par   \vspace*{3pt}
     \centerline{(b$_2$) $\delta\Delta\tau_1\,\varepsilon$ for $s^{+}=-1$ }
\end{minipage}

\caption{The magnitudes of $\delta\Delta\tau_0$ (in units of min) and $\delta\Delta\tau_1\,\varepsilon$ (in units of s) for various $w$ and $\beta$.  }   \label{Table4}
\end{table*}

Finally, we turn our attention to the velocity effects on the differential time delay between the primary and secondary images. The velocity effects on
the second- and third-order contributions to the differential time delay as the color-indexed functions of $w$ and $\beta$ are plotted on the bottom of Fig.~\ref{Figure8}, and the magnitudes of these velocity effects are given in Tab.~\ref{Table4}. For a fixed $\beta\in[0.01,~10]$ and the prograde motion the massive particle takes, both $\delta\Delta\tau_0$ and $\delta\Delta\tau_1\,\varepsilon$ increase monotonically with decreasing $w$. This conclusion holds for $\delta\Delta\tau_1\,\varepsilon$ when the particle takes retrograde motion, with $0.54\lesssim\beta\leq10$. However, it is surprising to find that the velocity effect on the third-order contribution to the differential time delay for the case of $s^{+}=-1$ decreases with the decrease of $w$, supposing $0.01\leq\beta\lesssim0.54$. It indicates the sign of $\hat{L}$ affects the behavior of $\delta\Delta\tau_1\,\varepsilon$. Furthermore, with the present differential VLBI accuracy ($\sim10^{-12}$s), it appears that the velocity effect on the second- or third-order contribution to the differential time delay is measurable, whether the value of $s^{+}$ is $+1$ or not.

\section{Summary} \label{sect7}
In this paper we have studied the weak-field gravitational lensing of a relativistic neutral massive particle induced by a Kerr-Newman black hole in detail. The explicit form of the equatorial gravitational deflection angle of the massive particle up to the 3PM order has been achieved and found to be in agreement with the result given in the previous literature. Based on the bending angle, the Virbhadra-Ellis lens equation has been solved.
The analytical expressions of the timelike lensing observables, which include the positions, magnifications, and gravitational time delays of the primary and secondary images, along with the differential time delay, the total magnification, and the magnification-weighted centroid position, have thus been obtained beyond the weak-deflection limit. The analytical forms of the correctional effects originated from the deviation of the particle's initial velocity $w$ from the speed of light on the lensing observables of the images have also been achieved.

The formalism has been applied to the supermassive black hole at the center of our galaxy by assuming Sagittarius A$^{\ast}$ to be a Kerr-Newman lens. In this situation, we have concentrated on the analysis of the velocity-induced effects on the angular position and flux of the positive-parity primary image, the sum and difference relations for the image positions and fluxes, the centroid, and the differential time delay. The behaviors of these velocity effects acting as the bivariate functions of $w$ and the scaled angular source position $\beta$ have been discussed systematically. Interestingly, it is found that for a given angular source position, the velocity effects on the zeroth- and second-order contributions to the primary-image position, the centroid position, and the positional sum relation increase monotonically with decreasing $w$. This trend holds for the velocity effects on the first-order contribution to the positional sum for retrograde motion of the particle or to the primary-image position, on the zeroth-order normalized primary-image flux, and on the zeroth-order sum and second-order difference of the normalized fluxes. This conclusion also applies to the velocity effects on the zeroth- and first-order contributions to the differential time delay for particle's prograde motion. The residual components of the velocity effects, such as the velocity effect on the first-order contribution to the normalized image flux, appear more complex or non intuitive. Taken overall, it is indicated that the observable image properties in the massive-particle lensing scenario are more evident than those in the null lensing scenario under the same circumstances. We have also analyzed the possibilities to detect these velocity effects briefly. It seems reasonable to conclude that the velocity effects on the zeroth-order contribution to the primary-image position, the positional sum relation, and the centroid, as well as the velocity effect on the second-order differential time delay, are feasible to be detected in current resolution for most cases. It is also likely to detect the velocity effects on the first-order contribution to the primary-image position, the positional sum and difference relations, and the centroid, as well as the velocity effect on the third-order differential time delay in many scenarios. This conclusion applies to the velocity effects on the zeroth-order contribution to the normalized primary-image flux and the normalized flux sum. The possibilities to observe the residual components of the velocity effects (e.g., the velocity effect on the second-order contribution to the normalized image flux) are relatively small or even not existed in present precision. We argue that the direction of the orbital angular momentum of the particle's motion relative to the lens' rotation has a relatively obvious influence on the behaviors and detection of the velocity effects on the first-order contribution to the positional difference relation and the centroid. It also applies to the velocity effect on the third-order contribution to the differential time delay.

\begin{acknowledgments}
G.H. would like to thank Prof. Yungui Gong for guide and support when some of the work was carried out at Huazhong University of Science and Technology. This work was supported in part by the National Natural Science Foundation of China under Grant Nos. 11973025 and 12147208.
\end{acknowledgments}

\end{document}